\documentclass[12pt,aps,showpacs]{revtex4}
\usepackage{amsmath,amssymb,amsfonts,epsfig}
\newcommand{\beeq}{\begin{equation}}
\newcommand{\eneq}{\end{equation}}
\newcommand{\be}{\begin{eqnarray}}
\newcommand{\ee}{\end{eqnarray}}
\newcommand{\bpic}{\begin{picture}}
\newcommand{\epic}{\end{picture}}
\newcommand{\bs}{\begin{scriptsize}}
\newcommand{\es}{\end{scriptsize}}

\def\dd{\partial}
\def\la{\raise.16ex\hbox{$\langle$} \, }
\def\ra{\, \raise.16ex\hbox{$\rangle$} }

\def\l{\lambda}

\def\Box{\kern1pt\vbox{\hrule height 1.2pt\hbox{\vrule width 1.2pt\hskip 3pt
   \vbox{\vskip 6pt}\hskip 3pt\vrule width 0.6pt}\hrule height 0.6pt}\kern1pt}
\def\gtwid{\mathrel{\raise.3ex\hbox{$>$\kern-.75em\lower1ex\hbox{$\sim$}}}}
\def\ltwid{\mathrel{\raise.3ex\hbox{$<$\kern-.75em\lower1ex\hbox{$\sim$}}}}

\begin{document}


\title{Quantum Effects of Mass\\ on\\ Scalar Field Correlations and Fluctuations\\ during Inflation}

\author{G. Karakaya}\email{gulaykarakaya@itu.edu.tr}

\author{V. K. Onemli}\email{onemli@itu.edu.tr}

\affiliation{Department of Physics, Istanbul Technical
University, Maslak, Istanbul 34469, Turkey}

\begin{abstract}
We consider an infrared truncated massive minimally coupled scalar field with a quartic self-interaction in the locally de Sitter background of an inflating universe. We compute the two-point correlation function of the scalar and the mean squared fluctuations (variance) of the field variation analytically, at tree, one- and two-loop order. The one-loop correlator at a fixed comoving separation asymptotes to zero in the massive case but grows, at late times, like $-\lambda\ln^2(a)$ in the massless limit, where $a$ is the cosmic scale factor. For a fixed physical distance, on the other hand, it grows, at late times, like $-\lambda\ln^3(a)$ in the massless limit. This growth is severely suppressed in the massive case. In fact, the one-loop correlator asymptotes effectively to zero for masses larger than half the expansion rate. We find out also that the tree-order variance of field variation decreases when quantum corrections are included. Hence, the actual effect that any local observer perceives in the field strength as fluctuations happen does not deviate from the average effect as much as the tree-order variance implies.

\end{abstract}


\maketitle \vskip 0.2in \vspace{.4cm}

\section{Introduction}
\label{sec:intro}

Two point correlation function and the mean squared fluctuations, i.e. the variance, of field variation of an infrared (IR) truncated {\it massless} minimally coupled (MMC) scalar field with a quartic self interaction on a locally de Sitter background of an inflating spacetime were computed at one- and two-loop order in Ref.~\cite{vacuum}. In this paper, we study how a mass insertion alters those results analytically. Our motivations for choosing the model and the formalism we apply to study it are given below.

The mechanism for large scale structure formation is the amplification of primordial vacuum fluctuations of a scalar quantum field by the inflationary dynamics of spacetime in the early universe. It is therefore essential to study the vacuum fluctuations to identify and understand the origin of structure formation. Quantum fields, by their quantum nature, fluctuate. The fluctuations correspond to virtual particle-antiparticle pairs continuously emerging from the vacuum, persisting for a time $\Delta t$, and then annihilating. Persistence time $\Delta t$ is governed by the energy-time uncertainty principle which gives the minimum time $\Delta t$ to resolve an energy change $\Delta E$, that is $\Delta E\Delta t \gtrsim 1$. Emergence of a particle-antiparticle pair, each of mass $m$ and wave vector $\pm\vec{k}$ (to conserve momentum), out of vacuum implies a change of energy from zero to $E\!\!=\!\!2\sqrt{m^2\!+\!k^2}$, where the wave number $k\!=\!\|\vec{k}\|\!\!=\!\!\sqrt{\vec{k}\!\cdot\!\vec{k}}$ is defined in terms of wavelength $\lambda$ as $k\!\!=\!\!2\pi/\lambda$.  Hence, {\it to not resolve} such a violation of energy conservation, the uncertainty principle demands that $\Delta E\Delta t\! \lesssim\! 1$. Therefore, a virtual particle that emerges from the vacuum should annihilate in $\Delta t\!\!\lesssim\! 1/\sqrt{m^2\!+\!k^2}$. Any quantum effect can be considered \cite{WR1,WR2,WR3} as a classical response to the virtual particles with a persistence time $\Delta t$. The lightest virtual particles with the largest wavelengths have the longest persistence time. They induce greatest field strength and, therefore, cause the strongest quantum effects.

In an expanding spacetime, energy of a particle becomes time dependent,
$E(t, \vec{k})\!\!=\!\!\sqrt{m^2\!+\!k^2/a^2(t)}$, where $a(t)$ is the cosmic scale factor. Therefore, a virtual particle of comoving wave number $k$ which pops up out of vacuum at time $t$ can persist \cite{WR3} for a time $\Delta t$ such that
\be
\int_t^{t+\Delta t}\!\!E(t'\!, \vec{k})\,dt'\!=\!\int_t^{t+\Delta t}\!\!\sqrt{m^2\!+\!\frac{k^2}{a^2(t')}}\,dt'\!\lesssim\!1\; .\nonumber
\ee Thus, in general, cosmic expansion causes the virtual particles to persist longer and therefore enhances their quantum effects. For fields with $m\!\!\neq\!\!0$ the integral grows with $\Delta t$ and eventually the inequality is violated. Thus, massive virtual particles have finite persistence time. The lighter the particle, the longer the $\Delta t$. Just as in flat spacetime, massless virtual particles that emerge with sufficiently small wave numbers can persist forever and therefore have the strongest quantum effects. This is most easily seen \cite{WR3} considering the energy-time uncertainty principle for virtual particles during de Sitter phase of inflation where the expansion rate $H$ is a constant,
\beeq
\lim_{m\rightarrow 0}\!\int_t^{t+\Delta t}\!\!E(t'\!, \vec{k})\,dt' \!=\! k\!\!\int_t^{t+\Delta t} \!\! \frac{dt'}{a(t')}\!=\!\frac{k}{Ha(t)}\!\left[1\!\!-\!e^{-H\Delta t}\right]\!\lesssim\!1\; .\nonumber
\eneq
Thus, for modes with $k\!\!<\!\!\!Ha(t)$, called super-horizon (IR) modes, the inequality is not violated even for persistence time $\Delta t\!\!\rightarrow\!\!\infty$; hence they engender strong quantum effects. Inflation continuously shifts many sub-horizon modes to the super-horizon scales stretching their physical wavelengths beyond the horizon size $1/H$. An IR truncated quantum field comprising of the modes whose physical wavelengths longer than the physical horizon size can be obtained from the free field mode expansion by introducing a time dependent cutoff via a convenient dynamical window function. Starobinsky obtains \cite{Star,StarYok} an IR truncated scalar free field $\bar\varphi_0(t,\vec{x})$ by introducing the Heaviside step function $\Theta(Ha(t)\!-\!k)$ in Fourier space and uses it to compute leading quantum contributions---at each perturbative order---to some observables in scalar potential models during inflation. The fluctuations on super-horizon scales ''freeze in'' after the modes exit the horizon and can be considered as classical. After the end of inflationary phase, the super-horizon modes start reentering the horizon as the physical horizon size successively reaches their physical wavelengths. The modes that reenter the horizon behaves quantum mechanically and play the key role in seeding the large scale structure formation in the early universe. The formalism was extended \cite{W1} to include scalars interacting with fermions \cite{W3,W4} and with gauge bosons \cite{Wstocsqed}. Inclusion of interactions which involve derivatives such as those of quantum gravity \cite{Wstocqgrav,MiaoWood} or of the nonlinear sigma model \cite{Kit1,Kit2} is still an unsolved problem.  We follow Starobinsky's formalism in this paper.

Being massless, however, is not a sufficient condition for the virtual particles to engender significant quantum effects. Most massless fields possess conformal symmetry which suppresses the emission rate of virtual particles by $1/a(t)$. Their number density remains small during inflation. Gravitons and MMC scalars, however, are the only two exceptional fields that are both massless and conformally noninvariant. Hence, their field strengths grow during inflation and can cause significant quantum effects. In fact, the observed scalar \cite{MC} and potentially observable tensor \cite{AS1} perturbations
are the amplified imprints of virtual inflatons and gravitons on the cosmic microwave anisotropy, respectively.

Fluctuations of interacting scalar fields in an inflating spacetime have been the focus of cosmologists \cite{sfl}. Recently, there has been a revival of interest on IR dynamics of scalar potential models with various approaches that include extending the stochastic formalism \cite{TT}, applying complementary series analysis \cite{AMPP}, computing effective actions \cite{MR,R1,DB1,CW1}, using Schwinger-Keldish formalism \cite{CWX}, implementing Fokker-Planck equation and $\delta N$ formalism \cite{AFNVW}, employing $1/N$ expansion \cite{GS1}, adopting reduced density matrix method \cite{DB2}, applying renormalization group analysis \cite{GS2} and computing effective potentials \cite{JSS}. Influence of fermions on scalar field fluctuations has been studied in Refs~\cite{O2,DB3}.

As pointed out earlier, following Starobinsky's approach \cite{Star, StarYok} and applying the techniques of perturbative quantum field theory, we consider, in this paper, an IR truncated {\it massive} minimally coupled scalar with a quartic self-interaction and study the effect of mass on the two-point correlation function and on the fluctuations of field variation. The model is of interest because it exhibits, in the massless limit, peculiar enhanced quantum effects: the renormalized energy density and pressure of the scalar violate \cite{OW1,OW2} the weak energy condition on cosmological scales at two-loop order and a phase of superacceleration is induced. As the inflationary particle production amplifies the field strength and therefore forces the scalar up its potential, the scalar develops \cite{BOW} a positive self-mass squared which, in turn, reduces the particle production. Furthermore, the classical restoring force pushes the scalar back down to the configuration where the potential is minimum. Thus, the scalar cannot continue to roll up its position and comes to a halt eventually. The process, therefore, is self-limiting and the model is stable \cite{KO}. For many quantum field theory computations in cosmology higher order quantum corrections necessarily involve changes in the initial state. Neglecting to correctly change the initial state can result in effective field equations that diverge on the initial value surface. The model provides \cite{KOW1} an example of how perturbative initial state corrections can absorb the initial value divergences. Moreover, the scalar makes \cite{KOW2} a time dependent contribution to the curvature power spectrum at two-loop order. The scalar's power spectrum itself is slightly red-tilted \cite{O1} hence the amplitudes of fluctuations grow toward the larger scales.

Two-point correlation function, $\langle\Omega|
\bar{\varphi}(t,\vec{x})\bar{\varphi}(t'\!,\vec{x}\,')|\Omega\rangle$, that we compute in this paper is a vacuum expectation value (VEV) hence, an average measure of how the amplitude of the field at one event (spacetime coordinate) is correlated with the amplitude at another event. As the field strength grows, due to the inflationary particle production, so does the two-point correlation function. Because the production process is self-limiting, as pointed out, the correlation function ought not grow forever. In fact, the growth is logarithmic in the massless limit, just as expected. Because the mass suppresses the particle production, the logarithmic growth ought to be suppressed in the massive case. As mass increases the suppression must get stronger; an expectation proven to be true in this paper.

Field strength is positive at some events; at others it is negative. As the field fluctuates in spacetime, the magnitude of field strength increases at some locations; at others it decreases. An expectation value is merely a measure of an average effect. The actual effect perceived \cite{W0, Tegmark} by any local observer is not an average value but either an increase or a decrease in the field strength. The mean squared deviation of field variation $\Delta \bar{\varphi}\!\!\equiv\!\!
\bar{\varphi}(t, \vec{x})\!-\!\bar{\varphi}(t'\!, \vec{x}\,')$ from its average $\langle\Omega|\Delta\bar{\varphi}|\Omega\rangle$, i.e. the variance $\sigma^2_{\Delta \bar{\varphi}}\!\!\equiv\!\langle\Omega|\!
\left[\Delta\bar{\varphi} \!-\!\langle\Omega|\Delta\bar{\varphi}|\Omega\rangle\right]^2\!|\Omega\rangle$ of field variation, provides information about the magnitude of actual effect. We use our two-point correlation function to compute the $\sigma^2_{\Delta \bar{\varphi}}$ for the IR truncated massive scalar in our model at tree and one-loop order and examine how the tree-order contrast between the VEV and the variance change when the quantum corrections are included.

The outline of the paper is as follows. In Sec.~\ref{sec:model} we present the background geometry and the Lagrangian of the model. In Sec.~\ref{sec:Quantum} we analyze the model following Starobinsky's approach and applying the techniques of pertubative quantum field theory. We compute the two-point correlation function of the IR truncated massive scalar in our model at tree, one- and two-loop order in Sec.~\ref{sec:twopointcorrelator}. Using the two-point correlation function, we obtain the variance of field variation at tree and one-loop order in Sec.~\ref{sec:variance}. We summarize our conclusions in Sec.~\ref{sec:conclusions}. The Appendixes comprise the details of some computations.

\section{The Model}
\label{sec:model}

We consider a massive, minimally coupled, self-interacting spectator scalar field in an inflating spacetime. The Lagrangian density of the model,
\begin{equation}
\mathcal{L} \!=\! -\frac{1}{2}\, \partial_{\mu} \phi \, \partial_{\nu}
\phi\, g^{\mu\nu}\sqrt{-g}\!-\!\frac{1}{2}m_0^2\phi^2\sqrt{-g} \!-\!\! V(\phi) \sqrt{-g} \; ,\label{lagden}
\end{equation}
where $\phi(x)$ represents the bare field and $m_0$ denotes the bare mass. The $V(\phi)$ is specified as the quartic self-interaction potential and the $g_{\mu\nu}$ is chosen as the metric of a locally de~Sitter background. The $g$ stands for the determinant of the metric. The invariant
line element \beeq ds^2\!\!=\! g_{\mu\nu} dx^{\mu}
dx^{\nu}\!\!=\!-dt^2\!\!+\!a^2(t) d\vec{x} \cdot d\vec{x}
\; ,\label{desitter}\eneq where the scale factor with a constant expansion rate $H$, \beeq a(t)\!=\!e^{H t} \; ,\eneq is normalized to unity at initial comoving time $t\!=\!t_I\!\!=\!0$. We work in a $D$-dimensional spacetime and use the open conformal coordinate patch of the full de Sitter manifold where the spatial coordinates range as $0\!\leq\!x^i\!\leq\!H^{-1}$, $i\!=\!1,2, \dots,(D\!-\!1)$. We adopt the convention where
a Greek index $\mu\!=\!0,1,2, \dots,(D\!-\!1)$, hence $x^\mu\!=\!(x^0\!,\vec x)$, $x^0\!\equiv\!t$, and $\partial_\mu\!=\!(\partial_0,\vec\nabla)$. In the next section, we study this model applying quantum field theory.

\section{Quantum Field Theoretical Analysis}
\label{sec:Quantum}

We first obtain the renormalized Lagrangian density of the model. The renormalized field $\varphi(x)$ is defined by the field strength renormalization of the bare field $\phi(x)$ as $\varphi(x)\equiv\frac{1}{\sqrt{Z}}\phi(x)$. This brings Lagrangian density~(\ref{lagden}) with a quartic self-interaction to the form\beeq \mathcal{L} \!=\! -\frac{1}{2}Z\,
\partial_{\mu} \varphi\,
\partial_{\nu} \varphi\, g^{\mu\nu}\sqrt{-g}\!-\!\frac{1}{2}m^2_0 Z
\varphi^2\sqrt{-g}\!-\!\frac{1}{4!}\lambda_0 Z^2\varphi^4 \sqrt{-g} \; .
\eneq
Expressing the bare mass squared $m_0^2$ and bare coupling strength $\lambda_0$ in terms of the renormalized parameters as
\beeq
m^2_0 Z\equiv m^2 Z \!+\!\delta
m^2\;\;\; {\rm and} \;\;\; \lambda_0 Z^2 \!\equiv\! \lambda\!+\!\delta \lambda \;
,\eneq
where $\delta m^2$ and $\delta \lambda$ are respectively the mass renormalization and coupling strength renormalization counterterms,
one obtains\beeq \mathcal{L} \!=\! -\frac{1}{2}Z\!\left[
\partial_{\mu} \varphi\,
\partial_{\nu} \varphi\, g^{\mu\nu}\!\!+\!m^2\varphi^2\right]\!\!\sqrt{-g}\!-\!\frac{1}{2}\delta m^2
\varphi^2\sqrt{-g}\!-\!\frac{1}{4!}(\lambda\!+\!\delta\lambda) \varphi^4 \sqrt{-g}  \; .
\eneq
Finally, defining
\beeq
Z\!\equiv\! 1\!\!+\!\delta Z \; ,\eneq where $\delta Z$ is the field strength renormalization counterterm,
the renormalized Lagrangian density is obtained as\beeq
\hspace{-0.6cm}\mathcal{L} \!=\! -\frac{1}{2}\left(1\!\!+\!\delta Z\right)\!\left[
\partial_{\mu} \varphi\,
\partial_{\nu} \varphi\, g^{\mu\nu}\!\!+\!m^2\varphi^2\right]\!\!\sqrt{-g}\!-\!\!V(\varphi)\sqrt{-g}  \; ,
\label{fulLag}\eneq
where the renormalized potential\beeq
V(\varphi)\!=\!\frac{1}{2}\delta m^2
\varphi^2\!+\!\frac{1}{4!}(\lambda\!+\!\delta\lambda) \varphi^4\; .\label{potential}
\eneq Varying the
Lagrangian with density (\ref{fulLag}) yields the scalar field
equation \beeq \dd_{\mu} \Bigl(\sqrt{-g} g^{\mu\nu} \dd_{\nu}
\varphi\Bigr)\!\!-\!\sqrt{-g}\,m^2\varphi\!=\!\sqrt{-g}\,\frac{V'(\varphi)}{1\!\!+\!\delta Z}\; . \label{phieqn} \eneq Here, prime denotes derivative with respect to the argument. Solution of Eq.~(\ref{potential}) can be given as
\beeq \varphi(x)\!=\!\varphi_0(x)\!+\!\frac{1}{1\!\!+\!\delta
Z}\!\int_{t_I}^t
\!\!\!dt'\!\sqrt{-g(t')}\!\int\!\!d^{D-1}x' \,
G(x;x')V'(\varphi(x')) \; , \label{phix} \eneq where
$\varphi_0(x)$ is the solution for the homogeneous equation \beeq
\partial_{\mu} \Bigl(\sqrt{-g} g^{\mu\nu} \partial_{\nu}\varphi_0(x)
\Bigr)\!\!-\!\sqrt{-g}m^2\varphi_0(x)\!=\!0\; , \label{phi0}\eneq and the Green's function
$G(x;x')$ is any solution of the equation
\begin{equation}
\partial_{\mu} \Bigl(\sqrt{-g} g^{\mu\nu} \partial_{\nu} G(x;x') \Bigr)\!\!-\!\sqrt{-g}m^2 G(x;x')
\!=\! \delta^D(x \!-\! x') \; ,
\end{equation} which obeys retarded boundary conditions. Specializing to de Sitter metric~(\ref{desitter}) and interaction potential~(\ref{potential}), field equation~(\ref{phieqn}) leads to
\be
\hspace{-0.3cm}\ddot{\varphi}(t,\vec{x})\!+\!(D\!-\!1) H \dot{\varphi}(t,\vec{x})\!-\!\!
\left[\frac{\nabla^2}{a^2}-\!m^2\right]\!\!\varphi(t,\vec{x})\!=\!-\frac{1}{1\!\!+\!\delta Z}\!\left[\delta m^2
\varphi(t,\vec{x})\!+\!\frac{1}{6}(\lambda\!+\!\delta\lambda) \varphi^3(t,\vec{x})
\right]\; , \label{feq}
\ee
where an overdot denotes derivative with respect to comoving time $t$. It has the solution
\be
\hspace{-0.4cm}\varphi(t,\vec{x})\!=\!\varphi_0(t,\vec{x})
-\!\frac{1}{1\!\!+\!\delta Z}\!\int_0^t\!\!dt'\!\!\int\!d^{D-1}x'G(t,
\vec{x};t'\!, \vec{x}\,')\!\left[\delta m^2
\varphi(t'\!,\vec{x}\,')\!+\!\frac{1}{6}(\lambda\!+\!\delta\lambda) \varphi^3(t'\!,\vec{x}\,')
\right]\; ,
\label{fullfield}
\ee
where the free field $\varphi_0(t, \vec{x})$ in Eq.~(\ref{fullfield}) obeys
the linearized field equation,
\beeq
\ddot{\varphi}_0(t,\vec{x})\!+\!(D\!-\!1) H \dot{\varphi}_0(t,\vec{x})\!-\!\!
\left[\frac{\nabla^2}{a^2}-\!m^2\right]\!\!\varphi_0(t,\vec{x})\!=\!0 \; , \label{freefieldeqposition}
\eneq
and agrees with the full field
at $t\!=\!t_I\!=\!0$. Equation~(\ref{freefieldeqposition}) has a well known solution in terms of Hankel functions of the first and second kind which we use in Sec.~\ref{subsect:freethry}.

The Green's function ${G}(t,
\vec{x};t'\!, \vec{x}\,')$ in Eq.~(\ref{fullfield}), on the other hand, solves of the linearized field equation with a Dirac-delta source term
\beeq
\ddot{G}(t,
\vec{x};t'\!, \vec{x}\,')\!+\!(D\!-\!1)H \dot{G}(t,
\vec{x};t'\!, \vec{x}\,')\!-\!\!
\left[\frac{\nabla^2}{a^2}-\!m_0^2\right]\!G(t,
\vec{x};t'\!, \vec{x}\,')\!=\!\delta(t\!-\!t')\delta^{D-1}(\vec{x}\!-\!\vec{x}\,')\; ,\label{greeneqn}\eneq
and satisfies the retarded boundary conditions. The commutator function $\left[\varphi_0(t,\vec{x}), \varphi_0(t'\!,\vec{x}\,')\right]$ furnishes a convenient representation for the Green's function,\beeq
G(t,\vec{x};t'\!, \vec{x}\,')\!=\!i\Theta(t\!-\!t')\left[\varphi_0(t,\vec{x}), \varphi_0(t'\!,\vec{x}\,')\right]\; .\label{commutatorgreen}
\eneq
Equation~(\ref{commutatorgreen}) is used to compute the Green's function and its IR limit in Sec.~\ref{subsect:interactingtheory}.

One can iterate Eq.~(\ref{fullfield}) to obtain the full field in terms of the free field and the Green's function, perturbatively. The first iteration, for example, yields
\be
&&\hspace{0.3cm}\varphi(t,\vec{x})\!=\!\varphi_0(t,\vec{x})
\!-\!\frac{1}{1\!\!+\!\delta Z}\!\!\int_0^t\!\!dt'\!\int\!d^{D-1}x'G(t,
\vec{x};t'\!, \vec{x}\,')\Bigg\{\delta m^2\Bigg[\varphi_0(t'\!,\vec{x}\,')\!-\!\frac{1}{1\!\!+\!\delta Z}\nonumber\\
&&\hspace{-1cm}\times\!\!\int_0^{t'}\!\!dt''\!\!\!\int\!d^{D-1}x''G(t'\!,
\vec{x}\,';t''\!, \vec{x}\,'')\!\left[\delta m^2\varphi(t''\!,\vec{x}\,'')\!+\!\frac{1}{6}(\lambda\!+\!\delta\lambda)\varphi^3(t''\!,\vec{x}\,'')\right]\!\!\Bigg]
\!\!+\!\frac{(\lambda\!+\!\delta\lambda) }{6}\!\Bigg[\!\varphi_0(t'\!,\vec{x}\,')\nonumber\\
&&\hspace{0.1cm}
\!-\frac{1}{1\!\!+\!\delta Z}\!\!\int_0^{t'}\!\!dt''\!\!\!\int\!d^{D-1}x''G(t'\!,
\vec{x}\,';t''\!, \vec{x}\,'')\!\left[\delta m^2\varphi(t''\!,\vec{x}\,'')\!+\!\frac{1}{6}(\lambda\!+\!\delta\lambda)\varphi^3(t''\!,\vec{x}\,'')\right]\!\Bigg]^3\Bigg\}
\; .
\label{iterate}
\ee
Successive iterations give the higher order corrections to the free field, producing the full field as an expansion. We obtain the mode expansion of the free massive scalar in Sec.~\ref{subsect:freethry} and use it to study the interacting theory in Sec.~\ref{subsect:interactingtheory}.

\subsection{Free Theory}
\label{subsect:freethry}

As pointed out earlier, we work on a conformal coordinate patch with a topology $\mathcal{T}^{D-1}\!\times\!\Re$. Because the manifold is spatially compact, the modes of fields are discrete. Thus, we define the wave vector as
\beeq\vec{k}\!\equiv\! 2\pi H\vec{n}\;,\label{wavevector}\eneq where $\vec{n}$ is a $(D\!\!-\!\!1)$-dimensional vector with integer components. Hence, the comoving wavenumber \beeq
k\!\equiv\!\|\vec{k}\|\!=\!2\pi H\|\vec{n}\|\; .\label{wavenumber}\eneq To express the scalar $\varphi_0(t,\vec{x})$ as a sum over its modes we, therefore, expand it in a spatial Fourier series.

Spatially Fourier transforming the linearized field equation~(\ref{freefieldeqposition}), on one hand, yields \be
{\widetilde{\ddot{\varphi}}}_0(t,\vec{k})\!+\!(D\!-\!1) H {\widetilde{\dot{\varphi}}}_0(t,\vec{k})\!+\!\!
\left[\frac{k^2}{a^2}\!+\!m^2\right]\!\!{\widetilde{\varphi}}_0(t,\vec{k})\!=\!0 \; ,\label{Fouriertransfieldeq}
\ee
and the canonical quantization condition $[{\varphi}_0(t,\vec{x}), \Pi_0(t,\vec{x}\,')]\!=\!i\delta(\vec{x}\!-\!\vec{x}\,')$, where  ${\Pi}_0(t,\vec{x}\,')\!=\!a^{D-1}(t){\dot{\varphi}}_0(t,\vec{x}\,')$ is the
conjugate momentum, on the other hand, implies
\be
[{\widetilde{\varphi}}_0(t,\vec{k}), {\widetilde{\dot{\varphi}}}_0(t,\vec{k}\,')]\!=\!\frac{i\delta_{\vec{n},-\vec{n}\,'}}{a^{D-1}(t)H^{D-1}}\; ,\label{quant}\ee
in mode space. To express ${\widetilde{\varphi}}_0(t,\vec{k})$ satisfying Eq.~(\ref{quant}), in terms of the amplitude for each mode $\vec{k}$ of the field---the so called mode function---and the mode coordinates, we must solve Eq.~(\ref{Fouriertransfieldeq}).
To solve it, however, it is necessary to distinguish between the zero and nonzero modes. First, let us consider the case of zero mode.

\subsubsection{The zero mode solution in momentum space}

Equation~(\ref{Fouriertransfieldeq}) for $\vec{k}=0$,
\be
{\widetilde{\ddot{\varphi}}}_0(t,0)\!+\!(D\!-\!1) H {\widetilde{\dot{\varphi}}}_0(t,0)\!+\!m^2{\widetilde{\varphi}}_0(t,0)\!=\!0 \; ,\label{eqnofmotkzero}
\ee
has two linearly independent solutions,
\be
u_1(t, 0)\!=\!a^{-\frac{D-1}{2}+\nu}(t) \;\;{\rm{and}}\;\; u_2(t, 0)\!=\!\frac{-i}{2\nu H}a^{-\frac{D-1}{2}-\nu}(t)\; ,
\ee
where
\be
\nu\!\equiv\!\sqrt{\left(\!\frac{D\!-\!1}{2}\!\right)^2\!\!\!\!-\!\!\left(\frac{m}{H}\right)^2}\;\; {\rm with} \;\; \frac{m}{H}\!\leq\!\frac{D\!-\!1}{2} \; ,\label{nu}
\ee
obeying the Wronskian normalization
\be
u_1(t, 0)\dot{u}_2(t, 0)\!-\!u_2(t, 0)\dot{u}_1(t, 0)\!=\!\frac{i}{a^{D-1}(t)} \; .\label{Wronskianzero}
\ee

In momentum space, the zero mode of the free field is represented in terms of the zero mode functions $u_1(t,0)$ and $u_2(t,0)$ and the self-adjoint zero mode coordinates $\hat{Q}$ and $\hat{P}$ as
\be
{\widetilde{\varphi}}_0(t,0)\!=\!H^{-\frac{D-1}{2}}\!\left[\hat{Q}u_1(t, 0)\!-\!i\hat{P}u_2(t, 0)\right]\; .
\ee
Canonical commutation relation~(\ref{quant}) and Wronskian normalization~(\ref{Wronskianzero}), together, imply
\be
[\hat{Q}, \hat{P}]\!=\!i\;.
\ee
Thus, the zero mode field in momentum space is
\be
\hspace{0.4cm}{\widetilde{\varphi}}_0(t,0)\!=\!H^{-\frac{D-1}{2}}\!\!
\left[\frac{\hat{Q}}{a^{\frac{D-1}{2}-\nu}(t)}\!-\!\frac{\hat{P}}{2\nu Ha^{\frac{D-1}{2}+\nu}(t)}\right]\; .\label{zeromodeinmom}
\ee
The operators $\hat{Q}$ and $\hat{P}$ can be expressed in terms of the initial values of the spatial Fourier transforms of the field and its first time~derivative for $\vec{k}\!=\!0$,
\be
\hat{Q}\!\!&=&\!\!H^{\frac{D-1}{2}}\!\Bigg[{{\widetilde{\varphi}}}_0(0,0)
\frac{1}{2}\!\left(\!1\!+\!\frac{D\!-\!1}{2\nu}\!\right)
\!\!+\!\frac{\widetilde{\dot\varphi}_0(0, 0)}{2\nu H}\Bigg]\; ,\\
\hat{P}\!\!&=&\!\!{H^{\frac{D-1}{2}}}\!\!
\left[{{\widetilde{\varphi}}}_0(0,0)\!\left(\!\frac{D\!-\!1}{2}\!-\!\nu\!\right)\!\!H
\!+\!\widetilde{\dot\varphi}_0(0,0)\right]\; .
\ee
The analysis of nonzero modes follows similar steps as we outline in the next section.

\subsubsection{The nonzero mode solution in momentum space}

Two linearly independent solutions of Eq.~(\ref{Fouriertransfieldeq}) for the nonzero modes ($\vec{k}\!\neq\! 0$) are well known. They can be given in terms of the Hankel function of the first kind $\mathcal{H}^{(1)}_{\nu}$ and its complex conjugate. The solution obeying the Wronskian normalization
\beeq
u(t,k)\,\dot{u}^*(t,k)\!-\!u^*(t,k)\,\dot{u}(t,k)\!=\!\frac{i}{a^{D-1}(t)}\; ,\label{Wronskiannonzero}
\eneq
is known as the Bunch-Davies mode function
\beeq
u(t, k) \!=\! i\sqrt{\frac{\pi}{4Ha^{D-1}(t)}}\,
\mathcal{H}^{(1)}_{\nu}\! \Bigl(\frac{k}{H a(t)}\Bigr)
\; .
\label{udef}\eneq
It is the amplitude function of the normal mode with wave vector $\vec{k}$.

In momentum space, a nonzero mode of the free field is represented in terms of the mode function $u(t,k)$, its complex conjugate, and mode coordinates $\hat{A}_{\vec{n}}$ and $\hat{A}^\dagger_{-\vec{n}}$ as
\be
{\widetilde{\varphi}}_0(t,\vec{k}\!\neq\! 0)\!=\!H^{-\frac{D-1}{2}}\!\left[u(t,k)\hat{A}_{\vec{n}}\!+\!u^*(t, k)\hat{A}^\dagger_{-\vec{n}}\right]\; .\label{nonzeromodeinmom}
\ee
Canonical commutation relation~(\ref{quant}) and Wronskian normalization~(\ref{Wronskianzero}) together imply that the operators $\hat{A}_{\vec{n}}$ and $\hat{A}^\dagger_{\vec{n}}$ satisfy the commutation relation
\be
\left[\hat{A}_{\vec{n}} ,
\hat{A}^{\dagger}_{\vec{m}} \right] \!\!=\! \delta_{\vec{n} , \vec{m}} \;.\label{normalaadagger}
\ee
They can be expressed in terms of the initial values of the spatial Fourier transforms of the field and its first time~derivative,
\be
\hat{A}_{\vec{n}}\!\!&=&\!\!-iH^{\frac{D-1}{2}}\!\left[\dot{u}^*(0, k)\widetilde{\varphi}_0(0, \vec{k})\!-\!u^*(0, k)\widetilde{\dot\varphi}_0(0, \vec{k})\right]\; ,\nonumber\\
\hat{A}^\dagger_{\vec{n}}\!\!&=&\!\!iH^{\frac{D-1}{2}}\!\left[\dot{u}(0, k)\widetilde{\varphi}_0(0, -\vec{k})\!-\!u(0, k)\widetilde{\dot\varphi}_0(0, -\vec{k})\right]\; .
\ee
Note that the state $\vert \Omega \rangle$ annihilated by all $\hat{A}_{\vec{n}}$
is known as Bunch-Davies~vacuum. Next, we discuss the free field expansion in position space.

\subsubsection{The free field expansion in position space}

Inverse Fourier transforming Eqs.~(\ref{zeromodeinmom}) and (\ref{nonzeromodeinmom}) back to position space and combining the outcomes we obtain the free field mode expansion as
\be
\hspace{-0.3cm}\varphi_0(t,\vec{x})\!=\!H^{\frac{D-1}{2}}\!\Bigg\{\!\frac{\hat{Q}}{a^{\frac{D-1}{2}-\nu}(t)}
\!-\!\frac{\hat{P}}{2\nu Ha^{\frac{D-1}{2}+\nu}(t)}\!+\!\!\sum_{\vec{n} \neq 0} \!\left[ u(t,k) e^{i \vec{k} \cdot \vec{x}}\!
\hat{A}_{\vec{n}} \!+\! u^*(t,k) e^{-i \vec{k} \cdot \vec{x}}\! \hat{A}^{\dagger}_{\vec{n}}
\right]\!\!\Bigg\} \; . \label{freefieldinposspace}
\ee

Free field expansion~(\ref{freefieldinposspace}) includes arbitrarily large wave numbers~(\ref{wavenumber}).
Recall that in any phase of an expanding universe physical size (or any given comoving scale) grows as the scale factor $a(t)$. An interesting fact about the inflationary phase of expansion that distinguishes it from the other phases of expansion is that the scale factor $a(t)$, hence the physical size, grows more rapidly than the horizon size. In fact, the scale factor grows exponentially whereas the horizon size remains constant during inflation. Hence, at comoving time $t\!=\!t_k$ the scale with comoving wave number $
k\!=\!2\pi H(t_k)\,a(t_k)\!=\!2\pi He^{Ht_k}$
exits the horizon. In other words, the mode $k$ is a subhorizon (ultraviolet) mode when $t\!\!<\!t_k$ but becomes a superhorizon (IR) mode at $t\!=\!t_k$.
During the other phases of expansion (radiation domination and matter domination) though, the horizon size
grows more rapidly than the scale factor $a(t)$, hence than the physical size. Thus, the comoving scales that exit the horizon during inflation reenter in a later phase of expansion. The scales that leave the horizon latest reenter earliest. To study the IR physics during inflation, one can cut out~\cite{Star,StarYok} the ultraviolet modes with wave number $k\!>\!Ha(t)$ in mode expansion~(\ref{freefieldinposspace}) by introducing a dynamical Heaviside step function $\Theta$ in Fourier space,
\beeq
\varphi_0(t,\vec{x}) \!=\! H^{\frac{D-1}{2}}\! \sum_{\vec{n} \neq 0} \Theta(Ha(t)\!-\!k)\!\left[ u(t,k) e^{i \vec{k} \cdot \vec{x}}
\hat{A}_{\vec{n}} \!+\! u^*(t,k) e^{-i \vec{k} \cdot \vec{x}} \hat{A}^{\dagger}_{\vec{n}}
\right]\; .\label{truncexpwithu}
\eneq
We neglected the zero mode which is just one mode as opposed to ever increasing number of nonzero modes. In Eq~(\ref{truncexpwithu}), one can further take the IR limit of the mode function~(\ref{udef}) given in terms of the Hankel function\beeq
\mathcal{H}^{(1)}_{\nu}(z)\!=\!{J}_{\nu}(z)\!+\!i{Y}_{\nu}(z)\; .\label{HankelinBessel}
\eneq
Here $z\!\equiv\!\frac{k}{H a(t)}$, ${J}_{\nu}(z)$ is the Bessel function of order $\nu$ and
\beeq
{Y}_{\nu}(z)\!=\!
{J}_{\nu}(z)\cot(\nu\pi)
\!-\!{J}_{-\nu}(z)\csc(\nu\pi)\; ,\label{Neumann}
\eneq
is the Neumann function of order $\nu$. The right side of Eq.~(\ref{Neumann}) is replaced by its limiting value if $\nu$ is an integer or zero. Using the series expansion of the Bessel function\beeq
{J}_{\nu}\!(z)\!=\!\left(\frac{z}{2}\right)^\nu\!\!\frac{1}{\nu\,\Gamma(\nu)}\!
\left[1\!-\!\frac{1}{\nu\!+\!\!1}\left(\frac{z}{2}\right)^2
\!\!\!+\!\mathcal{O}(z^4)\right]\; ,\label{BesselSeries}
\eneq
the reflection formula for the Gamma function
\beeq
\pi\csc(p\pi)\!=\!\Gamma(1\!-\!p)
\Gamma(p)\; ,\label{gammareflection}
\eneq
for $p\!=\!\nu$ and $p\!=\!\nu\!+\!\frac{1}{2}$, and the identity
\beeq
\Gamma(\nu)
\!=\!\frac{\sqrt{\pi}}{2^{2\nu-1}}\frac{\Gamma(2\nu)}{\Gamma(\nu\!+\!\frac{1}{2})}\; ,\label{gammaidentity}
\eneq
one gets the leading IR limit of the mode function
\beeq
u(t, k)\longrightarrow\frac{1}{\sqrt{Ha^{D-1}(t)}}
\frac{\Gamma(2\nu)}{\Gamma(\nu\!+\!\frac{1}{2})}  \left(\!\frac{2k}{Ha(t)}\!\right)^{-\nu}\!\!\Biggl\{\!1\!\!+\!
\mathcal{O}\left(\Bigl(\frac{k}{H a(t)}\Bigr)^{2\beta}\right)\!\Biggr\}\; ,\label{modeleadingorder}
\eneq
where $\beta\!=\!\nu$ if $\frac{1}{2}\sqrt{(D\!-\!3)(D\!+\!1)}\!<\!\frac{m}{H}\!<\!\frac{1}{2}(D\!-\!1)$ or $\beta\!=\!1$ if $\frac{m}{H}\!<\!\!\frac{1}{2}\sqrt{(D\!-\!3)(D\!+\!1)}\!<\!\frac{1}{2}(D\!-\!1)$.  Then, inserting $u(t, k)$ in Eq.~(\ref{modeleadingorder}) into Eq.~(\ref{truncexpwithu}) we obtain the IR truncated massive free field in $D$-dimensions,
\be
{\bar\varphi}_0(t, \vec{x})\!=\!H^{\nu+\frac{D}{2}-1}\!\frac{\Gamma(2\nu)}{\Gamma(\nu\!+\!\frac{1}{2})2^\nu}\,
a^{\nu-\frac{D-1}{2}}(t)\!\sum_{\vec{n} \neq 0} \!\frac{\Theta\left(H a(t) \!-\! k\right)}{k^{\nu}}\!\left[e^{i \vec{k} \cdot \vec{x}}\hat{A}_{\vec{n}} \!+\! e^{-i \vec{k} \cdot \vec{x}} \hat{A}^{\dagger}_{\vec{n}}
\right] \; .\label{massivefree}
\ee
Note that, in the massless limit, Eq.~(\ref{massivefree}) reduces to the form obtained in Ref.~\cite{vacuum}. We express the IR truncated massive full field $\bar\varphi$ in terms of the IR truncated massive free field $\bar\varphi_0$ in the next section.

\subsection{Interacting Theory}
\label{subsect:interactingtheory}

The full field $\varphi(t,\vec{x})$ with potential
$V(\varphi)$ satisfying equation of motion~(\ref{phieqn}) is given in Eq.~(\ref{phix}). In background~(\ref{desitter}) it takes the form
\beeq
\varphi(t,\vec{x}) \!=\! \varphi_0(t,\vec{x})
\!-\!\!\!\int_0^t \!\!dt' a^{D-1}(t')\!\!\int\! d^{D-1}x'G(t,
\vec{x};t'\!, \vec{x}\,')\frac{V'(\varphi)(t'\!,\vec{x}\,')}{1\!\!+\!\delta Z} \; .\label{fullfieldIR}
\eneq
As can be inferred from iteration~(\ref{iterate}), obtaining the full field---even perturbatively---for potential~(\ref{potential}), Green's function~(\ref{commutatorgreen}), mode function~(\ref{udef}) and free field~(\ref{truncexpwithu}) is a formidable task. Amputating the ultraviolet modes, however, to study the IR limit of the theory, simplifies the computation. In this limit, the mode function and the free field are already obtained in Eqs.~(\ref{modeleadingorder}) and (\ref{massivefree}), respectively. To get the IR truncated full field $\bar{\varphi}(t,\vec{x})$, the last ingredient we need is the IR limit of Green's function~(\ref{commutatorgreen})  given in terms of the commutator function
\be
&&\hspace{-0.7cm}\left[\varphi_0(t,\vec{x}), \varphi_0(t'\!,\vec{x}\,')\right]\!=\!H^{D-1}\! \sum_{\vec{n}\neq 0}\Bigl\{ u(t,k) u^*(t'\!,k)
\!-\!u^*(t,k) u(t'\!,k) \Bigr\}e^{i \vec{k} \cdot (\vec{x}-\vec{x}'\!)}  \; ,\\
&&\hspace{-0.6cm}\!=\!i\frac{\pi}{2}H^{D-2}\frac{\csc\!\left(\nu\pi\right)}
{\sqrt{a^{D-1}(t)\,a^{D-1}(t')}}\! \sum_{\vec{n}\neq 0}\!\Big[{J}_{\nu}(z)\,
{J}_{-\nu}(z')\!-\!{J}_{-\nu}(z)\,{J}_{\nu}(z')\Big]e^{i \vec{k} \cdot (\vec{x}-\vec{x}'\!)}  \; .\label{commutfunc}
\ee
We combined  Eqs.~(\ref{udef}) and (\ref{truncexpwithu})-(\ref{Neumann}) in obtaining Eq.~(\ref{commutfunc}). In leading order, Eq.~(\ref{BesselSeries}) and
\be
H^{D-1}\!\sum_{\vec{n}}e^{i \vec{k} \cdot (\vec{x}-\vec{x}'\!)}\!=\!\delta^{D-1}(\vec{x}\!-\!\vec{x}'\!)\; ,
\ee
gives
\be
&&\hspace{-0.7cm}\left[\varphi_0(t,\vec{x}), \varphi_0(t'\!,\vec{x}\,')\right]\longrightarrow\!-\frac{i}{2H\nu}\Biggl[\frac{a^{2\nu}(t)\!-\!a^{2\nu}(t')}
{[a(t)\,a(t')]^{\nu+\frac{D-1}{2}}}\Biggr]\delta^{D-1}(\vec{x}\!-\!\vec{x}\,')\; .\label{Greeninu}
\ee
Hence, the retarded Green's function
\be
G(t,\vec{x};t'\!, \vec{x}\,')\!\longrightarrow\!\frac{\Theta(t\!-\!t')}{2H\nu}\Biggl[\frac{a^{2\nu}(t)\!-\!a^{2\nu}(t')}
{[a(t)\,a(t')]^{\nu+\frac{D-1}{2}}}\Biggr]\delta^{D-1}(\vec{x}\!-\!\vec{x}\,')\; ,\label{GreenIRlimit}
\ee
in leading order. Substituting limit~(\ref{GreenIRlimit}) into Eq.~(\ref{fullfieldIR}) yields
\be
\bar\varphi(t,\vec{x}) \!=\! \bar\varphi_0(t,\vec{x})
\!-\!\frac{1}{2\nu H}\!\int_0^t \!\!dt' \!\left[\left(\frac{a(t)}{a(t')}\right)^{\nu-\frac{D-1}{2}}
\!\!-\!\left(\frac{a(t')}{a(t)}\right)^{\nu+\frac{D-1}{2}}\right]\!\frac{V'(\bar\varphi)(t'\!,\vec{x})}{1\!\!+\!\delta Z} \; .\label{fullfieldgeneral}
\ee
The latter term in the square brackets can be neglected next to the former which dominates throughout the range of integration. Moreover, the counterterms in potential~(\ref{potential}) cannot contribute \cite{Wstocsqed} in the leading order we consider. One can see this by comparing the powers of the fields and orders of $\lambda$ involved in various counter terms in the model: $\delta \lambda \!\sim\! \mathcal{O}(\lambda^2)$, $\delta m^2 \!\sim\! \mathcal{O}(\lambda)$ and $\delta Z \!\sim\! \mathcal{O}(\lambda^2)$ \cite{BOW}. Firstly, compare the contributions involving $\lambda\varphi^4$ and the contributions involving $\delta\lambda\varphi^4$ terms. Powers of the fields are the same, so the former and the latter have the same structure of leading terms. The latter, however, are suppressed by at least one extra factor of $\lambda$ (with $\delta \lambda \!\sim\! \mathcal{O}(\lambda^2)$), they can never be in leading order. Secondly, compare the contributions involving $\lambda\varphi^4$ and the contributions involving $\delta m^2\varphi^2$ terms. Although the former and the latter are linear in $\lambda$ (with $\delta m^2 \!\sim\! \mathcal{O}(\lambda)$), the former are quartic in field whereas the latter are quadratic in field. Therefore, at a given order in $\lambda$, the latter can never have as high order leading terms as the former. Finally, the field strength counterterm $\delta Z$ appears in the field equations in the form
\be
\frac{V'(\bar\varphi)(t'\!,\vec{x})}{1\!+\!\delta Z}\!=\!V'(\bar\varphi)\left[1\!-\!\delta Z\!+\!(\delta Z)^2\!-\!\cdots\right]\; ,
\ee
with $\delta Z \!\sim\! \mathcal{O}(\lambda^2)$. Hence, exactly the same leading order contributions that Eq.~(\ref{fullfieldgeneral}) would yield are obtained from its simplified version without the counterterms, i.e., from
\be
\bar\varphi(t,\vec{x}) \!=\! \bar\varphi_0(t,\vec{x})
\!-\!\frac{\lambda}{6(2\nu)H}\,a^{-\frac{\delta}{2}}(t)\!\!\int_0^t \!\!dt'a^{\frac{\delta}{2}}(t')\,\bar\varphi^3(t',\vec{x})
 \; ,\label{freefieldsimple}
\ee
where we define \beeq
D\!-\!1\!-\!\!2\nu\!\equiv\delta\; .\label{delta}
\eneq
(Note that $\delta\!\!\rightarrow\!0$ as the mass $m\!\!\rightarrow\!0$.) Infrared truncated full field $\bar\varphi(t,\vec{x})$ can be expressed in terms of the IR truncated free field $\bar\varphi_0(t,\vec{x})$, at any order of $\lambda$, by iterating Eq.~(\ref{freefieldsimple}) successively. Iterating it twice, for example, yields
\be
&&\hspace{0.2cm}\bar\varphi(t, \vec{x})\!=\! \bar\varphi_0(t,\vec{x})
\!-\!\frac{\lambda}{6(2\nu) H}\,a^{-\frac{\delta}{2}}(t)\!\!\!\int_0^t \!\!dt'a^{\frac{\delta}{2}}(t')\,\bar\varphi_0^3(t',\vec{x})
\!+\!\frac{\l^2}{12(2\nu)^2H^2}\,a^{-\frac{\delta}{2}}(t)\!\!\int_0^t\!\!
dt'\bar\varphi^2_0(t'\!,\vec{x})\nonumber\\
&&\hspace{-0.4cm}\times\!\!\!\int_0^{t'}\!\!\!\!dt''a^{\frac{\delta}{2}}(t'')
\bar\varphi^3_0(t''\!,\vec{x})
\!-\!\frac{\l^3}{24(2\nu)^3H^3}\,a^{-\frac{\delta}{2}}(t)\Bigg\{\!\!\int_0^t\!\!\!
dt'\bar\varphi^2_0(t'\!,\vec{x})\!\!\int_0^{t'}
\!\!\!\!dt''\bar\varphi^2_0(t''\!,\vec{x})\!\!\int_0^{t''}
\!\!\!\!\!dt'''a^{\frac{\delta}{2}}(t''')\bar\varphi^3_0(t'''\!,\vec{x})\nonumber\\
&&\hspace{2.3cm}+\frac{1}{3}\!\int_0^t\!\!\!
dt'a^{-\frac{\delta}{2}}(t')\,\bar\varphi_0(t'\!,\vec{x})
\!\left[\int_0^{t'}\!\!\!\!dt''a^{\frac{\delta}{2}}(t'')\,\bar\varphi^3_0(t''\!,\vec{x})\right]^2\!\Bigg\}
\!+\!\mathcal{O}(\l^4)\; .\label{freefieldexpfulfield}\ee
In the next section, we use Eq.~(\ref{freefieldexpfulfield}) to compute the two-point correlation function of the IR truncated full field, at two-loop order.

\section{Two-point correlation function}
\label{sec:twopointcorrelator}

The two-point correlation function of the IR truncated full field for two distinct events
\be
\langle\Omega|\!
\bar{\varphi}(t,\vec{x})\bar{\varphi}(t'\!,\vec{x}\,')|\Omega\rangle\!\!&=&\!\!\langle\Omega|
\bar{\varphi}(t,\vec{x})\bar{\varphi}(t'\!,\vec{x}\,')|\Omega\rangle_{\rm tree}\!+\!\langle\Omega|
\bar{\varphi}(t,\vec{x})\bar{\varphi}(t'\!,\vec{x}\,')|\Omega\rangle_{\rm 1-loop}\nonumber\\
\!&+&\!\!\!\langle\Omega|
\bar{\varphi}(t,\vec{x})\bar{\varphi}(t'\!,\vec{x}\,')|\Omega\rangle_{\rm 2-loop}\!+\!\mathcal{O}(\lambda^3)\; ,\label{genelfullexpect}
\ee
with $t'\!\leq\!t$ and $\vec{x}\,'\!\!\neq\!\vec{x}$, can be obtained for the field with a quartic self-interaction using Eq.~(\ref{freefieldexpfulfield}). It yields, at tree-order,
\beeq
\langle\Omega|
\bar{\varphi}(t,\vec{x})\bar{\varphi}(t'\!,\vec{x}\,')|\Omega\rangle_{\rm tree}\!=\!\langle\Omega|
\bar{\varphi}_0(t,\vec{x})\bar{\varphi}_0(t'\!,\vec{x}\,')|\Omega\rangle\; .
\eneq
The leading (one-loop) quantum correction
\be
&&\hspace{0cm}\langle\Omega|
\bar{\varphi}(t,\vec{x})\bar{\varphi}(t'\!,\vec{x}\,')|\Omega\rangle_{\rm 1-loop}\!=\!-\frac{\lambda}{6(2\nu)H}\Bigg[a^{-\frac{\delta}{2}}(t')\langle\Omega|
\bar{\varphi}_0(t,\vec{x})\!\!\!\int_0^{t'}\!\!\!\!d\tilde{t}\,a^{\frac{\delta}{2}}(\tilde{t})
\,{\bar{\varphi}}^3_0(\tilde{t},\vec{x}\,')
|\Omega\rangle\nonumber\\
&&\hspace{3cm}+a^{-\frac{\delta}{2}}(t)\langle\Omega|\!\!
\int_0^{t}\!\!\!dt''a^{\frac{\delta}{2}}(t''){\bar{\varphi}}^3_0(t''\!,\vec{x})
\bar{\varphi}_0(t'\!,\vec{x}\,')|\Omega\rangle\Bigg]\; ,\label{1loopcorr}
\ee
is not hard to compute. Computation of the next to leading order (two-loop) quantum correction
\be
&&\hspace{-0.5cm}\langle\Omega|\bar{\varphi}(t,\vec{x})\bar{\varphi}(t'\!,\vec{x}\,')|\Omega\rangle_{\rm 2-loop}\!=\!\frac{\lambda^2}{12(2\nu)^2H^2}\!\Bigg[a^{-\frac{\delta}{2}}(t')\langle\Omega|
\bar{\varphi}_0(t,\vec{x})\!\!\!\int_0^{t'}\!\!\!\!d\tilde{t}\,{\bar{\varphi}}^2_0(\tilde{t},\vec{x}\,')
\!\!\!\int_0^{\tilde{t}}\!\!\!d\tilde{\tilde{t}}\,a^{\frac{\delta}{2}}(\tilde{\tilde{t}}) {\bar{\varphi}}^3_0(\tilde{\tilde{t}},\vec{x}\,')
|\Omega\rangle\nonumber\\
&&\hspace{2cm}+a^{-\frac{\delta}{2}}(t)\langle\Omega|\!\!
\int_0^{t}\!\!dt''{\bar{\varphi}}^2_0(t''\!,\vec{x})\!\!
\int_0^{t''}\!\!\!\!dt'''\,a^{\frac{\delta}{2}}(t'''){\bar{\varphi}}^3_0(t'''\!,\vec{x})
\bar{\varphi}_0(t'\!,\vec{x}\,')
|\Omega\rangle\nonumber\\
&&\hspace{1.6cm}+\frac{1}{3}\Big[a(t)\,a(t')\Big]^{-\frac{\delta}{2}}\langle\Omega|\!\!
\int_0^{t}\!\!dt''\,a^{\frac{\delta}{2}}(t''){\bar{\varphi}}^3_0(t''\!,\vec{x})\!\!
\int_0^{t'}\!\!\!d\tilde{t}\,a^{\frac{\delta}{2}}(\tilde{t}){\bar{\varphi}}^3_0(\tilde{t},\vec{x}\,')
|\Omega\rangle\Bigg]\; ,\label{correlation}
\ee
where $0\!\leq\!t'''\!\leq\!t''\!\leq\!t$, $0\!\leq\!\tilde{\tilde{t}}\!\leq\!\tilde{t}\!\leq\!t'$ and $t'\!\leq\!t$, however, is demanding because it involves three VEVs each of which has a double time integral without a definite time ordering in the integrand.

Note that the two-point correlation function for equal time, equal space and equal spacetime events can be obtained as limits of Eq.~(\ref{genelfullexpect}). Note also that perturbation theory breaks down when $\ln(a(t))\!=\!Ht\!\sim\!1/\sqrt{\lambda}$ \cite{OW1, KO}. Thus, for $\lambda\!\ll\!1$, the perturbation theory and hence Eq.~(\ref{genelfullexpect}) are accurate for an arbitrarily long period of time. The correlation function, at tree, one- and two-loop order is computed in Secs.~\ref{subsect:treecorr}, \ref{subsect:1loopcorr} and \ref{subsect:2loopcorr}, respectively.

\subsection{Tree-order correlator}
\label{subsect:treecorr}

The tree-order two-point correlation function of the IR truncated massive scalar is obtained
using Eqs.~(\ref{normalaadagger}), (\ref{massivefree}) and (\ref{delta}) as
\be
&&\hspace{2cm}\langle\Omega|
\bar{\varphi}(t,\vec{x})\bar{\varphi}(t'\!,\vec{x}\,')|\Omega\rangle_{\rm tree}\!=\!\langle\Omega|{\bar\varphi}_0(t,\vec{x}) {\bar\varphi}_0(t'\!,\vec{x}\,')|\Omega\rangle\nonumber\\
&&\hspace{-1.3cm}=\!\frac{\Gamma^2(2\nu)}{\Gamma^2(\nu\!+\!\frac{1}{2})}
\frac{H^{D-2}}{(4\pi)^{2\nu}}\!\left[\,a(t)\,a(t')\right]^{-\frac{\delta}{2}}\!\sum_{\vec{n}\neq 0}\!\frac{\Theta(H\!a(t)\!-\!\!H2\pi n)\Theta(H\!a(t')\!-\!\!H2\pi n)}{n^{2\nu}}\,e^{i2\pi H\vec{n}\cdot(\vec{x}-\vec{x}\,')}
\; ,\label{sumtheta}
\ee
where $t'\!\!\leq\!t$ and  $\vec{x}\,'\!\neq\!\vec{x}$. The discrete sum over $\vec{n}$ in Eq.~(\ref{sumtheta}) can be approximated, in the continuum limit, as a $(D\!-\!\!1)$-dimensional integral
\be
&&\hspace{0cm}\sum_{\vec{n}\neq 0}\!\frac{\Theta(H\!a(t)\!-\! H2\pi n)\,\Theta(H\!a(t')\!-\!H2\pi n)}{n^{2\nu}}\,e^{i2\pi H\vec{n}\cdot(\vec{x}-\vec{x}\,')}\nonumber\\
&&\hspace{-1.2cm}\simeq\!\frac{1}{(2\pi H)^{\delta}}\!\!\int\!d\Omega_{D-1}\!\!\int_0^\infty\!\!\!\frac{d k}{k^{1-\delta}}\,\Theta(Ha(t)\!-\!k)\,\Theta(Ha(t')\!-\!k)\,e^{ik \Delta x \cos{(\theta)}}\; ,\label{angkeyeqn}
\ee
where $\Delta x\equiv\parallel\!\Delta\vec{x}\!\parallel=\parallel\!\vec{x}\!-\!\vec{x}\,'\!\!\parallel$ and $\theta$ is the angle between the vectors $\vec{n}$ and $\Delta\vec{x}$. Evaluating the angular integrations on the right side of Eq.~(\ref{angkeyeqn}) yields
\be
&&\hspace{-1cm}\frac{\Gamma\left(\frac{D}{2}\right)}{\Gamma(D\!-\!1)}2^{2\nu}\pi^{2\nu-\frac{D}{2}}
H^{-\delta}\!\!\int_0^\infty\!\!\!\frac{d k}{k^{1-\delta}}\Theta(Ha(t)\!-\!k)\,\Theta(Ha(t')\!-\!k)\frac{\sin(k\Delta x)}{k\Delta x}\, .\label{sumthetaintegral}
\ee
Because $0\!\leq\!t'\!\leq\!t$, we have $Ha(t')\!\leq\!Ha(t)$ which implies,
\beeq
\Theta(H\!a(t)\!-\!k)\,\Theta(H\!a(t')\!-\!k)\!=\!\Theta(H\!a(t')\!-\!k)\; .
\label{thetaatatprime}
\eneq
Therefore, the integral over $k$ in Eq.~(\ref{sumthetaintegral}),
\be
\int_0^\infty\!\!\frac{d k}{k^{1-\delta}}\Theta(Ha(t')\!-\!k)\frac{\sin(k\Delta x)}{k\Delta x}\!=\!\!\int_H^{Ha(t')}\!\!\!\frac{d k}{k^{1-\delta}}\frac{\sin(k\Delta x)}{k\Delta x}
\!=\!(\Delta x^2)^{-\frac{\delta}{2}}\!\!\int_{H\Delta x}^{Ha(t')\Delta x}\!\!d y\frac{\sin(y)}{y^{2-\delta}}\, .\label{logar}
\ee
The remaining integral in Eq.~(\ref{logar}) is
\be
\int\!d y\frac{\sin(y)}{y^{2-\delta}}\!=\!-\frac{i}{2}y^{-1+\delta}[E_{2-\delta}(iy)\!-\!E_{2-\delta}(-iy)]\; ,\label{intexpint}
\ee
where $E_\beta(z)$ is the exponential integral function defined in Eq.~(\ref{expintdefn}). Substituting Eqs.~(\ref{thetaatatprime})-(\ref{intexpint}) into Eq.~(\ref{sumthetaintegral}) yields sum~(\ref{angkeyeqn}) and hence, tree-order correlator~(\ref{sumtheta}) in terms of
the exponential integral function. However, expressing
$E_\beta(z)$ in terms of the incomplete gamma function $\Gamma(\beta, z)$, defined in Eq.~(\ref{incompgammadefn}), as
\be
E_\beta(z)\!=\!z^{\beta-1}\Gamma(1\!-\!\beta, z)\; ,\label{expintincompgamma}
\ee
gives a more useful result for sum~(\ref{angkeyeqn}); see Eq.~(\ref{sumthetaincompGamma}). Employing it in Eq.~(\ref{sumtheta}) yields the tree-order correlator for a massive scalar as
\beeq
\langle\Omega|{\bar\varphi}_0(t,\vec{x}) {\bar\varphi}_0(t'\!,\vec{x}\,')|\Omega\rangle\!\simeq\!
\mathcal{A}\,\emph{f}_0(t,t'\!,\Delta x)\; ,\label{treecorrgamma}
\eneq
where we define the constant
\beeq
\mathcal{A}\!\equiv\!\frac{\Gamma^2\!\left(2\nu\right)}{\Gamma^2\!\left(\nu\!+\!\frac{1}{2}\right)}
\frac{\Gamma\!\left(\frac{D}{2}\right)}{\Gamma\!\left(D\!-\!\!1\right)}
\frac{H^{D-2}}{2^{2\nu}\pi^{\frac{D}{2}}}\;,\label{Acons}
\eneq
and the spacetime dependent function\be
&&\hspace{-0.5cm}\emph{f}_0(t,t'\!,\Delta x)\!=\!\!\frac{1}{2}\!\left[-\alpha(t)\,\alpha(t')\right]^{-\frac{\delta}{2}}
\!\Bigg\{\!\Gamma\!\left(-\!1\!\!+\!\delta, i\alpha(t')\right)\!-\!\Gamma\!\left(-\!1\!\!+\!\delta, iH\!\Delta x\right)\!+\!(-1)^{-\delta}\Big[\Gamma\!\left(-\!1\!\!+\!\delta, -i\alpha(t')\right)\nonumber\\
&&\hspace{5.6cm}-\Gamma\!\left(-\!1\!\!+\!\delta, -iH\!\Delta x\right)\!\Big]\!\Bigg\}\; ,\label{treecorrgammaf}
\ee
where
\beeq
\alpha(t)\!\equiv\!\alpha(t,\Delta x)\!\equiv\!a(t)H\!\Delta x\; .
\eneq

The tree-order correlator for a massless scalar can be obtained by taking the $m\!\!\rightarrow\!0$ limit of Eq.~(\ref{treecorrgamma}); see Appendix~\ref{App:zeromasscorr}. Equation~(\ref{treecorrgamma}) can also be used to infer the behavior of the tree-order correlator for a {\it fixed comoving separation} at late $t'$ when $a(t')$ [hence $a(t)$] becomes large; see Appendix~\ref{App:asympttreecorr}. Asymptotic form~(\ref{treecorrgasym}) of the tree-order correlator implies that it asymptotes to {\it zero} for the massive case whereas to a {\it nonzero constant} in the massless limit.

Using power series representation~(\ref{sumaspowerseries}) of sum~(\ref{angkeyeqn}) in Eq.~(\ref{sumtheta}), on the other hand, yields
\be
\emph{f}_0(t,t'\!,\Delta x)\!=\!\Big[a(t)\,a(t')\Big]^{-\frac{\delta}{2}}\!\sum_{n=0}^\infty\!\frac{(-1)^n (H\!\Delta x)^{2n}}{(2n\!\!+\!\!1\!)!} \frac{a^{2n+\delta}(t')\!-\!\!1}{2n\!\!+\!\delta}\; .
\label{treecorrseries}
\ee
Alternatively, summing up the infinite sum in Eq.~(\ref{treecorrseries}) we find
\be
\hspace{-0.25cm}\emph{f}_0(t,t'\!,\Delta x)\!=\!a^{-\frac{\delta}{2}}(t)\Bigg\{\!\frac{a^{\frac{\delta}{2}}(t')}{\delta}
{}_1\mathcal{F}_2\Big(\!\frac{\delta}{2};\!\frac{3}{2},\!1\!\!+\!\frac{\delta}{2};\!-\frac{\alpha^2(t')}{4}\Big)
\!\!-\!\frac{a^{-\frac{\delta}{2}}(t')}{\delta}
{}_1\mathcal{F}_2\Big(\!\frac{\delta}{2};\!\frac{3}{2},\!1\!\!+\!\frac{\delta}{2};\!-\frac{(H\!\Delta x)^2}{4}\!\Big)\!\!\Bigg\}\; ,\label{treecorrgenhyp}
\ee
where ${}_1\mathcal{F}_2$ is the generalized hypergeometric function whose power series representation is given in Eq.~(\ref{hypergpq}). Form~(\ref{treecorrseries}) or~(\ref{treecorrgenhyp}) of $\emph{f}_0$ can be used to figure out the behaviors of tree-order correlator (i) at a {\it fixed comoving separation} at late times---as in Eq.~(\ref{treecorrgasym})--- and (ii) for a {\it fixed physical distance}.

In case (i), $\emph{f}_0$ asymptotes to \be
&&\hspace{-0.7cm}\emph{f}_0(t,t'\!,\Delta x)\!\rightarrow\!\Big[a(t)\,a(t')\Big]^{-\frac{\delta}{2}}\!
\Bigg\{\!\frac{\sqrt{\pi}\,\Gamma(\frac{\delta}{2})}{2^{2-\delta}\Gamma(\frac{3-\delta}{2})}\frac{1}{(H\!\Delta x)^\delta}-\!\frac{1}{\delta}
{}_1\mathcal{F}_2\Big(\!\frac{\delta}{2};\!\frac{3}{2},\!1\!\!+\!\frac{\delta}{2};\!-\frac{(H\!\Delta x)^2}{4}\!\Big)\!\Bigg\}\;,
\ee
which implies, as pointed out earlier, that the tree-order correlator asymptotes to zero as $\mathcal{O}((a(t)\,a(t'))^{-\frac{\delta}{2}})$ for a massive scalar. In the massless ($\delta\!\!\rightarrow\! 0$) limit though the tree-order correlator asymptotes to a nonzero constant. To see this, it is useful to rewrite Eq.~(\ref{treecorrseries}) singling out $n\!=\!0$ term in the infinite sum
\be
\emph{f}_0(t,t'\!,\Delta x)\!=\!
\Big[a(t)\,a(t')\Big]^{-\frac{\delta}{2}}\Bigg\{\!\frac{a^{\delta}(t')\!-\!1}{\delta}
\!+\!\!\sum_{n=1}^\infty\frac{(-1)^n (H\!\Delta x)^{2n}}{(2n\!\!+\!\!1\!)!}\frac{a^{2n+\delta}(t')\!-\!1}{2n\!\!+\!\!\delta}\!\Bigg\}
\; .\label{singledouttree}
\ee
The $n\!=\!0$ term in the curly brackets causes a ${0}/{0}$ ambiguity in the $\delta\!\!\rightarrow\! 0$ limit. Expanding the $a^{\delta}(t')\!=\!e^{\delta\ln(a(t'))}$ in this term around $\delta\!=\!0$ and then taking the massless limit of Eq.~(\ref{singledouttree}) we find\be
\hspace{0cm}\lim_{m\rightarrow 0}\!\emph{f}_0(t,t'\!,\Delta x)
\!=\!\ln(a(t'))\!+\!\!\sum_{n=1}^\infty\frac{\!(-1)^n (H\!\Delta\! x)^{2n}}{(2n\!\!+\!\!1\!)!}\frac{a^{2n}(t')\!-\!\!1}{2n}\; .\label{treemasslesssingled}
\ee
Summing up the terms in the infinite sum in Eq.~(\ref{treemasslesssingled}) yields \be
\hspace{0cm}\lim_{m\rightarrow 0}\!\emph{f}_0(t,t'\!,\Delta x)
\!=\!{\rm ci}(\alpha(t'))\!-\!\frac{\sin(\alpha(t'))}{\alpha(t')}\!-\!{\rm ci}(H\!\Delta x)
\!+\!\frac{\sin(H\!\Delta x)}{H\!\Delta x}\;,\label{treemasslesspower}
\ee
where ${\rm ci}(z)$ is the cosine integral function defined in Eq.~(\ref{cosintdefn}). [We used Eq.~(\ref{cieksisin}) in obtaining Eq.~(\ref{treemasslesspower}).] Employing Eq.~(\ref{cisasym}) in the time dependent terms in Eq.~(\ref{treemasslesspower}) implies that\be
{\rm ci}(\alpha(t'))\!-\!\frac{\sin(\alpha(t'))}{\alpha(t')}\!\longrightarrow 0\!+\!\mathcal{O}\!\left(\alpha^{-2}(t')\right)\; .\label{asymassless}
\ee
Thus, the tree-order correlator, in the massless case, asymptotes to a nonzero constant
\be
&&\hspace{-1cm}\lim_{m\rightarrow 0}\langle\Omega|{\bar\varphi}_0(t,\vec{x}) {\bar\varphi}_0(t'\!,\vec{x}\,')|\Omega\rangle\longrightarrow
\frac{\Gamma\!\left(D\!-\!\!1\right)}{\Gamma\!\left(\frac{D}{2}\right)}
\frac{H^{D-2}}{2^{D-1}\pi^{\frac{D}{2}}}
\Bigg\{\!\frac{\sin(H\!\Delta x)}{H\!\Delta x}\!-\!{\rm ci}(H\!\Delta x)\!\Bigg\}\; .\label{zeromasstreeaympi}
\ee

\begin{figure}
\includegraphics[width=10cm,height=6.3cm]{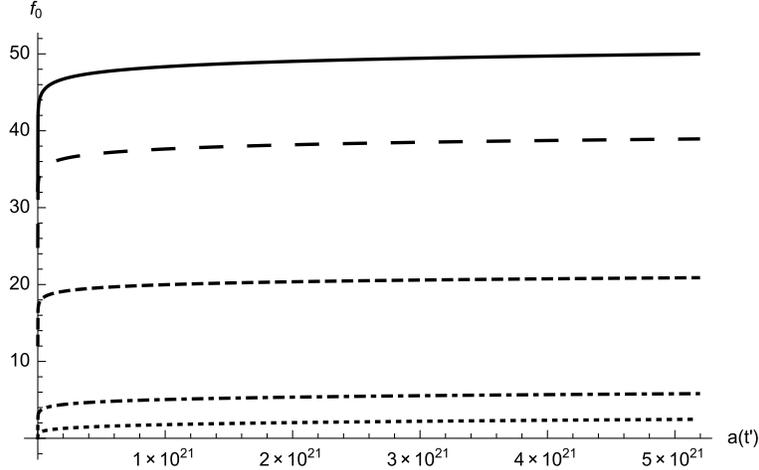}
\caption{Plots of the function $f_0(t,t', K)$, defined in Eq.~(\ref{treecorrfixeddist}), versus $a(t')$ for  different mass values. The scale factor $a(t)$ and the fraction $K$ are chosen to be $e^{50}$ and $1/2$, respectively. $a(t')$ runs from $1$ to $a(t)$. The solid curve is the plot for a massless scalar. The large-dashed, dashed, dot-dashed and dotted curves are plots for massive scalars with $m\!=\!H/8, H/4, H/2$ and $3H/4$, respectively.}
\label{fig:tree-corr}
\end{figure}

In case (ii), choosing the comoving separation in Eq.~(\ref{treecorrgenhyp}) as\beeq
\Delta x\!=\!\frac{K}{Ha(t')}\; ,\label{fixedphysdist}
\eneq
to keep the physical distance $a(t')\Delta x$ a constant fraction $K$ of the Hubble length, we obtain\be
&&\hspace{-1.1cm}\emph{f}_0(t,t'\!,K)\!=\!a^{-\frac{\delta}{2}}(t)
\Bigg\{\!\frac{a^{\frac{\delta}{2}}(t')}{\delta}
{}_1\mathcal{F}_2\Big(\frac{\delta}{2};\frac{3}{2},1\!\!+\!\frac{\delta}{2};-\frac{K^2}{4}\Big)
\!\!-\!\frac{a^{-\frac{\delta}{2}}(t')}{\delta}
{}_1\mathcal{F}_2\Big(\frac{\delta}{2};\frac{3}{2},1\!\!+\!\frac{\delta}{2};-\frac{K^2}{4a^2(t')}\Big)\!\!\Bigg\}
\; .\label{treecorrfixeddist}
\ee
The function $\emph{f}_0(t,t'\!,K)$ grows logarithmically for a massless scalar. This growth, however, is suppressed for massive scalars; see Fig.~\ref{fig:tree-corr}. The larger the mass, the stronger the suppression.
To see the logarithmic growth and the effect of mass analytically, consider massless form~(\ref{treemasslesssingled}) of $\emph{f}_0$ with $\Delta x$ is as given in Eq.~(\ref{fixedphysdist}), \be
\lim_{m\rightarrow 0}\emph{f}_0(t,t'\!,K)
\!=\!\ln(a(t'))
\!+\!\!\sum_{n=1}^\infty\!\frac{(-1)^{n}K^{2n}}{(2n\!\!+\!\!1\!)!}\,\frac{1\!\!-\!a^{-2n}(t')}{2n} \; ,\label{treemasslesscompare}
\ee
and compare it with the corresponding massive form of $\emph{f}_0$ given in Eq.~(\ref{singledouttree}) \be
\emph{f}_0(t,t'\!,K)\!=\!\Big[a(t)\,a(t')\Big]^{-\frac{\delta}{2}}\Bigg\{\!\frac{a^{\delta}(t')\!-\!1}{\delta}
\!+\!\!\sum_{n=1}^\infty\frac{(-1)^n K^{2n}}{(2n\!\!+\!\!1\!)!}\frac{a^{\delta}(t')\!-\!a^{-2n}(t')}{2n\!\!+\!\!\delta}\!\Bigg\}
\; .\label{treemassivecompare}
\ee
Notice that the time dependent terms in the infinite sum in Eq.~(\ref{treemasslesscompare}) decay rapidly, at least as fast as $\mathcal{O}(a^{-2}(t'))$. The time independent part of the infinite sum adds up to a constant,
\beeq
\sum_{n=1}^\infty\!\frac{(-1)^{n}K^{2n}}{(2n\!\!+\!\!1\!)!2n}\!=\!{\rm ci}(K)
\!-\!\frac{\sin(K)}{K}\!-\!\ln(K)\!-\!\gamma\!+\!1\;.
\eneq Hence, the tree-order correlator at a fixed physical distance grows logarithmically with the scale factor $a(t')$ (linearly with the comoving time $t'$) in the massless limit. This is because more and more scalar particles are created out of vacuum which increases the local field strength. In the massive case, on the other hand, the time dependent terms that are also mass ($\delta$) dependent in Eq.~(\ref{treemassivecompare}) imply that the mass suppresses the logarithmic growth in Eq.~(\ref{treemasslesscompare}). This is because $\frac{a^{\delta}(t')-1}{\delta}\leq\ln(a(t'))$, where the equality holds in the massless limit and $a^\delta(t')\leq\left[a(t)\,a(t')\right]^{\frac{\delta}{2}}$, where the equality holds in the equal time limit.

Physical reason behind this suppression in the tree-order correlator is that the {\it persistence time} for virtual particles which emerge out of vacuum is {\it suppressed by mass}, as pointed out in Sec.~\ref{sec:intro}. Particles with smallest masses remain coherent longest during inflation. In fact, massless sufficiently long wavelength modes can persist forever and therefore become real. As more and more particles are created the local field strength increases. Because the particle production is suppressed by mass, the growth in tree-order correlator for massive scalars is suppressed as well.

In Secs.~\ref{subsubsect:equaltimecorr}-3, we obtain the various limits of the tree-order correlator which will be needed in computing the one- and two-loop corrections in Secs.~\ref{subsect:1loopcorr}-C.

\subsubsection{Equal time limit of the tree-order correlator}
\label{subsubsect:equaltimecorr}

Equal time limit of the tree-order correlator is trivial to read off either from Eq.~(\ref{treecorrgamma}) or from Eq.~(\ref{treecorrseries}). The latter, for example, gives\beeq
\langle\Omega|{\bar\varphi}_0(t,\vec{x}) {\bar\varphi}_0(t,\vec{x}\,')|\Omega\rangle\!\simeq\!\mathcal{A}\,
a^{-\delta}(t)\!\!\sum_{n=0}^\infty\!\frac{(-1)^n (H\!\Delta x)^{2n}}{(2n\!\!+\!\!1)!}\frac{ a^{2n+\delta}(t)\!-\!\!1}{2n\!\!+\!\delta}\; ,\label{massiveequaltimeVEV}
\eneq
the required limit as a power series expansion.

\subsubsection{Equal space limit of the tree-order correlator}
\label{subsubsect:equalspacecorr}

The equal space limit of Eq.~(\ref{treecorrseries}) yields\beeq
\langle\Omega|{\bar\varphi}_0(t,\vec{x}) {\bar\varphi}_0(t'\!,\vec{x})|\Omega\rangle\!\simeq\!\mathcal{A}\!
\left[\,a(t)\,a(t')\right]^{-\frac{\delta}{2}}\frac{a^{\delta}(t')\!-\!1}
{\delta}\; .\label{massiveequalspaceVEV}
\eneq
Equal spacetime limit of the tree-order correlator, which we evaluate in the following section, can be  obtained either from the equal time~(\ref{massiveequaltimeVEV}) or from the equal space~(\ref{massiveequalspaceVEV}) limit.

\subsubsection{Equal spacetime limit of the tree-order correlator}
\label{subsubsect:equalspacetimecorr}

Taking the equal space limit of Eq.~(\ref{massiveequaltimeVEV}) [or the equal time limit of Eq.~(\ref{massiveequalspaceVEV})] leads to \beeq
\langle\Omega|\bar{\varphi}^2_0(t,\vec{x})|\Omega\rangle\!\simeq\!
\mathcal{A}
\frac{1\!\!-\!a^{-\delta}(t)}
{\delta}\; .\label{phisquare}
\eneq In the limit of vanishing mass, Eqs.~(\ref{treecorrgamma})-(\ref{treecorrseries})
yield the tree-order correlator for a {\it massless} IR truncated scalar in analytic function and power series forms; see Eqs.~(\ref{C1mless1})-(\ref{treemasslessseriesrep}).

Tree-order correlator is the basic entity in the computation of correlation function perturbatively in a self-interacting theory. We employ results~(\ref{treecorrgamma}), (\ref{Acons}), (\ref{treecorrseries}) and (\ref{massiveequaltimeVEV})-(\ref{phisquare}) in Secs.~\ref{subsect:1loopcorr}-C to compute one- and two-loop correlator in the $\lambda \bar\varphi^4$ theory.

\subsection{One-loop correlator}
\label{subsect:1loopcorr}

Computation of one-loop contribution~(\ref{1loopcorr}) to two-point correlation~(\ref{genelfullexpect}) involves evaluations of two VEVs. The first one is
\beeq
\langle\Omega|
\bar{\varphi}_0(t,\vec{x})\!\!\int_0^{t'}\!\!\!d\tilde{t}\,a^{\frac{\delta}{2}}(\tilde{t})
\,{\bar{\varphi}}^3_0(\tilde{t},\vec{x}\,')
|\Omega\rangle\; .
\eneq
It can be reexpressed, in terms of topologically inequivalent field parings, as
\beeq
\int_0^{t'}\!\!d\tilde{t}\,a^{\frac{\delta}{2}}(\tilde{t})
\,3\cdot\!1\langle\Omega|
\bar{\varphi}_0(t,\vec{x})\, \bar{\varphi}_0(\tilde{t},\vec{x}\,')|\Omega\rangle
\langle\Omega|{\bar{\varphi}}^2_0(\tilde{t},\vec{x}\,')|\Omega\rangle\; .\label{1loopfirstvev}
\eneq
Using Eq.~(\ref{treecorrseries}) with $\tilde{t}\leq t$ and Eq.~(\ref{phisquare}) in Eq.~(\ref{1loopfirstvev}) yields
\beeq
-\mathcal{A}^2\,\frac{a^{-\frac{\delta}{2}}(t)}{\delta}\,
3\!\!\int_0^{t'}\!\!\!d\tilde{t}\Big[1\!\!-\!a^{-\delta}(\tilde{t})\Big]\!\sum_{n=0}^\infty\!\frac{(-1)^{n}(H\!\Delta x)^{2n}}{(2n\!\!+\!\!1)!}\frac{ [1\!\!-\!a^{2n+\delta}(\tilde{t})]}
{2n\!\!+\!\delta}\; .\label{pairequaltime}
\eneq
Evaluating the integral in Eq.~(\ref{pairequaltime}) we obtain the first VEV as
\be
&&\hspace{-0.4cm}\langle\Omega|
\bar{\varphi}_0(t,\vec{x})\!\!\int_0^{t'}\!\!\!d\tilde{t}\,a^{\frac{\delta}{2}}(\tilde{t})
\,{\bar{\varphi}}^3_0(\tilde{t},\vec{x}\,')
|\Omega\rangle\!=\!-\frac{{\mathcal{A}}^2}{H}\,\frac{a^{-\frac{\delta}{2}}(t)}{\delta}\,
3\!\sum_{n=0}^\infty
\!\frac{(-1)^{n}(H\!\Delta x)^{2n}}{(2n\!\!+\!\!1\!)!(2n\!\!+\!\delta)}\Big[\!\ln(a(t'))\!+\!\frac{1\!\!-\!a^{2n+\delta}(t')}{2n\!\!+\!\!\delta}
\!\nonumber\\
&&\hspace{5.3cm}-\frac{1\!\!-\!a^{2n}(t')}{2n}\!-\!\frac{1\!\!-\!a^{-\delta}(t')}{\delta}
\Big]
\; .\label{1loop1result}
\ee

The second VEV that contributes at one-loop order in Eq.~(\ref{1loopcorr}) is
\be
&&\hspace{-1.15cm}\langle\Omega|\!\!\!
\int_0^{t}\!\!\!dt''a^{\frac{\delta}{2}}(t''){\bar{\varphi}}^3_0(t''\!,\vec{x})
\bar{\varphi}_0(t'\!,\vec{x}\,')|\Omega\rangle\!=\!\!\!\int_0^{t}\!\!\!dt''a^{\frac{\delta}{2}}(t'') 3\!\cdot\!\!1\!\langle\Omega|
\bar{\varphi}_0(t''\!,\vec{x})\bar{\varphi}_0(t'\!,\vec{x}\,')|\Omega\rangle
\langle\Omega|{\bar{\varphi}}^2_0(t''\!,\vec{x})|\Omega\rangle\; , \label{1loopsecondvevilk}
\ee
where we have $0\leq\!t''\!\leq\!t$ and $t'\!\leq\!t$. The coincident tree-order correlator in the integrand is as in Eq.~(\ref{phisquare}). To evaluate the remaining VEV in the integrand we need to break up the integral into two as $\int_0^{t}\!dt''\!=\!\int_0^{t'}\!dt''\!+\!\int_{t'}^t\!dt''$. In the first integral $t''\!\leq\!t'$, whereas in the second $t'\!\leq\!t''$. Hence, appropriate use of Eq.~(\ref{treecorrseries}) in  Eq.~(\ref{1loopsecondvevilk}) and evaluating the integrals give
\be
&&\hspace{-0.4cm}\langle\Omega|\!\!
\int_0^{t}\!\!dt''a^{\frac{\delta}{2}}(t''){\bar{\varphi}}^3_0(t''\!,\vec{x})
\bar{\varphi}_0(t'\!,\vec{x}\,')|\Omega\rangle\!=\!-\frac{{\mathcal{A}}^2}{H}\,
\frac{a^{-\frac{\delta}{2}}(t')}{\delta}\,
3\!\sum_{n=0}^\infty
\!\frac{(-1)^{n}(H\!\Delta x)^{2n}}{(2n\!\!+\!\!1\!)!(2n\!\!+\!\!\delta)}\Bigg\{\!\!\ln(a(t))\nonumber\\
&&\hspace{-1.1cm}-a^{2n}(t')\!
\Bigg[a^{\delta}(t')\!\left(\!\ln(a(t))\!-\!\ln(a(t'))
\!+\!\frac{a^{-\delta}(t)}{\delta}\!\right)\!\!-\!\!\frac{1}{\delta}\Bigg]
\!\!-\!\frac{1\!\!-\!a^{-\delta}(t)}
{\delta}\!+\!\frac{1\!\!-\!a^{2n+\delta}(t')}{2n\!\!+\!\!\delta}
\!-\!\frac{1\!\!-\!a^{2n}(t')}{2n}\Bigg\}
\; .\label{quant1loopsecnd}
\ee Combining Eqs.~(\ref{1loop1result}) and (\ref{quant1loopsecnd}) in Eq.~(\ref{1loopcorr})
yields the one-loop correlator, for the massive scalar, as a power series expansion
\beeq
\langle\Omega|
\bar{\varphi}(t,\vec{x})\bar{\varphi}(t'\!,\vec{x}\,')|\Omega\rangle_{\rm 1-loop}\!\simeq\!-\frac
{\lambda}{2\nu}\frac{{\mathcal{A}}^2}{H^2}\,\emph{f}_{1}(t,t'\!,\Delta x)\;,\label{1loopmassivefinal}
\eneq
where we define the spacetime and mass dependent function\be
&&\hspace{-0.45cm}\emph{f}_1(t,t'\!,\Delta x)\!=\!\!\frac{\left[a(t)\,a(t')\right]^{-\frac{\delta}{2}}}
{2\,\delta}\!\!\sum_{n=0}^\infty
\!\frac{(-1)^{n}(H\!\Delta x)^{2n}}{(2n\!\!+\!\!1\!)!(2n\!\!+\!\delta)}
\Bigg\{\!a^{2n}(t')\!
\Bigg[a^{\delta}(t')\!\Bigg(\!\!\ln(a(t))\!-\!\ln(a(t'))\!\!+\!\frac{a^{-\delta}(t)}{\delta}\!\Bigg)
\!\!-\!\!\frac{1}{\delta}\Bigg]\nonumber\\
&&\hspace{0.65cm}-\ln(a(t))\!-\!\ln(a(t'))
\!+\!\frac{1\!\!-\!a^{-\delta}(t)}{\delta}\!+\!\frac{1\!\!-\!a^{-\delta}(t')}{\delta}
\!-\!2\!\left[\frac{1\!-\!a^{2n+\delta}(t')}{2n\!\!+\!\delta}\right]\!\!+\!
\frac{1\!\!-\!a^{2n}(t')}{n}\!\Bigg\}\; .\label{fmassive}
\ee
Using the equal spacetime limit of Eq.~(\ref{fmassive}) in Eq.~(\ref{1loopmassivefinal}) yields the VEV of the field strength squared
\beeq
\langle\Omega|
\bar{\varphi}^2(t,\vec{x})|\Omega\rangle_{\rm 1-loop}\!\simeq\!-\frac
{\lambda}{2\nu}\frac{{\mathcal{A}}^2}{H^2}
\frac{a^{-\delta}(t)}{\delta^2}\left[-2\ln(a(t))
\!+\!\frac{a^{\delta}(t)\!-\!a^{-\delta}(t)}{\delta}\right]\;.\label{1loopeqlspctm}
\eneq
The $m\!\!\rightarrow\!0$ limit of Eq.~(\ref{fmassive}), on the other hand, yields the one-loop correlator for the massless scalar; see Appendix~\ref{App:oneloopcorrmassless}.

Summing up the infinite sum in Eq.~(\ref{fmassive}) gives an analytic form for the one-loop correlator; see Appendix~\ref{App:oneloopcorrexpanalyt}. Equation~(\ref{1loopanalyt}) can be used to study the behaviors of one-loop correlator  (i) at a {\it fixed comoving separation} at late times and (ii) for a {\it fixed physical distance} as in the tree-order computation in Sec.~\ref{subsect:treecorr}.

In case (i) asymptotic limit of Eq.~(\ref{1loopanalyt}), at a fixed comoving separation $\Delta x$, is
\be
&&\hspace{-0.9cm}\emph{f}_1(t,t'\!,\Delta x)\!\rightarrow\!\frac{1}{\delta}\!
\frac{\sqrt{\pi}}{2^{3-\delta}}
\frac{\Gamma\!\left(\frac{\delta}{2}\right)}{\Gamma\!\left(\frac{3-\delta}{2}\right)}\!\left[ \alpha(t)\,\alpha(t')\right]^{-\frac{\delta}{2}}\!\Bigg\{\!\!
\ln\!\Big(\!a(t)a(t')\!\Big)\!-\!\frac{1\!\!-\!a^{-\delta}(t)}{\delta}\!+\!\frac{1\!\!-\!a^{-\delta}(t')}{\delta}
\!+\!\mathcal{C}_1\!\Bigg\}
\nonumber\\
&&\hspace{-1.25cm}
-\frac{1}{2\delta^2}\!\left[a(t)a(t')\right]^{-\frac{\delta}{2}}
\!\Bigg\{\!\!\!\left[\ln\!\Big(\!a(t)a(t')\!\Big)\!\!-\!\!\frac{1\!\!-\!a^{-\delta}(t)}{\delta}
\!-\!\frac{1\!\!-\!a^{-\delta}(t')}{\delta}\right]\!\!
{}_1\mathcal{F}_2\Big(\!\frac{\delta}{2};\!\frac{3}{2},\!1\!\!+\!\!\frac{\delta}{2};\!-\frac{(H\!\Delta x)^2}{4}\!\Big)\!\!+\!\!2\,\mathcal{C}_2\!\Bigg\},\label{1loopasympt}
\ee
where\be
&&\hspace{1.5cm}\mathcal{C}_1(\Delta x)\!=\!\ln\!\left(\!\frac{(H\!\Delta x)^2}{4}\!\right)\!\!-\!\psi\Big(\frac{\delta}{2}\Big)\!\!-\!\psi\Big(\frac{3\!-\!\delta}{2}\Big)\; ,\\
&&\hspace{0.4cm}\mathcal{C}_2(\Delta x)\!=\!\ln(H\!\Delta x)\!-\!{\rm ci}(H\!\Delta x)
\!+\!\frac{1\!\!+\!\delta}{\delta}\!\left[\frac{\sin(H\!\Delta x)}{H\!\Delta x}\!-\!1\right]\!\!+\!\gamma\nonumber\\
&&\hspace{-1.8cm}+\frac{1}{\delta}
{}_2\mathcal{F}_3\Big(\!\frac{\delta}{2}, \frac{\delta}{2};\frac{3}{2},\!1\!\!+\!\frac{\delta}{2},\!1\!\!+\!\frac{\delta}{2};-\frac{(H\!\Delta x)^2}{4}\!\Big)\!+\!\frac{(H\!\Delta x)^2}{3\delta(2\!+\!\delta)}
{}_1\mathcal{F}_2\Big(\!1\!\!+\!\frac{\delta}{2};\frac{5}{2},\!2\!\!+\!\frac{\delta}{2};-\frac{(H\!\Delta x)^2}{4}\!\Big)\;.
\ee
The digamma function $\psi(z)$ is defined in Eq.~(\ref{defndigamma}). Equation~(\ref{1loopasympt}) implies that the one-loop correlator of a massive scalar for a fixed comoving separation asymptotes to zero as $\mathcal{O}({\lambda\ln\!\left(a(t)\,a(t')\right)}/{\left[a(t)\,a(t')\right]^{\frac{\delta}{2}}})$ during inflation. For the massless scalar, Eq.~(\ref{1loopmassivefinal}) and the asymptotic limit of Eq.~(\ref{1loopmassless2}),\be
&&\hspace{0cm}\lim_{m\rightarrow 0}\!\emph{f}_1\!\rightarrow\!\frac
{1}{4}\Big[\!\ln^2(a(t))\!+\!\ln^2(a(t'))\Big]\!\!\left[\frac{\sin(H\!\Delta x)}{H\!\Delta x}\!-\!{\rm ci}(H\!\Delta x)\right]\!\!+\!\mathcal{C}_0\; , \label{1loopmasslessasymp}
\ee
where
\be
&&\hspace{0.1cm}
\mathcal{C}_0(\Delta x)\!=\!\frac{1}{6}\ln^3(H\!\Delta x)\!-\!\frac{1}{2}\!\left[1\!-\!\gamma\right]\ln^2(H\!\Delta x)\!+\!\!\left[1\!\!-\!\gamma\!+\!\!\frac{\gamma^2}{2}\!-\!\frac{\pi^2}{24}\right]\!\ln(H\!\Delta x)\nonumber\\
&&\hspace{-1.7cm}-\frac{(H\!\Delta x)^2}{48}{}_4\mathcal{F}_5\Big(\!1,1,1,1
;2,2,2,2,\frac{5}{2};-\frac{(H\!\Delta x)^2}{4}\Big)\!\!-\!\frac{5}{6}\!+\!\frac{\pi^2}{24}(1\!-\!\gamma)
\!+\!\frac{1}{3}\zeta(3)\!+\!\frac{\gamma}{2}\!-\!\frac{1}{6}(1\!-\!\gamma)^3\; ,\label{1loopmasslessasymp2}
\ee
imply that the one-loop correlator grows negatively and proportional to $\ln^2(a)$ during inflation even for a $\Delta x$ of order the horizon size $1/H$ since $\frac{\sin(H\!\Delta x)}{H\!\Delta x}\!-\!{\rm ci}(H\!\Delta x)\!>\!0$. Thus, tree-order result~(\ref{zeromasstreeaympi}) decreases when quantum corrections are included.

\begin{figure}
\includegraphics[width=10cm,height=6.3cm]{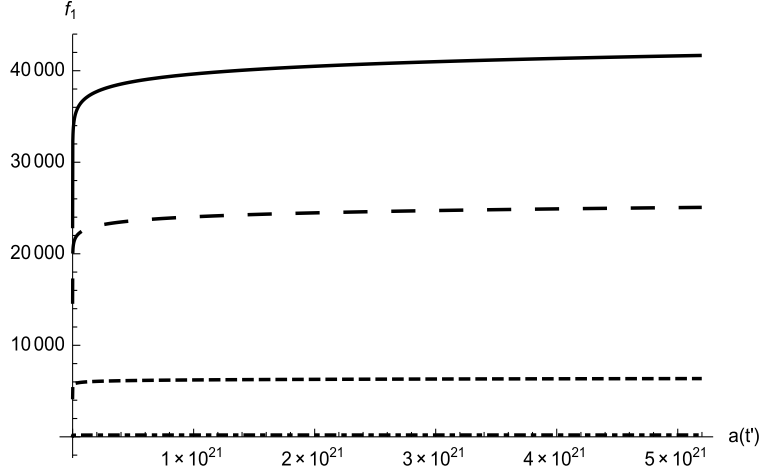}
\caption{Plots of the function $f_1(t,t', K)$, defined in Eq.~(\ref{1loopcorrfixeddist}), versus $a(t')$ for  different mass values. The scale factor $a(t)$ and the fraction $K$ are chosen to be $e^{50}$ and $1/2$, respectively. $a(t')$ runs from $1$ to $a(t)$. The solid curve is the plot for a massless scalar. The large-dashed and dashed curves are for massive scalars with $m\!=\!H/8$ and $m\!=\!H/4$, respectively. The dot-dashed curve, which is barely seen on top of the horizontal axis, corresponds to $m\!=\!H/2$.}
\label{fig:oneloop-corr}
\end{figure}

In case (ii) to obtain the one-loop correlator for a fixed physical distance, we choose $a(t')\Delta x$ as a constant fraction $K$ of the Hubble length, as in the corresponding tree-order computation in Sec.~\ref{subsect:treecorr}. Using Eq.~(\ref{fixedphysdist}) in Eq.~(\ref{fmassive}) yields a power series representation of the one-loop correlator. Alternatively, employing Eq.~(\ref{fixedphysdist}) in Eq.~(\ref{1loopanalyt}) we find\be
&&\hspace{-1cm}\langle\Omega|
\bar{\varphi}(t,\vec{x})\bar{\varphi}(t'\!,\vec{x}\,')|\Omega\rangle_{\rm 1-loop}\!\simeq\!-\frac
{\lambda}{2\nu}\frac{{\mathcal{A}}^2}{H^2}\,\emph{f}_1(t, t'\!, K)\; ,\label{oneloopcorrK}
\ee
where the time and $K$ dependent function
\be
&&\hspace{-0.6cm}\emph{f}_1(t, t'\!, K)\!=\!\frac{\left[a(t)\,a(t')\right]^{-\frac{\delta}{2}}}{\delta^2}
\Bigg\{\!\frac{1}{2}\!\left\{\!a^{\delta}(t')\!\!\left[\ln(a(t))\!-\!\ln(a(t'))
\!+\!\frac{a^{-\delta}(t)}{\delta}\right]
\!\!-\!\!\frac{1}{\delta}\!\right\}\!
{}_1\mathcal{F}_2\Big(\frac{\delta}{2};\frac{3}{2},1\!\!+\!\frac{\delta}{2};-\frac{K^2}{4}\Big)\nonumber\\
&&\hspace{1cm}-\frac{1}{2}\!\left\{\ln(a(t))\!+\!\ln(a(t'))
\!-\!\frac{1\!-\!a^{-\delta}(t)}{\delta}\!-\!\frac{1\!-\!a^{-\delta}(t')}{\delta}\right\}\!
{}_1\mathcal{F}_2\Big(\frac{\delta}{2};\frac{3}{2},1\!\!+\!\frac{\delta}{2};\!\frac{-K^2}{4a^2(t')}\Big)\nonumber\\
&&\hspace{0.8cm}+\frac{1}{\delta}
\left\{a^{\delta}(t')\,{}_2\mathcal{F}_3\Big(\frac{\delta}{2},\frac{\delta}{2}
;\frac{3}{2},\!1\!\!+\!\frac{\delta}{2},\!1\!\!+\!\frac{\delta}{2};\!-\frac{K^2}{4}\!\Big)
\!-\!{}_2\mathcal{F}_3\Big(\frac{\delta}{2},\frac{\delta}{2}
;\frac{3}{2},\!1\!\!+\!\frac{\delta}{2},\!1\!\!+\!\frac{\delta}{2};\!\frac{-K^2}{4a^2(t')}\Big)\!\right\}\nonumber\\
&&\hspace{2.4cm}-\Big\{{\rm ci}(K)\!-\!{\rm ci}(Ka^{-1}(t'))\Big\}\!\!+\!\!\left[1\!\!+\!\frac{1}{\delta}\right]\!\!\!\left[\frac{\sin(K)}{K}\!-\!
\frac{\sin(Ka^{-1}(t'))}{Ka^{-1}(t')}\right]\nonumber\\
&&\hspace{1cm}+\frac{K^2}{3\delta(2\!+\!\delta)}
\Bigg[{}_1\mathcal{F}_2\Big(\!1\!\!+\!\frac{\delta}{2};
\frac{5}{2},2\!+\!\!\frac{\delta}{2};-\frac{K^2}{4}\!\Big)\!-\!a^{-2}(t')
{}_1\mathcal{F}_2\Big(\!1\!\!+\!\frac{\delta}{2};\frac{5}{2},
2\!+\!\!\frac{\delta}{2};\!\frac{-K^2}{4a^2(t')}\Big)\!\Bigg]\!\Bigg\}\; ,\label{1loopcorrfixeddist}
\ee
is positive definite and grows logarithmically in the massless limit; see Eq.~(\ref{1loopmassless2}). Hence one-loop correlator~(\ref{oneloopcorrK}), at fixed physical distance, grows negatively during inflation. This growth, however, is suppressed for massive scalars; see Fig.~\ref{fig:oneloop-corr}. As can be seen in the plots of the function $\emph{f}_1(t\!=\!\frac{50}{H}, t'\!, K\!\!=\!\frac{1}{2})$ for $m\!=\!0,H/8,H/4$ and $H/2$, the suppression is very sensitive to an increase in mass. In fact, for masses $m\!\geq\!\! H/2$ the one-loop correlator is effectively zero at late times.

One can also infer the late time behavior of one-loop correlator~(\ref{oneloopcorrK}) from the asymptotic form of $f_1(t,t', K)$ which can be obtained taking the late time limit of Eq.~(\ref{1loopcorrfixeddist}). Up to order $a^{-2-\delta}(t')$, we find
\be
&&\hspace{-0.9cm}\emph{f}_1\!\rightarrow\! \frac{1}{\delta^2}\! \left[\frac{a(t')}{a(t)}\right]^\frac{\delta}{2}\!\!\Bigg\{\!\frac{1}{2}\!
\left[\ln(a(t))\!-\!\ln(a(t'))\!+\!\frac{a^{-\delta}(t)}{\delta}\!-\!\frac{a^{-\delta}(t')}{\delta}\right]\!
{}_1\mathcal{F}_2\Big(\frac{\delta}{2};\!
\frac{3}{2},\!1\!\!+\!\frac{\delta}{2};\!-\frac{K^2}{4}\!\Big)\nonumber\\
&&\hspace{-1.5cm}+\frac{1}{\delta}\,{}_2\mathcal{F}_3\Big(\frac{\delta}{2},\frac{\delta}{2}
;\frac{3}{2},\!1\!\!+\!\frac{\delta}{2},\!1\!\!+\!\frac{\delta}{2};\!-\frac{K^2}{4}\!\Big)
\!\Bigg\}\!-\!\frac{\left[\,a(t)\,a(t')\right]^{-\frac{\delta}{2}}}{\delta^2}
\Bigg\{\!\frac{1}{2}
\Bigg[\!\ln(a(t))\!+\!3\ln(a(t'))\!+\!\frac{1\!\!+\!a^{-\delta}(t)}{\delta}\nonumber\\
&&\hspace{-1.35cm}+\frac{1\!\!+\!a^{-\delta}(t')}{\delta}\!\Bigg]\!\!\!+\!{\rm ci}(K)\!-\!\!\left[1\!\!+\!\frac{1}{\delta}\right]\!\frac{\sin(K)}{K}\!-\!\ln(K)\!-\!\!\gamma
\!\!+\!1\!\!-\!\frac{K^2}{3\delta(2\!+\!\delta)}{}_1\mathcal{F}_2\Big(\!1\!\!+\!\frac{\delta}{2};\!
\frac{5}{2},\!2\!+\!\frac{\delta}{2};\!-\frac{K^2}{4}\!\Big)\!\!
\Bigg\}\; .
\ee
Thus, the growth is very slight for a massive ($\delta\!>\!0$) scalar, at late times. For the massless scalar, on the other hand, the asymptotic form up to order $\frac{\ln^2(a(t))}{a^2(t)}$,\be
&&\hspace{-0.4cm}\lim_{m\rightarrow 0}\!\emph{f}_1\!\!\rightarrow\!\!\frac
{1}{4}
\Bigg\{\!\!\ln^2(a(t))\ln(a(t'))\!+\!\frac{\ln^3(a(t'))}{3}
\!+\!\!\Big[\!\ln^2(a(t))\!-\!\ln^2(a(t'))\Big]\!\!\!\left[{\rm ci}(\!K)
\!-\!\frac{\sin(\!K)}{K}\!-\!\ln(\!K)\!-\!\!\gamma\!\!+\!\!1\!\right]\nonumber\\
&&\hspace{0.5cm}
-\frac{K^2}{6}
{}_{3}\mathcal{F}_{4}\!\!\left(\!1,\!1,\!1;2,\!2,\!2,\!\frac{5}{2};
-\frac{K^2}{4}\!\right)\!\ln(a(t'))\!+\!\frac{K^2}{12}
{}_{4}\mathcal{F}_{5}\!\!\left(\!1,\!1,\!1,\!1;2,\!2,\!2,\!2,\!\frac{5}{2};
-\frac{K^2}{4}\!\right)\!\!\Bigg\}\; ,\label{massless1loopasym}
\ee
implies a stronger---though logarithmic--- growth compared to the massive case. [Note that Eq.~(\ref{massless1loopasym}) is also the late time limit of Eq.~(\ref{1loopmassless2}), for $\Delta x\!=\!{K}/{Ha(t')}$.]

In the next section, we compute the two-loop contribution to the correlation function of the IR truncated scalar field in our model.

\subsection{Two-loop correlator}
\label{subsect:2loopcorr}

Computation of two-loop contribution~(\ref{correlation}) in two-point correlation function~(\ref{genelfullexpect}) involves evaluations of three VEVs each of which has a double time integral without a definite time ordering in the integrand. Hence the computation is more intricate than that of the one-loop correlator which required evaluations of two VEVs each of which had a single time integral with a trivial time ordering in the integrand.

Let us start evaluating the first VEV in Eq.~(\ref{correlation}),
\be
&&\hspace{2.5cm}\langle\Omega|
\bar{\varphi}_0(t,\vec{x})\!\!\int_0^{t'}\!\!\!\!d\tilde{t}\,{\bar{\varphi}}^2_0(\tilde{t}\!,\vec{x}\,')
\!\!\int_0^{\tilde{t}}\!\!d\tilde{\tilde{t}}\,a^\frac{\delta}{2}(\tilde{\tilde{t}}) {\bar{\varphi}}^3_0(\tilde{\tilde{t}},\vec{x}\,')
|\Omega\rangle\nonumber\\
&&\hspace{-1cm}\!\!=\!\!\!\int_0^{t'}\!\!\!\!d\tilde{t}\!\!\int_0^{\tilde{t}}\!\!\!d\tilde{\tilde{t}}
\,a^\frac{\delta}{2}(\tilde{\tilde{t}})\Bigg\{\!2\!\cdot\!3\!\cdot\!1\,\langle\Omega|
\bar{\varphi}_0(t,\vec{x})\bar{\varphi}_0(\tilde{t},\vec{x}\,')|\Omega\rangle\langle\Omega|
\bar{\varphi}_0(\tilde{t},\vec{x}\,')\bar{\varphi}_0(\tilde{\tilde{t}}\!,\vec{x}\,')|\Omega\rangle\langle\Omega|
\bar{\varphi}^2_0(\tilde{\tilde{t}},\vec{x}\,')|\Omega\rangle\nonumber\\
&&\hspace{1cm}+3\!\cdot\!1\!\cdot\!1\,\langle\Omega|
\bar{\varphi}_0(t,\vec{x})\bar{\varphi}_0(\tilde{\tilde{t}},\vec{x}\,')|\Omega\rangle\langle\Omega|
\bar{\varphi}^2_0(\tilde{t},\vec{x}\,')|\Omega\rangle\langle\Omega|
\bar{\varphi}^2_0(\tilde{\tilde{t}},\vec{x}\,')|\Omega\rangle\nonumber\\
&&\hspace{1.1cm}+3\!\cdot\!2\!\cdot\!1\,\langle\Omega|
\bar{\varphi}_0(t,\vec{x})\bar{\varphi}_0(\tilde{\tilde{t}},\vec{x}\,')|\Omega\rangle\!\left[\langle\Omega|
\bar{\varphi}_0(\tilde{t},\vec{x}\,')\bar{\varphi}_0(\tilde{\tilde{t}}\!,\vec{x}\,')|\Omega\rangle\right]^2
\!\!\Bigg\}\; ,\label{twoloopfirstexpandequaltime}
\ee
where we have $0\!\leq\!\tilde{\tilde{t}}\!\leq\!\tilde{t}\!\leq\!t'$ and $t'\!\leq\!t$. This time ordering makes it easy to read off the VEVs in the integrand from Eqs.~(\ref{treecorrseries}), (\ref{massiveequalspaceVEV}) and (\ref{phisquare}). Substituting the outcomes into Eq.~(\ref{twoloopfirstexpandequaltime}) turns the computation into the evaluations of three double integrals, \be
&&\hspace{1.75cm}\langle\Omega|
\bar{\varphi}_0(t,\vec{x})\!\!\int_0^{t'}\!\!\!d\tilde{t}{\bar{\varphi}}^2_0(\tilde{t}\!,\vec{x}\,')
\!\!\int_0^{\tilde{t}}\!\!d\tilde{\tilde{t}}\,a^\frac{\delta}{2}(\tilde{\tilde{t}}) {\bar{\varphi}}^3_0(\tilde{\tilde{t}},\vec{x}\,')
|\Omega\rangle\nonumber\\
&&\hspace{-1.4cm}\!=\!{\mathcal{A}}^3\frac{a^{-\frac{\delta}{2}}(t)}{\delta^2}6\Bigg\{\!\!\sum_{n=0}^\infty\!
                 \frac{(-1)^{n}(H\!\Delta x)^{2n}}{(2n\!\!+\!\!1)!(2n\!\!+\!\delta)}\!\Bigg[\!\!\int_0^{t'}\!\!\!d\tilde{t}
                 \Big[a^{2n}\!(\tilde{t})\!-\!a^{-\delta}\!(\tilde{t})\Big]\!\!\int_0^{\tilde{t}}\!\!d\tilde{\tilde{t}}
                 \Big[a^{\delta}(\tilde{\tilde{t}})\!+\!a^{-\delta}(\tilde{\tilde{t}})\!-\!2\Big]\nonumber\\
&&\hspace{1cm}+\frac{1}{2}\!\!\int_0^{t'}\!\!\!d\tilde{t}
                 \Big[1\!-\!a^{-\delta}(\tilde{t})\Big]
                 \!\!\int_0^{\tilde{t}}\!\!d\tilde{\tilde{t}}
                 \Big[a^{2n+\delta}(\tilde{\tilde{t}})\!-\!a^{2n}(\tilde{\tilde{t}})
                 \!+\!a^{-\delta}(\tilde{\tilde{t}})\!-\!1\Big]\nonumber\\
&&\hspace{1.4cm}+\!\!\int_0^{t'}\!\!\!d\tilde{t}a^{-\delta}(\tilde{t})
                 \!\!\int_0^{\tilde{t}}\!\!d\tilde{\tilde{t}}
                 \Big[a^{2n}(\tilde{\tilde{t}})\!-\!a^{-\delta}(\tilde{\tilde{t}})\Big]\!
                 \Big[a^{\delta}(\tilde{\tilde{t}})\!-\!1\Big]^2\Bigg]\!\Bigg\}\; .\label{2loop1stequaltime}
\ee
The results of the three double integrals in Eq.~(\ref{2loop1stequaltime}) are given in Appendix \ref{App:VEV1}. Employing Eqs.~(\ref{2loop1st1stintequaltime})-(\ref{2loop1st3rdintequaltime}) in Eq.~(\ref{2loop1stequaltime}) we express the first VEV in the two-loop correlator as\be
&&\hspace{-0.45cm}\langle\Omega|
\bar{\varphi}_0(t,\vec{x})\!\!\int_0^{t'}\!\!\!d\tilde{t}\,{\bar{\varphi}}^2_0(\tilde{t},\vec{x}\,')
\!\!\int_0^{\tilde{t}}\!\!d\tilde{\tilde{t}}\,a^\frac{\delta}{2}(\tilde{\tilde{t}}) {\bar{\varphi}}^3_0(\tilde{\tilde{t}},\vec{x}\,')
|\Omega\rangle\!=\!-\frac{{\mathcal{A}}^3}{H^2}\frac{a^{-\frac{\delta}{2}}(t)}{\delta^2}6\!\sum_{n=0}^\infty\!
                 \frac{(-1)^{n}(H\!\Delta x)^{2n}}{(2n\!\!+\!\!1)!(2n\!\!+\!\delta)}\Bigg\{\!\frac{\ln^2(a(t'))}{4}\nonumber\\
&&\hspace{0cm}-\!\!\left[\frac{1\!-\!a^{2n}(t')}{n}\!+\!\frac{9}{2}
\Big[\frac{1\!-\!a^{-\delta}(t')}{\delta}\Big]
\!\!-\!\frac{6}{\delta}\!-\!\frac{3}{4n}\!-\!\frac{1}{2(2n\!\!+\!\delta)}\right]\!\ln(a(t'))
\!+\!\left[\frac{1}{\delta}\!+\!\frac{1}{2(n\!\!+\!\delta)}\!+\!\frac{1}{2(2n\!\!+\!\delta)}\right]\nonumber\\
&&\hspace{1.1cm}\times\!\!\left[\frac{1\!-\!a^{2n+\delta}(t')}{2n\!\!+\!\delta}\right]
\!\!+\!\!\left[\frac{3}{4n}\!-\!\frac{5}{2(2n\!\!+\!\delta)}\right]\!\!\left[\frac{1\!-\!a^{2n}(t')}{2n}\right]
\!\!-\!\left[\frac{1}{\delta}\!-\!\frac{3}{4n}\right]\!
\!\left[\frac{1\!-\!a^{2n-\delta}(t')}{2n\!\!-\!\delta}\right]\nonumber\\
&&\hspace{1.6cm}-\!\!\left[\frac{7}{2\delta}
\!-\!\frac{3}{4n}\!-\!\frac{1}{2(n\!\!+\!\delta)}\!+\!\frac{5}{2(2n\!\!+\!\delta)}\right]
\!\!\left[\frac{1\!-\!a^{-\delta}(t')}{\delta}\right]
\!\!-\!\frac{5}{2\delta}\!\left[\frac{1\!-\!a^{-2\delta}(t')}{2\delta}\right]\!\!\Bigg\}\; .\label{2loop1stequaltimeresult}
\ee
Note that the $n=0$ term in the infinite sum in Eq.~(\ref{2loop1stequaltimeresult}) is finite---though it includes terms proportional to $n^{-1}$ and $n^{-2}$; see Eq.~(\ref{n0term1stvev}).

The second VEV that contributes at two-loop order in Eq.~(\ref{correlation}) is evaluated similarly,
\be
&&\hspace{3cm}\langle\Omega|\!\!
\int_0^{t}\!\!dt''{\bar{\varphi}}^2_0(t''\!,\vec{x})\!\!
\int_0^{t''}\!\!\!\!dt'''a^\frac{\delta}{2}(t''') {\bar{\varphi}}^3_0(t'''\!,\vec{x})
\bar{\varphi}_0(t'\!,\vec{x}\,')
|\Omega\rangle\nonumber\\
&&\hspace{-0.5cm}\!=\!\!\int_0^{t}\!\!dt''\!\!\!\int_0^{t''}\!\!\!\!dt'''a^\frac{\delta}{2}(t''')\Bigg\{\!2\!\cdot\!3\!\cdot\!1\,\langle\Omega|
\bar{\varphi}_0(t''\!,\vec{x})\bar{\varphi}_0(t'\!,\vec{x}\,')|\Omega\rangle\,\langle\Omega|
\bar{\varphi}_0(t''\!,\vec{x})\bar{\varphi}_0(t'''\!,\vec{x})|\Omega\rangle\,\langle\Omega|
\bar{\varphi}^2_0(t'''\!,\vec{x})|\Omega\rangle\nonumber\\
&&\hspace{1.8cm}+1\!\cdot\!3\!\cdot\!1\,\langle\Omega|
\bar{\varphi}^2_0(t''\!,\vec{x})|\Omega\rangle\,\langle\Omega|
\bar{\varphi}_0(t'''\!,\vec{x})\bar{\varphi}_0(t'\!,\vec{x}\,')|\Omega\rangle\,\langle\Omega|
\bar{\varphi}^2_0(t'''\!,\vec{x})|\Omega\rangle\nonumber\\
&&\hspace{2cm}+3\!\cdot\!2\!\cdot\!1\Big[\langle\Omega|
\bar{\varphi}_0(t''\!,\vec{x})\bar{\varphi}_0(t'''\!,\vec{x})|\Omega\rangle\Big]^2\,\langle\Omega|
\bar{\varphi}_0(t'''\!,\vec{x})\bar{\varphi}_0(t'\!,\vec{x}\,')|\Omega\rangle
\!\Bigg\}\; ,\label{quant2loopsecnd}
\ee
where $0\!\leq\!t'''\!\leq\!t''\!\leq\!t$ and $t'\!\leq\!t$. At this stage we employ Eqs.~(\ref{massiveequalspaceVEV}) and (\ref{phisquare}) in Eq.~(\ref{quant2loopsecnd}) to bring it into a simpler form which includes integrals each with a single VEV in the integrands,
\be
&&\hspace{2cm}\langle\Omega|\!\!
\int_0^{t}\!\!\!dt''{\bar{\varphi}}^2_0(t''\!,\vec{x})\!\!
\int_0^{t''}\!\!\!\!dt'''a^\frac{\delta}{2}(t'''){\bar{\varphi}}^3_0(t'''\!,\vec{x})
\bar{\varphi}_0(t'\!,\vec{x}\,')
|\Omega\rangle\nonumber\\
&&\hspace{-0.9cm}\!=\!\frac{{\mathcal{A}}^2}{\delta^2}6\Bigg\{\!\!\int_0^{t}\!\!\!dt''\!\!\!\int_0^{t''}\!\!\!\!dt'''
a^\frac{\delta}{2}(t''')\langle\Omega|
\bar{\varphi}_0(t''\!,\vec{x})\bar{\varphi}_0(t'\!,\vec{x}\,')|\Omega\rangle\!\Big[a(t'')\,a(t''')\Big]
^{-\frac{\delta}{2}}\Big[a^\delta(t''')\!-\!\!1\Big]\!\Big[1\!\!-\!a^{-\delta}(t''')\Big]\nonumber\\
&&\hspace{0.5cm}+\frac{1}{2}\!\int_0^{t}\!\!\!dt''\!\!\!\int_0^{t''}\!\!\!\!dt'''a^\frac{\delta}{2}(t''') \langle\Omega|
\bar{\varphi}_0(t'''\!,\vec{x})\bar{\varphi}_0(t'\!,\vec{x}\,')|\Omega\rangle\!\Big[1\!-\!a^{-\delta}(t'')\Big]
\!\Big[1\!\!-\!a^{-\delta}(t''')\Big]\nonumber\\
&&\hspace{0.4cm}+\!\!\int_0^{t}\!\!\!dt''\!\!\!\int_0^{t''}\!\!\!\!dt'''a^\frac{\delta}{2}(t''')\langle\Omega|
\bar{\varphi}_0(t'''\!,\vec{x})\bar{\varphi}_0(t'\!,\vec{x}\,')|\Omega\rangle\!\Big[a(t'')\,a(t''')\Big]
^{-\delta}\Big[a^\delta(t''')\!-\!\!1\Big]^2
\!\Bigg\}\; .\label{twoloopsecondexpandequaltime}
\ee
Because the orderings between $t'$ and $t''$ and between $t'$ and $t'''$ are unknown in the double integrals, we cannot just read off the VEVs from Eq.~(\ref{treecorrseries}). We can, however, decompose the double integrals so that a definite ordering between the time parameters exists. That makes the evaluation of the integrals, using Eq.~(\ref{treecorrseries}), possible; see Appendix \ref{App:VEV2}. Substituting Eqs.~(\ref{firstint}), (\ref{secondint}) and (\ref{thirdint}) into Eq.~(\ref{twoloopsecondexpandequaltime}), we find
\be
&&\hspace{-0.35cm}\langle\Omega|\!\!
\int_0^{t}\!\!dt''{\bar{\varphi}}^2_0(t''\!,\vec{x})\!\!
\int_0^{t''}\!\!\!\!dt'''a^\frac{\delta}{2}(t'''){\bar{\varphi}}^3_0(t'''\!,\vec{x})
\bar{\varphi}_0(t'\!,\vec{x})
|\Omega\rangle\!=\!-\frac{{\mathcal{A}}^3}{H^2}\frac{a^{-\frac{\delta}{2}}(t')}{\delta^2}3\!\sum_{n=0}^\infty\!
                 \frac{(-1)^{n}(H\!\Delta x)^{2n}}{(2n\!\!+\!\!1)!(2n\!\!+\!\delta)}\nonumber\\
&&\hspace{-0.3cm}\times\!\Bigg\{\!\!\left[\frac{1\!-\!a^{2n+\delta}(t')}{2}\right]\!
                 \ln^2(a(t))\!+\!a^{2n+\delta}(t')\!\left[\ln(a(t))\!-\!\frac{\ln(a(t'))}{2}\right]\!\ln(a(t'))
                 \!+\!\!\Bigg[\!\!\left[\frac{1\!-\!a^{2n+\delta}(t')}{\delta}\right]\nonumber\\
&&\hspace{-0.1cm}\times\!\!\left[9a^{-\delta}(t)\!+\!\frac{\delta}{2n\!\!+\!\delta}\!+\!4\right]
\!\!-\!\!\left[\frac{1\!-\!a^{2n}(t')}{\delta}\right]\!\!\left[1\!+\!\frac{\delta}{2n}\right]\!\Bigg]
\!\ln(a(t))\!+\!\Bigg[\frac{a^{2n}(t')}{\delta}\!
\left[2a^{\delta}(t')\!+\!\frac{3\delta}{2n}\!+\!3\right]\nonumber\\
&&\hspace{-0.5cm}
+\frac{a^{2n+\delta}(t')}{\delta}\!\left[5a^{-\delta}(t)\!+\!\frac{\delta}{2n\!\!+\!\delta}\!+\!2\right]\!\Bigg]
\!\ln(a(t'))\!-\!\!\left[\frac{1\!-\!a^{2n+2\delta}(t')}{\delta}\right]\!\!
\left[\frac{a^{-\delta}(t)}{\delta}\!\left[\frac{\delta}{n\!\!+\!\delta}\!-\!2\right]
\!\!-\!\frac{a^{-\delta}(t')}{n\!\!+\!\delta}\right]\nonumber\\
&&\hspace{-0.4cm}+\!\left[\frac{1\!-\!a^{2n+\delta}(t')}{\delta}\right]\!\!
\Bigg[\frac{a^{-\delta}(t)}{\delta}\!\!\left[\frac{5a^{-\delta}(t)}{2}\!+\!\frac{5\delta}{2n\!\!+\!\delta}\!+\!8\right]
\!\!+\!\frac{a^{-\delta}(t')}{\delta}\!\!\left[a^{-\delta}(t')\!-\!\frac{4\delta}{2n\!\!+\!\delta}\right]
\!\!+\!\frac{\delta}{2n\!\!+\!\delta}\!\Bigg[\frac{1}{2n\!\!+\!\delta}
\nonumber\\
&&\hspace{-0.25cm}
+\frac{1}{n\!\!+\!\delta}\!+\!\frac{2}{\delta}\Bigg]
\!\!-\!\frac{2}{\delta}\Bigg]\!\!-\!\!\left[\frac{1\!-\!a^{2n}(t')}{\delta}\right]\!\!\Bigg[\frac{a^{-\delta}(t)}{\delta}
\!\left[\frac{3\delta}{2n}\!+\!3\right]\!\!-\!\frac{3a^{-\delta}(t')}{2n}
\!+\!\frac{1}{2n\!\!+\!\delta}\!\!\left[\frac{5\delta}{2n}\!+\!1\right]\!\!-\!
\frac{3\delta}{4n^2}\!+\!\frac{8}{\delta}\Bigg]\nonumber\\
&&\hspace{-0.9cm}+\!\!\left[\frac{1\!\!-\!a^{2n-\delta}(t')}{\delta}\right]\!\!\left[
\frac{\delta}{2n\!\!-\!\delta}\!\!\left[\frac{3}{2n}\!-\!\frac{2}{\delta}\right]\!\!-\!\frac{1}{2\delta}\right]
\!\!+\!\!\left[\frac{1\!\!-\!a^{-\delta}(t')}{\delta}\right]\!\!\left[\frac{a^{-\delta}(t')}{\delta}
\!+\!\frac{3}{2n}\!+\!\frac{1}{n\!\!+\!\delta}\!-\!\frac{4}{2n\!\!+\!\delta}\!+\!\frac{1}{\delta}\right]
\!\!\Bigg\}\; .\label{2loop2ndresult}
\ee
Note that the $n=0$ term in the infinite sum in Eq.~(\ref{2loop2ndresult}) is finite---though it includes terms proportional to $n^{-1}$ and $n^{-2}$; see Eq.~(\ref{n0term2ndvev}).

Finally, the third VEV that contributes at two-loop order in Eq.~(\ref{correlation})~is
\be
&&\hspace{3cm}\langle\Omega|\!\!
\int_0^{t}\!\!dt'' a^{\frac{\delta}{2}}(t''){\bar{\varphi}}^3_0(t''\!,\vec{x})\!\!
\int_0^{t'}\!\!\!d\tilde{t} \, a^{\frac{\delta}{2}}(\tilde{t}){\bar{\varphi}}^3_0(\tilde{t},\vec{x}\,')
|\Omega\rangle\nonumber\\
&&\hspace{-0.5cm}\!=\!\!\int_0^{t}\!\!dt''\, a^{\frac{\delta}{2}}(t'')\!\!\int_0^{t'}\!\!\!d\tilde{t}\,a^{\frac{\delta}{2}}(\tilde{t})\Bigg\{3\!\cdot\!1\!\cdot\!1\,\langle\Omega|
\bar{\varphi}_0(t''\!,\vec{x})\bar{\varphi}_0(\tilde{t},\vec{x}\,')|\Omega\rangle\,\langle\Omega|
\bar{\varphi}^2_0(t''\!,\vec{x})|\Omega\rangle\,\langle\Omega|
\bar{\varphi}^2_0(\tilde{t}\!,\vec{x}\,')|\Omega\rangle\nonumber\\
&&\hspace{1.5cm}+2\!\cdot\!3\!\cdot\!1\,\langle\Omega|
\bar{\varphi}^2_0(t''\!,\vec{x})|\Omega\rangle\,\langle\Omega|
\bar{\varphi}_0(t''\!,\vec{x})\bar{\varphi}_0(\tilde{t},\vec{x}\,')|\Omega\rangle\,\langle\Omega|
\bar{\varphi}^2_0(\tilde{t}\!,\vec{x}\,')|\Omega\rangle\nonumber\\
&&\hspace{4cm}+3\!\cdot\!2\!\cdot\!1\Big[\langle\Omega|
\bar{\varphi}_0(t''\!,\vec{x})\bar{\varphi}_0(\tilde{t},\vec{x}\,')|\Omega\rangle\Big]^3
\Bigg\}\; ,\label{2loop3rd}
\ee
where $0\!\leq\!t''\!\leq\!t$, $0\!\leq\!\tilde{t}\!\leq\!t'$, and $t'\!\leq\!t$. The integrand on the right of Eq.~(\ref{2loop3rd}) consists of three terms that are added together. Notice that the second term is twice the first term. Using Eq.~(\ref{phisquare}) in Eq.~(\ref{2loop3rd}) we get
\be
&&\hspace{-0.6cm}\langle\Omega|\!\!\!
\int_0^{t}\!\!\!dt'' a^{\frac{\delta}{2}}(t''){\bar{\varphi}}^3_0(t''\!,\vec{x})\!\!\!
\int_0^{t'}\!\!\!\!d\tilde{t}\, a^{\frac{\delta}{2}}(\tilde{t}){\bar{\varphi}}^3_0(\tilde{t},\vec{x}\,')
|\Omega\rangle\!=\!\!\!\int_0^{t}\!\!\!dt'' a^{\frac{\delta}{2}}(t'')\!\!\!\int_0^{t'}\!\!\!\!d\tilde{t}\, a^{\frac{\delta}{2}}(\tilde{t})\Bigg\{\!9\langle\Omega|
\bar{\varphi}_0(t''\!,\vec{x})\bar{\varphi}_0(\tilde{t},\vec{x}\,')|\Omega\rangle\nonumber\\
&&\hspace{2cm}\times\!{\mathcal{A}}^2\!\!\left[\frac{1\!\!-\!a^{-\delta}(t'')}{\delta}\right]\!\!
\left[\frac{1\!\!-\!a^{-\delta}(\tilde{t})}{\delta}\right]\!\!+\!6\Big[\langle\Omega|
\bar{\varphi}_0(t''\!,\vec{x})\bar{\varphi}_0(\tilde{t},\vec{x}\,')|\Omega\rangle\Big]^3
\!\Bigg\}\; .\label{2loop3rdmidequaltime}
\ee
At this stage, we break up the double integral into two parts: $\int_0^{t}\!dt''\!\!\int_0^{t'}\!d\tilde{t}\!=\!\int_0^{\tilde{t}}\!dt''\!\!\int_0^{t'}\!d\tilde{t}
\!+\!\int_{\tilde{t}}^{t}\!dt''\!\!\int_0^{t'}\!d\tilde{t}$, where in the first part---on the right side of the equation--- we have $t''\!\leq\!\tilde{t}$ whereas $\tilde{t}\!\leq\!t''$ in the second, and employ Eq.~(\ref{treecorrseries}) to obtain\be
&&\hspace{0.6cm}\langle\Omega|\!\!
\int_0^{t}\!\!dt''\, a^{\frac{\delta}{2}}(t''){\bar{\varphi}}^3_0(t''\!,\vec{x})\!\!
\int_0^{t'}\!\!\!d\tilde{t}\, a^{\frac{\delta}{2}}(\tilde{t}){\bar{\varphi}}^3_0(\tilde{t},\vec{x}\,')
|\Omega\rangle\!=\!{\mathcal{A}}^3
\Bigg\{\!\frac{9}{\delta^2}\!\sum_{n=0}^\infty
\!\frac{(-1)^{n}(H\!\Delta x)^{2n}}{(2n\!\!+\!\!1)!(2n\!\!+\!\delta)}\nonumber\\
&&\hspace{-0.8cm}\times\!\Bigg[\!\!\int_0^{t'}\!\!\!d\tilde{t}
\!\left[1\!-\!a^{-\delta}(\tilde{t})\right]\!\!\left[\!\int_0^{\tilde{t}}\!\!\!dt''\!\left[1\!-\!a^{-\delta}(t'')\right]\!
\left[a^{2n+\delta}(t'')\!-\!1\right]\!+\!\left[a^{2n+\delta}(\tilde{t})\!-\!1\right]\!\!
\int_{\tilde{t}}^t\!\!\!dt''\!\left[1\!-\!a^{-\delta}(t'')\right]\!\right]\!\Bigg]\nonumber\\
&&\hspace{1cm}+\,6\!\!\int_0^{t'}\!\!\!d\tilde{t}\,a^{-\delta}(\tilde{t})
\Bigg[\!\int_0^{\tilde{t}}\!\!\!dt''a^{-\delta}(t'')\!\left[\sum_{n=0}^\infty
\!\frac{(-1)^{n}(H\!\Delta x)^{2n}}{(2n\!\!+\!\!1)!(2n\!\!+\!\delta)}\!\left[a^{2n+\delta}(t'')\!-\!1\right]\right]^3\nonumber\\
&&\hspace{2cm}+\!\!
\int_{\tilde{t}}^t\!\!dt''a^{-\delta}(t'')\!\left[\sum_{n=0}^\infty
\!\frac{(-1)^{n}(H\!\Delta x)^{2n}}{(2n\!\!+\!\!1)!(2n\!\!+\!\delta)}\!\left[a^{2n+\delta}(\tilde{t})\!-\!1\right]\right]^3\Bigg]\!\Bigg\}\; .\label{2loop3rdaftmidequaltime}
\ee
We need to evaluate the four double integrals in Eq.~(\ref{2loop3rdaftmidequaltime}) to get the VEV. Results of the integrals are given in Appendix \ref{App:VEV3}. Using Eqs.~(\ref{firstdbint})-(\ref{thirddbint}) and (\ref{fourthdbint}) in Eq.~(\ref{2loop3rdaftmidequaltime}), we find
\be
&&\hspace{0.9cm}\langle\Omega|\!\!
\int_0^{t}\!\!dt''\, a^{\frac{\delta}{2}}(t''){\bar{\varphi}}^3_0(t''\!,\vec{x})\!\!
\int_0^{t'}\!\!\!d\tilde{t}\, a^{\frac{\delta}{2}}(\tilde{t}){\bar{\varphi}}^3_0(\tilde{t},\vec{x}\,')
|\Omega\rangle\!=\!-\frac{{\mathcal{A}}^3}{H^2}9
\Bigg\{\!\frac{1}{\delta^2}\!\sum_{n=0}^\infty
\!\frac{(-1)^{n}(H\!\Delta x)^{2n}}{(2n\!\!+\!\!1)!(2n\!\!+\!\delta)}\nonumber\\
&&\hspace{0.6cm}\times\!\Bigg[\ln(a(t)) \ln(a(t'))\!+\!\!\left[\frac{1\!\!-\!a^{2n+\delta}(t')}{2n\!\!+\!\delta}
\!-\!\frac{1\!\!-\!a^{2n}(t')}{2n}\!-\!\frac{1\!\!-\!a^{-\delta}(t')}{\delta}\right]\!\ln(a(t))
\!-\!\!\Bigg[\!\frac{1\!\!-\!a^{-\delta}(t)}{\delta}\nonumber\\
&&\hspace{-0.3cm}-\frac{1\!\!+\!a^{2n+\delta}(t')}
{2n\!\!+\!\delta}
\!+\!\frac{1\!\!+\!a^{2n}(t')}{2n}\Bigg]\!\ln(a(t'))
\!+\!\frac{1\!\!-\!a^{2n+\delta}(t')}{2n\!\!+\!\delta}\!
\left[\frac{a^{-\delta}(t)}{\delta}\!+\!\frac{2}
{2n\!\!+\!\delta}\right]\!\!-\!\frac{1\!\!-\!a^{2n}(t')}
{2n}\!\Bigg[\!\frac{1\!\!+\!a^{-\delta}(t)}{\delta}\nonumber\\
&&\hspace{1cm}+\frac{1}
{n}\!+\!\frac{1}{2n\!\!+\!\delta}\Bigg]\!\!+\!\frac{1\!\!-\!a^{2n-\delta}(t')}
{2n\!\!-\!\delta}\!\left[\frac{1}{\delta}\!+\!\frac{1}
{2n}\right]\!\!+\!\frac{1\!\!-\!a^{-\delta}(t')}
{\delta}\!\left[\frac{1\!\!-\!a^{-\delta}(t)}{\delta}\!+\!\frac{1}
{2n}\!-\!\frac{1}{2n\!\!+\!\delta}\right]\Bigg]\nonumber\\
&&\hspace{-0.4cm}-\frac{2}
{3}\!\sum_{q=0}^\infty\sum_{p=0}^q\sum_{n=0}^p
\Gamma_{qp\,n}(H\!\Delta x)^{2q}\Bigg[\frac{a^{-\delta}(t)}{\delta}\!\left[\frac{1\!\!-\!a^{2(q+\delta)}(t')}
{2(q\!+\!\delta)}\!-\!3\!\left[\frac{1\!\!-\!a^{2p+\delta}(t')}
{2p\!+\!\delta}\right]\!\!+\!3\!\left[\frac{1\!\!-\!a^{2(q-p)}(t')}
{2(q\!-\!p)}\right]\right]\nonumber\\
&&\hspace{-0.1cm}-\frac{1\!\!-\!a^{2q+\delta}(t')}
{2q\!\!+\!\delta}\!\left[\frac{1}{2(q\!\!+\!\delta)}\!+\!\frac{1}{\delta}\right]
\!\!+\!3\!\left[\frac{1\!\!-\!a^{2p}(t')}
{2p}\right]\!\!\left[\frac{1}{2p\!\!+\!\delta}\!+\!\frac{1}{\delta}\right]
\!\!-\!3\!\left[\frac{1\!\!-\!a^{2(q-p)-\delta}(t')}
{2(q\!-\!p)\!-\!\delta}\right]\!\!\left[\frac{1}{2(q\!-\!p)}\!+\!\frac{1}{\delta}\right]\nonumber\\
&&\hspace{2.8cm}-\frac{1\!\!-\!a^{-\delta}(t')}
{\delta}\!\left[\frac{1\!\!-\!a^{-\delta}(t)}{\delta}\!+\!\frac{1}
{2(q\!\!+\!\delta)}\!-\!\frac{3}{2p\!\!+\!\delta}\!+\!\frac{3}{2(q\!-\!p)}\right]\!\Bigg]\!\Bigg\}\; , \label{2loop3rdequaltimeresult}
\ee
where the $\delta$-dependent coefficients $\Gamma_{qp\,n}$ are defined in Eq.~(\ref{coeffgamma}) which we copy here, \be\Gamma_{qp\,n}\!\equiv\!
\frac{(-1)^q}{[2(q\!-\!p)\!+\!1]![2(q\!-\!p)\!+\!\delta][2(p\!-\!n)\!+\!1]!
[2(p\!-\!n)\!+\!\delta][2n\!\!+\!\!1]![2n\!\!+\!\delta]}\; .\nonumber\ee

Note that the $n=0$ term in the first infinite sum in Eq.~(\ref{2loop3rdequaltimeresult}) is finite---though the sum includes terms proportional to $n^{-1}$ and $n^{-2}$; see Eq.~(\ref{n0term3rdvev}). Note also that the $p\!=\!0$ and $p\!=\!q$ terms in the triple infinite sum in Eq.~(\ref{2loop3rdequaltimeresult}) are finite---though the sum includes terms proportional to $p^{-1}$ and $(q\!-\!p)^{-1}$; see Eqs.~(\ref{p0term3rdvev}) and~(\ref{pqterm3rdvev}).

We have completed evaluating each of the three VEVs that yields an $\mathcal{O}(\lambda^2)$ contribution to two-point correlation function~(\ref{genelfullexpect}). Hence, substitution of VEVs~(\ref{2loop1stequaltimeresult}), (\ref{2loop2ndresult}) and (\ref{2loop3rdequaltimeresult}) into Eq.~(\ref{correlation}) yields the two-loop correlator which can be expressed as\beeq
\langle\Omega|
\bar{\varphi}(t,\vec{x})\bar{\varphi}(t',\vec{x}\,')|\Omega\rangle_{\rm 2-loop}\!\simeq\!\frac
{\lambda^2}{(2\nu)^2}\frac{{\mathcal{A}}^3}{H^4}\,\emph{f}_2(t,t'\!,\Delta x)\;.\label{2loopmassivefinal}
\eneq
To save space we don't write the explicit form of the function $\emph{f}_2(t,t'\!,\Delta x)$.

The reader can add up tree, one- and two-loop correlators~(\ref{treecorrseries}), (\ref{1loopmassivefinal}) and (\ref{2loopmassivefinal}) and obtain one- and two-loop corrected two-point correlation function~(\ref{genelfullexpect}) of the IR truncated full field. Its equal spacetime limit yields
\be
&&\hspace{-0.8cm}\langle\Omega|
\bar{\varphi}^2(t,\vec{x})|\Omega\rangle\!=\!\mathcal{A}\frac{1\!-\!a^{-\delta}(t)}{\delta}\!-\!\frac
{\lambda}{2\nu}\frac{{\mathcal{A}}^2}{H^2}\frac{a^{-\delta}(t)}{\delta^2}\!
\left[\frac{a^{\delta}(t)\!-\!a^{-\delta}(t)}{\delta}\!-\!2\ln(a(t))\right]\!\!+\!\frac
{\lambda^2}{(2\nu)^2}\frac{{\mathcal{A}}^3}{H^4}\frac{a^{-\delta}(t)}{\delta^3}2\nonumber\\
&&\hspace{-0.9cm}\times\!\Bigg\{\!\frac{4}{3}\!\!
\left[\!\frac{a^{\delta}(t)\!-\!1}{\delta^2}\right]\!\!-\!\ln^2(a(t))
\!-\!\!\left[\!\frac{3\!+\!4a^{-\delta}(t)}{\delta}\right]\!\!\ln(a(t))\!+\!4\!\!
\left[\!\frac{1\!\!-\!a^{-\delta}(t)}{\delta^2}\right]\!\!\!+\!\frac{5}{6}\!\!
\left[\!\frac{1\!\!-\!a^{-2\delta}(t)}{\delta^2}\right]\!\!\!\Bigg\}\!\!+\!\mathcal{O}(\lambda^3)\;.
\label{eqlspctmexpect}
\ee
We employed Eqs.~(\ref{phisquare}) and~(\ref{1loopeqlspctm}) to find Eq.~(\ref{eqlspctmexpect}) from which one can obtain the leading contributions to the induced self-mass by the curvature of the potential, $\lambda\langle\Omega|
\varphi^2(t,\vec{x})|\Omega\rangle/2$, at each perturbative order up to $\mathcal{O}(\lambda^4)$. Let us note that the massless limit of Eq.~(\ref{eqlspctmexpect}),
\be
&&\hspace{-0.8cm}\lim_{m\rightarrow 0}\langle\Omega|
\bar{\varphi}^2(t,\vec{x})|\Omega\rangle\!=\!\frac{H^{D-2}}{2^{D-1}\pi^{{D}/{2}}}
\frac{\Gamma\!\left(D\!-\!1\right)}{\Gamma\!\left(\frac{D}{2}\right)}\ln(a(t))\!-\!\frac{\lambda}{(D\!-\!1)}
\frac{H^{2D-6}}{2^{2D-2}\,3\,\pi^{D}}\frac{\Gamma^2\!\left(D\!-\!1\right)}{\Gamma^2\!\left(\frac{D}{2}\right)}
\ln^3(a(t))\nonumber\\
&&\hspace{2.5cm}+\frac{\lambda^2}{(D\!-\!1)^2}\frac{H^{3D-10}}{2^{3D-3}\,5\,\pi^{3D/2}}
\frac{\Gamma^3\!\left(D\!-\!1\right)}{\Gamma^3\!\left(\frac{D}{2}\right)}\ln^5(a(t))\!+\!\mathcal{O}(\lambda^3)
\; ,\label{limeqst}
\ee
reproduces the result found in Ref.~\cite{vacuum}. Note also that, the renormalized $\langle\Omega|
\varphi^2(t,\vec{x})|\Omega\rangle$ for the massless full field $\varphi(t,\vec{x})$ with a quartic self interaction in de Sitter background was obtained in Ref.~\cite{KO} using the results computed in Ref.~\cite{BOW} applying the Schwinger-Keldish (in-in) formalism. Limit (\ref{limeqst}) agrees with it in leading order, as it should.

In the next section, we use Eqs.~(\ref{genelfullexpect}) and (\ref{eqlspctmexpect}) to obtain the variance of the IR truncated scalar at one-loop order.

\section{Fluctuations in Spacetime}
\label{sec:variance}

The variation field $\Delta \bar{\varphi}$ of the IR truncated scalar $\bar{\varphi}$ is defined as the difference of fields at two events,
\beeq
\Delta \bar{\varphi}(t, t' ; \vec{x}, \vec{x}\,') \!\equiv\!
\bar{\varphi}(t, \vec{x})\!-\!\bar{\varphi}(t'\!, \vec{x}\,') \; .
\eneq
Because the mean field $\langle\Omega|\bar{\varphi}(t, \vec{x})|\Omega\rangle$ is zero at any event, the VEV of variation
$\Delta\bar{\varphi}$ vanishes. Therefore, the fluctuation field\be\delta\bar{\varphi}(t, t' ; \vec{x}, \vec{x}\,')\!\equiv\!\Delta\bar{\varphi}(t, t' ; \vec{x}, \vec{x}\,')
\!-\!\langle\Omega|\Delta\bar{\varphi}(t, t' ; \vec{x}, \vec{x}\,')|\Omega\rangle\!=\!\Delta\bar{\varphi}(t, t' ; \vec{x}, \vec{x}\,')\;.\ee Thus, the mean squared fluctuations, or variance,
\be
&&\hspace{1.2cm}\sigma^2_{\Delta \bar{\varphi}}(t, t' ; \vec{x}, \vec{x}\,')\!\equiv\!\langle\Omega|\!\left(\delta\bar{\varphi}\right)^2\!|\Omega\rangle\!=\!\langle\Omega|
\!\left(\Delta\bar{\varphi}\right)^2\!|\Omega\rangle\nonumber\\
&&\hspace{-2cm}=\langle\Omega|{\bar{\varphi}}^2(t, \vec{x})|\Omega\rangle \!-\!\langle\Omega|\bar{\varphi}(t, \vec{x})\bar{\varphi}(t'\!, \vec{x}\,')|\Omega\rangle\!-\!\langle\Omega|\bar{\varphi}(t'\!, \vec{x}\,')\bar{\varphi}(t, \vec{x})|\Omega\rangle \!+\!\langle\Omega|\bar{\varphi}^2(t'\!, \vec{x}\,')|\Omega\rangle\; ,\label{vargeneral}
\ee
does not necessarily vanish. We find, the tree-order variance
\beeq
\sigma^2_{{\Delta \bar{\varphi}_{\rm tree}}}(t, t' ; \vec{x}, \vec{x}\,')\!=\!\mathcal{A}
\emph{g}_0(t, t' ; \vec{x}, \vec{x}\,')\;,\label{vartree}
\eneq
where
\be
\hspace{-0.6cm}\emph{g}_0(t, t' ; \vec{x}, \vec{x}\,')\!=\!\frac{1\!\!-\!a^{-\delta}(t)}{\delta}\!+\!\frac{1\!\!-\!a^{-\delta}(t')}{\delta}
\!+\!2\Big[a(t)\,a(t')\Big]^{-\frac{\delta}{2}}\!\sum_{n=0}^\infty\frac{(-1)^n (H\!\Delta x)^{2n}}{(2n\!\!+\!\!1)!}\frac{1\!\!-\!a^{2n+\delta}(t')}{2n\!\!+\!\delta}\;.\label{gzero}
\ee
The one-loop variance
\beeq
\sigma^2_{{\Delta \bar{\varphi}_{\rm 1-loop}}}(t, t' ; \vec{x}, \vec{x}\,')\!=\!-\frac
{\lambda}{2\nu}\frac{{\mathcal{A}}^2}{H^2}\emph{g}_1(t, t' ; \vec{x}, \vec{x}\,')\;,\label{var1loop}
\eneq
where
\be
&&\hspace{-0.45cm}\emph{g}_1(t,t';\vec{x},\vec{x}\,')\!\!=\!\frac{1}{\delta}\Bigg\{\!\!
\Big[a(t)\,a(t')\Big]^{-\frac{\delta}{2}}\!\sum_{n=0}^\infty
\!\frac{(-1\!)^{n}(H\!\Delta x\!)^{2n}}{(2n\!\!+\!\!1\!)!(2n\!\!+\!\delta)}\Bigg\{\!\!\ln(a(t))\!+\!\ln(a(t'))\!-\!\frac{1\!\!-\!a^{-\delta}(t)}{\delta}
\!-\!\frac{1\!\!-\!a^{-\delta}(t')}{\delta}\nonumber\\
&&\hspace{0.4cm}-a^{2n}(t')
\Bigg[a^{\delta}(t')\Bigg(\!\!\ln(a(t))\!-\!\ln(a(t'))
\!+\!\frac{a^{-\delta}(t)}{\delta}\!\Bigg)\!\!-\!\frac{1}{\delta}\Bigg]\!\!+\!2\!\left[\frac{1\!\!-\!a^{2n+\delta}(t')}{2n\!\!+\!\delta}\right]
\!\!-\!\frac{1\!\!-\!a^{2n}(t')}{n}\Bigg\}\nonumber\\
&&\hspace{1.5cm}-\!\left\{\!\Bigg[\frac{a^{-\delta}(t)}{\delta}
\Bigg(\!2\ln(a(t))\!-\!\frac{a^{\delta}(t)\!-\!a^{-\delta}(t)}{\delta}\Bigg)\!\Bigg]
\!\!+\!\!
\Bigg[a(t)\!\rightarrow\!a(t')\Bigg]\!\right\}\!\!\Bigg\}
\!\!+\!\mathcal{O}(\lambda^2)\; .\label{g1loop}
\ee
In the next two sections, we consider its equal time and equal space limits, respectively.

\subsection{Fluctuations in Space}

The spatial variation of the IR truncated scalar is the difference of field strengts at equal time events
$\Delta \bar{\varphi}(t, \vec{x}, \vec{x}\,')\!=\!
\bar{\varphi}(t, \vec{x})\!-\!\bar{\varphi}(t, \vec{x}\,')$. The
field strength is positive at some places, at others it is negative. Moreover, the inflationary particle production is a quantum mechanical random process. While the {\it magnitude}
of the field strength increases at some places, at others it decreases. Hence, the {\it expectation value} of the spatial variation $\Delta \bar{\varphi}(t, \vec{x}, \vec{x}\,')$  vanishes, but
the spatial variance $\sigma^2_{\Delta \bar{\varphi}}(t, \vec{x}, \vec{x}\,')$, which we compute next,  does not.

\subsubsection{Tree-order variance in space}
\label{subsect:varytree}

For a massive scalar, we employ
equal time limit of Eq.~(\ref{genelfullexpect}) and Eq.~(\ref{eqlspctmexpect}) in Eq.~(\ref{vargeneral}) to obtain\beeq
\sigma^2_{{\Delta \bar{\varphi}_{\rm tree}}}(t,\vec{x},\vec{x}\,')\!=\!\mathcal{A}
\emph{g}_0(t,\vec{x},\vec{x}\,')\; ,\label{spatvartree}
\eneq
where
\beeq
\emph{g}_0(t, \vec{x}, \vec{x}\,')\!=\!2a^{-\delta}(t)\!\sum_{n=1}^\infty\!\frac{(-1)^n (H\!\Delta x)^{2n}}{(2n\!\!+\!\!1\!)!} \frac{1\!-\!a^{2n+\delta}(t)}{2n\!\!+\!\delta}\; .\label{spatvarg0}
\eneq
In the late time limit, for a fixed comoving separation, Eq.~(\ref{spatvarg0}) asymptotes to \be
&&\hspace{-1.3cm}\emph{g}_0(t, \vec{x}, \vec{x}\,')\!\!\rightarrow\!2\!\left\{\!\!\frac{1\!\!-\!a^{-\delta}(t)}{\delta}
\!+\!\frac{a^{-\delta}(t)}{\delta}\!\!\left[\frac{\sin(H\!\Delta x)}{H\!\Delta x}\!+\!\frac{(H\!\Delta x)^2}{3(2\!\!+\!\delta)}{}_{1}\mathcal{F}_{2}\!\left(\!\!1\!\!+\!\frac{\delta}{2}; \frac{5}{2},\! 2\!\!+\!\frac{\delta}{2}; -\frac{(H\!\Delta x)^2}{4}\!\right)\!\right]\!\right\}.
\ee

For a massless scalar, on the other hand,
\be
&&\hspace{-2.1cm}\lim_{m\rightarrow 0}\!\emph{g}_0(t, \vec{x}, \vec{x}\,')\!=\!\!\sum_{n=1}^\infty\!\frac{(-1)^n (H\!\Delta x)^{2n}}{(2n\!\!+\!\!1\!)!} \frac{1\!-\!a^{2n}(t)}{n}\; ,\label{vartreemassless}
\ee
asymptotes to
\be
&&\hspace{-1.5cm}\lim_{m\rightarrow 0}\!\emph{g}_0(t, \vec{x}, \vec{x}\,')\!\rightarrow\!
2\!\left\{\!\ln(a(t))\!+{\rm ci}(H\!\Delta x)
\!-\!\frac{\sin(H\!\Delta x)}{H\!\Delta x}\!\right\}\; .\label{varfreeasymp}
\ee
The growth in tree-order variance implied by Eqs.~(\ref{spatvarg0}) and (\ref{vartreemassless}) means that the {\it magnitude} of spatial variation increases in time. Hence, the average effect measured by a VEV may be misleading because the actual effect observed by any local observer is either a decrease or an increase in field strength as fluctuations happen. As we show in Sec.~\ref{subsect:varyoneloop}, however, the dominant quantum correction to the tree-order variance comes with a minus sign. Thus, the quantum corrected variance does not grow as large as the tree-order result implies, meaning that the contrast between the average and the actual effect is not as large as the tree-order variance indicates.

\begin{figure}
\includegraphics[width=10cm,height=6.3cm]{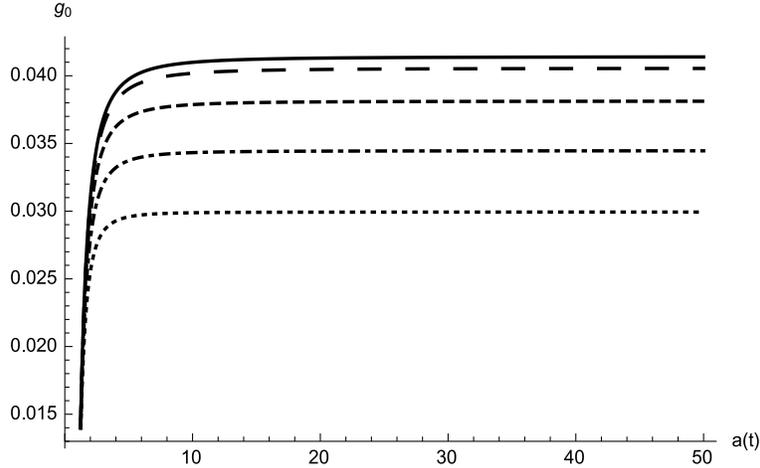}
\caption{Plots of the function $g_0(t, K)$, defined in Eq.~(\ref{gtreevar}), versus $a(t)$ for  different mass values. The fraction $K$ is chosen to be $1/2$ and the scale factor $a(t)$ runs from $1$ to $50$. Extending it to $e^{50}$ does not alter the plots any significantly. The solid curve is the plot for a massless scalar. The large-dashed, dashed, dot-dashed and dotted curves are for massive scalars with $m\!=\!H/4, H/2, 3H/4$ and $H$, respectively.}
\label{fig:tree-variance}
\end{figure}

To infer result~(\ref{spatvartree}) for a fixed physical distance we take comoving spatial separation $\Delta x\!=\!K/Ha(t)$. Hence, Eq.~(\ref{spatvarg0}) yields
\be
&&\hspace{0.7cm}\emph{g}_0(t,\! K)\!=\!\frac{2}{\delta}\Bigg\{\!1\!-\!a^{-\delta}(t)
\!-\!\frac{\sin\left(K\right)}{K}\!+\!a^{-\delta}(t)\frac{\sin(Ka^{-1}(t))}{Ka^{-1}(t)}\nonumber\\
&&\hspace{-1.5cm}-\frac{K^2}{3(2\!+\!\delta)}\!\!\left[{}_{1}\mathcal{F}_{2}\!\left(\!1\!\!+\!\frac{\delta}{2}; \frac{5}{2}, 2\!+\!\frac{\delta}{2}; -\frac{K^2}{4}\!\right)\!\!-\!a^{-2-\delta}(t){}_{1}\mathcal{F}_{2}\!\left(\!1\!\!+\!\frac{\delta}{2}; \frac{5}{2}, 2\!+\!\frac{\delta}{2}; -\frac{K^2}{4a^2(t)}\!\right)\!\right]\!\!\Bigg\}\; .\label{gtreevar}
\ee
Numerical values of $g_0(t, K)$, for a given $K$ and $\delta$, can be obtained from Eq.~(\ref{gtreevar}) at any time $t$ during inflation.
Figure~\ref{fig:tree-variance} shows that it rapidly grows and asymptotes to a constant during inflation. 
One can see this behavior of $g_0(t, K)$ analytically by computing its asymptotic form\beeq
\emph{g}_0(t,\! K)\!\!\rightarrow\!\frac{2}{\delta}\Bigg\{\!1\!\!-\!\frac{\sin\left(K\right)}{K}
\!-\!\frac{K^2}{3(2\!+\!\delta)}{}_{1}\mathcal{F}_{2}\!\left(\!\!1\!\!+\!\frac{\delta}{2}; \frac{5}{2}, 2\!+\!\frac{\delta}{2}; -\frac{K^2}{4}\!\right)\!\!\!\Bigg\}\; ,
\eneq
which is a constant up to $\mathcal{O}(a^{-2-\delta})$.

\subsubsection{One-loop variance in space}
\label{subsect:varyoneloop}

One loop equal time variance of a massive scalar,
\beeq
\hspace{0.4cm} \sigma^2_{{\Delta \bar{\varphi}_{\rm 1-loop}}}(t, \vec{x}, \vec{x}\,')\!=\!-\frac
{\lambda}{2\nu}\frac{{\mathcal{A}}^2}{H^2}\emph{g}_1(t, \vec{x}, \vec{x}\,')\; ,\label{spatvar1loop}
\eneq
where
\be
&&\hspace{-1.35cm}\emph{g}_1(t, \vec{x}, \vec{x}\,')\!=\!2\frac{a^{-\delta}(t)}
{\delta}\!\sum_{n=1}^\infty
\!\frac{(-1)^{n}(H\!\Delta x)^{2n}}{(2n\!\!+\!\!1\!)!(2n\!\!+\!\delta)}
\Bigg\{\!\!\ln(a(t))\!-\!\frac{1\!\!-\!a^{-\delta}(t)}{\delta}\!+\!\frac{1\!\!-\!a^{2n+\delta}(t)}{2n\!\!+\!\delta}
\!-\!\frac{1\!\!-\!a^{2n}(t)}{2n}\!\Bigg\}\; .\label{spatvarg1}
\ee
Hence, the one-loop spatial variance $\sigma^2_{{\Delta \bar{\varphi}_{\rm 1-loop}}}(t, \vec{x}, \vec{x}\,')\!<\!0$ contrary to the positive tree-order spatial variance, as commented in Sec.~\ref{subsect:varytree}.

In the late time limit, for a fixed comoving separation, $\emph{g}_1(t, \vec{x}, \vec{x}\,')$ asymptotes to
\be
&&\hspace{-0.45cm}2\frac{a^{-\delta}(t)}
{\delta^2}\Bigg\{\!\!\!\left[\ln(a(t))\!+\!\frac{1\!\!+\!a^{-\delta}(t)}
{\delta}\right]\!\!\!\left[\frac{\sin(H\!\Delta x)}{H\!\Delta x}\!-\!\!1\!\!+\!
\frac{(H\!\Delta x)^2}{3(2\!\!+\!\delta)}{}_{1}\mathcal{F}_{2}\!
\left(\!\!1\!\!+\!\frac{\delta}{2};\! \frac{5}{2},\! 2\!+\!\frac{\delta}{2};\!
-\frac{(H\!\Delta x)^2}{4}\!\right)\!\right]\!\!-\!\!\ln(a(t))\nonumber\\
&&\hspace{-0.25cm}+\frac{1\!\!+\!a^{\delta}(t)}
{\delta}\!-\!{\rm ci}(H\!\Delta x)\!+\!\frac{\sin(H\!\Delta x)}{H\!\Delta x}\!+\!
\frac{(H\!\Delta x)^2}{3(2\!\!+\!\delta)^2}{}_{2}\mathcal{F}_{3}\!
\left(\!\!1\!\!+\!\frac{\delta}{2},\! 1\!\!+\!\frac{\delta}{2};\! \frac{5}{2},\! 2\!+\!\frac{\delta}{2},\! 2\!+\!\frac{\delta}{2};\!
-\frac{(H\!\Delta x)^2}{4}\!\right)\!\!\!\Bigg\}\;.
\ee Its massless limit,
\be
\lim_{m\rightarrow 0}\!\emph{g}_1(t, \vec{x}, \vec{x}\,')\!=\!\!\sum_{n=1}^\infty
\!\frac{(-1)^{n}(H\!\Delta x)^{2n}}{(2n\!\!+\!\!1\!)!2n}
\Bigg\{\!\!\ln^2(a(t))\!-\!\frac{a^{2n}(t)}{n}\ln(a(t))
\!-\!\frac{1\!\!-\!a^{2n}(t)}{2n^2}\!\Bigg\}\;,
\ee
on the other hand, asymptotes to
\be
&&\hspace{-0.2cm}\lim_{m\rightarrow 0}\!\emph{g}_1(t, \vec{x}, \vec{x}\,')\!\rightarrow\!\frac{2}{3}\ln^3(a(t))\!+\!\!\left[{\rm ci}(H\!\Delta x)\!-\!\frac{\sin(H\!\Delta x)}{H\!\Delta x}\right]\!\ln^2(a(t))\!-\!\frac{1}{3}\Big[\!\ln(H\!\Delta x)\!+\!\gamma\!\!-\!\!1\Big]^3\nonumber\\
&&\hspace{3.8cm}-\!\!\left[1\!\!-\!\!\frac{\pi^2}{12}\right]\!\!\Big[\!\ln(H\!\Delta x)\!+\!\gamma\!-\!\!1\Big]\!\!-\!\frac{2}{3}\Big[\zeta(3)\!-\!\!1\Big]\; .
\ee

\begin{figure}
\includegraphics[width=10cm,height=6.3cm]{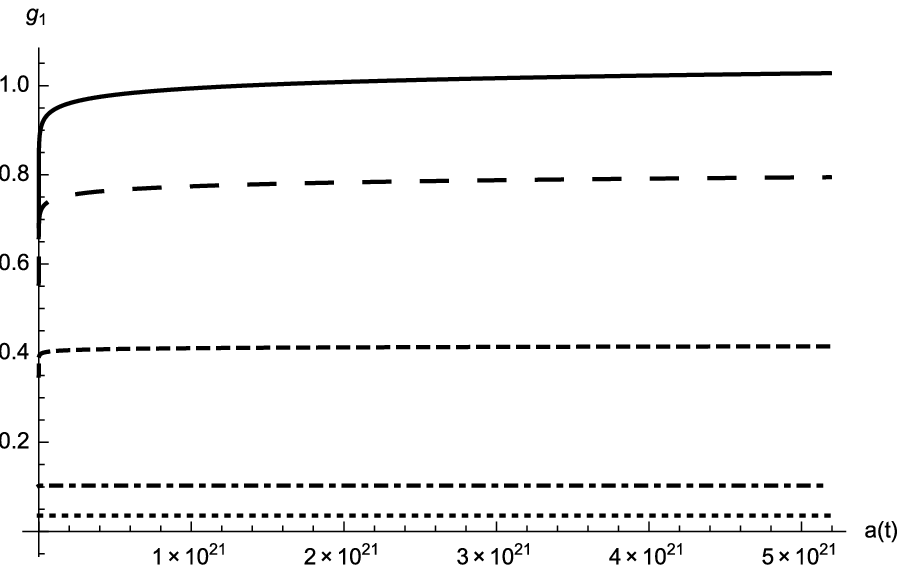}
\caption{Plots of the function $g_1(t, K)$, defined in Eq.~(\ref{spatvarg1}), versus $a(t)$ for different mass values. The fraction $K$ is chosen to be $1/2$. $a(t)$ runs from $1$ to $e^{50}$. The solid curve is the plot for a massless scalar. The large-dashed, dashed, dot-dashed and dotted curves are plots for massive scalars with $m\!=\!H/8, H/4, H/2$ and $3H/4$, respectively.}
\label{fig:oneloopvar}
\end{figure}

To interpret one-loop equal time variance~(\ref{spatvar1loop}) for a fixed physical distance we take $\Delta x\!=\!K/Ha(t)$ in Eq.~(\ref{spatvarg1}) and obtain
\be
&&\hspace{-0.45cm}\emph{g}_1\!(t,\!K)\!\!=\!2\frac{a^{-\delta}(t)}{\delta^2}
\Bigg\{\!\Bigg[\!\!\ln(a(t))\!-\!\frac{1\!\!-\!\!a^{-\delta}(t)}{\delta}\Bigg]
\!\Bigg[\!\frac{\sin(Ka^{-1}(t))}{Ka^{-1}(t)}\!-\!1\!\!+\!\frac{K^2a^{-2}(t)}{3(2\!\!+\!\delta)}
{}_{1}\mathcal{F}_{2}\!\left(\!\!1\!\!+\!\frac{\delta}{2};\! \frac{5}{2},\! 2\!\!+\!\!\frac{\delta}{2};\! \frac{-K^2}{4a^2(t)}\!\right)\!\!\!\Bigg]\nonumber\\
&&\hspace{-0.45cm}-\frac{a^{\delta}(t)}{\delta}\!\Bigg[\!\Bigg\{\!\frac{\sin(\!K)}{K}
\!-\!\!1\!\!+\!\frac{K^2}{3(2\!\!+\!\!\delta)}\!\Bigg[\!{}_{1}\mathcal{F}_{2}
\!\left(\!\!1\!\!+\!\!\frac{\delta}{2}; \frac{5}{2},\! 2\!\!+\!\!\frac{\delta}{2};\! -\frac{K^2}{4}\!\right)\!\!+\!\!\frac{\delta}{2\!\!+\!\delta}\,
{}_{2}\mathcal{F}_{3}\!\left(\!\!1\!\!+\!\!\frac{\delta}{2},\! 1\!\!+\!\!\frac{\delta}{2};\! \frac{5}{2},\! 2\!\!+\!\!\frac{\delta}{2},\! 2\!\!+\!\!\frac{\delta}{2};\! -\frac{K^2}{4}\!\right)\!\!\Bigg]\!\Bigg\}\nonumber\\
&&\hspace{-0.45cm}-a^{-\delta}(t)\Bigg\{\!\!K\!\!\rightarrow\! \frac{K}{a(t)}\!\Bigg\}\!\Bigg]\!\!\!+\!\!\Bigg[\!\Bigg\{\!{\rm ci}(\!K)
\!-\!\frac{\sin(\!K)}{K}\!-\!\ln(\!K)
\!-\!\!\frac{1}{\delta}\!\Bigg[\!\frac{\sin(\!K)}{K}\!+\!\frac{K^2}{3(2\!\!+\!\!\delta)}\,
{}_{1}\mathcal{F}_{2}
\!\left(\!\!1\!\!+\!\frac{\delta}{2};\! \frac{5}{2},\! 2\!\!+\!\frac{\delta}{2}; -\frac{K^2}{4}\!\right)\!\!\Bigg]\!\Bigg\}\nonumber\\
&&\hspace{6.5cm}-\Bigg\{\!\!K\!\!\rightarrow\! \frac{K}{a(t)}\!\Bigg\}\!\Bigg]\!\Bigg\}\;.\label{gdeltaKa}
\ee
Figure~\ref{fig:oneloopvar} shows that $g_1(t, K)$ grows logarithmically for the massless scalar and that the growth is suppressed for massive scalars. As the mass increases, the suppression gets stronger. 
One can infer this late time behavior from the asymptotic form of $g_1(t, K)$. For the massless scalar, up to order $\frac{\ln^2(a(t))}{a^2(t)}$, we have
\be
\hspace{-0.5cm}\lim_{m\rightarrow 0}\!\emph{g}_1\!(t,\!K)\!\rightarrow\!\frac{K^2}{12}
{}_{3}\mathcal{F}_{4}\!\!\left(\!1,\!1,\!1;2,\!2,\!2,\!\frac{5}{2};
-\frac{K^2}{4}\!\right)\!\ln(a(t))\!-\!\frac{K^2}{24}
{}_{4}\mathcal{F}_{5}\!\!\left(\!1,\!1,\!1,\!1;2,\!2,\!2,\!2,\!\frac{5}{2};
-\frac{K^2}{4}\!\right) ,
\ee
which implies a logarithmic growth at late times. For a massive scalar, on the other hand, $\emph{g}_1\!(t,\!K)$ asymptotes to
\be
&&\hspace{-0.4cm}\frac{2}{\delta^2}\Bigg\{\!a^{-\delta}(t)\!\Bigg[{\rm ci}(\!K)
\!-\!\frac{\sin(\!K)}{K}\!-\!\ln(\!K)
\!-\!\gamma\!+\!1\!\!-\!\!\frac{1}{\delta}\!\!\left[\frac{\sin(\!K)}{K}\!-\!1
\!\!+\!\frac{K^2}{3(2\!\!+\!\!\delta)}\,{}_{1}\mathcal{F}_{2}\!\left(\!\!1\!\!+\!\frac{\delta}{2};\! \frac{5}{2},\! 2\!\!+\!\frac{\delta}{2};\! -\frac{K^2}{4}\!\right)\!\right]\!\!\Bigg]\nonumber\\
&&\hspace{-0.6cm}-\frac{1}{\delta}\!\Bigg[\!\frac{\sin(\!K)}{K}\!-\!\!1\!\!+\!\frac{K^2}{3(2\!\!+\!\!\delta)}\!\!
\left[{}_{1}\mathcal{F}_{2}\!\left(\!\!1\!\!+\!\frac{\delta}{2};\! \frac{5}{2},\! 2\!\!+\!\frac{\delta}{2};\! -\frac{K^2}{4}\!\right)\!\!+\!\frac{\delta}{2\!\!+\!\!\delta}\, {}_{2}\mathcal{F}_{3}\!\left(\!\!1\!\!+\!\frac{\delta}{2},\! 1\!\!+\!\frac{\delta}{2};\! \frac{5}{2},\! 2\!\!+\!\frac{\delta}{2},\! 2\!\!+\!\frac{\delta}{2};\!-\frac{K^2}{4}\!\right)\!\right]\!\!\Bigg]\!\!\Bigg\},\label{g1asymp}\nonumber\\
\ee
up to order $\frac{\ln(a(t))}{a^{2+\delta}(t)}$. Equation~(\ref{g1asymp}) implies, as pointed out, a suppression of some constant terms by the factor $\frac{a^{-\delta}(t)}{\delta^2}$ which gets stronger as the mass increases.

\subsection{Fluctuations in Time}

The temporal variation is the difference of fields at equal space events
$\Delta\bar{\varphi}(t, t'\!, \vec{x})\!\equiv\!
\bar{\varphi}(t, \vec{x})\!-\!\bar{\varphi}(t'\!,\vec{x})$ whose VEV vanishes.
The temporal variance \be
\sigma^2_{\Delta \bar{\varphi}}(t, t')\!=\!\langle\Omega|\!
\left[\Delta\bar{\varphi}(t, t'\!, \vec{x})\!-\!\langle\Omega|\Delta\bar{\varphi}(t, t'\!, \vec{x})|\Omega\rangle\right]^2\!|\Omega\rangle\!=\!\langle\Omega|\!\left[\bar{\varphi}(t, \vec{x})\!-\!\bar{\varphi}(t'\!, \vec{x})\right]^2\!|\Omega\rangle\; ,\ee
can be obtained by taking the equal space limit of Eq.~(\ref{vargeneral}).

\subsubsection{Tree-order variance in time}

Using Eqs.~(\ref{vartree})-(\ref{gzero}) we find, the tree-order temporal variance
\beeq
\sigma^2_{{\Delta \bar{\varphi}_{\rm tree}}}(t, t')\!=\!\mathcal{A}
\emph{g}_0(t, t')\;,\label{vartreeeqsp}
\eneq
where
\beeq
\emph{g}_0(t, t')\!=\!\frac{1\!-\!a^{-\delta}(t)}{\delta}\!+\!\frac{1\!-\!a^{-\delta}(t')}{\delta}
\!+\!2\Big[a(t)\,a(t')\Big]^{-\frac{\delta}{2}}\,\frac{1\!-\!a^{\delta}(t')}{\delta}\;.\label{gzeroeqsp}
\eneq
Notice that $\lim_{t'\rightarrow t}\!\emph{g}_0(t, t')\!=\!0$,
hence $\sigma^2_{{\Delta \bar{\varphi}_{\rm tree}}}(t, t)$ vanishes, as it should. For $t'\!<\!t$, on the other hand, we have $\emph{g}_0(t, t')\!>\!0$, hence the tree-order temporal variance $\sigma^2_{{\Delta \bar{\varphi}_{\rm tree}}}(t, t')\!>\!0$.

Note also that the massless limit
\beeq
\lim_{m\rightarrow 0}\!\emph{g}_0(t, t')\!=\!\ln(a(t))\!-\!\ln(a(t'))\;,
\eneq
yields
\beeq
\lim_{m\rightarrow 0}\!\sigma^2_{{\Delta \bar{\varphi}_{\rm tree}}}(t, t')\!=\!\frac{\Gamma\!\left(D\!-\!\!1\right)}{\Gamma\!\left(\frac{D}{2}\right)}
\frac{H^{D-2}}{2^{D-1}\pi^{\frac{D}{2}}}\Big[\!\ln(a(t))\!-\!\ln(a(t'))\Big]\;,
\eneq
which is positive and grows logarithmically. Next, we consider the one-loop quantum correction
to the tree-order temporal variance.

\subsubsection{One-loop variance in time}

Using Eqs.~(\ref{var1loop})-(\ref{g1loop}) we find the one-loop temporal variance
\beeq
\hspace{0.3cm}\sigma^2_{{\Delta \bar{\varphi}_{\rm 1-loop}}}(t, t')\!=\!-\frac
{\lambda}{2\nu}\frac{{\mathcal{A}}^2}{H^2}\emph{g}_1(t, t')\;,\label{var1loopeqsp}
\eneq
where
\be
&&\hspace{-1.2cm}\emph{g}_1(t, t')\!=\!\Big[a(t)\,a(t')\Big]^{-\frac{\delta}{2}}\frac{1}{\delta^2}\Bigg\{\!\!\ln(a(t))\!+\!3\ln(a(t'))
\!-\!a^{\delta}(t')\Bigg[\!\ln(a(t))\!-\!\ln(a(t'))
\!+\!\frac{1\!\!+\!a^{-\delta}(t)}{\delta}\Bigg]\nonumber\\
&&\hspace{-1.4cm}+\frac{1\!\!-\!a^{\delta}(t')}{\delta}
\!+\!\frac{a^{-\delta}(t)\!+\!a^{-\delta}(t')}{\delta}\!\Bigg\}\!-\!\!\left\{\!\Bigg[\frac{a^{-\delta}(t)}{\delta^2}
\Bigg(\!\!2\ln(a(t))\!-\!\frac{a^{\delta}(t)\!-\!a^{-\delta}(t)}{\delta}\Bigg)\!\Bigg]
\!\!+\!\!
\Bigg[a(t)\!\!\rightarrow\! a(t')\Bigg]\!\right\}\; .\label{g1loopeqsp}
\ee

At equal time, we have $\lim_{t'\rightarrow t}\!\emph{g}_1(t, t')\!=\!0$,
hence $\sigma^2_{{\Delta \bar{\varphi}_{\rm 1-loop}}}(t, t)$ vanishes. For $t'\!\!<\!\!t$, however, $\emph{g}_1(t, t')\!>\!0$, hence the one-loop temporal variance $\sigma^2_{{\Delta \bar{\varphi}_{\rm 1-loop}}}(t, t')\!<\!0$ as opposed to positive tree-order temporal variance~(\ref{vartreeeqsp}). Thus, the temporal variation does not grow as fast as the $\sigma^2_{{\Delta \bar{\varphi}_{\rm tree}}}$ implies.

Let us note finally that the massless limit,
\beeq
\lim_{m\rightarrow 0}\!\emph{g}_1(t, t')\!=\!\frac{1}{2}\!\left[\!\frac{2\ln^3(a(t))}{3}\!-\!\ln^2(a(t))\ln(a(t'))\!+\!\frac{\ln^3(a(t'))}{3}\right]\;,
\eneq
yields
\beeq
\hspace{-0.1cm}\lim_{m\rightarrow 0}\!\sigma^2_{{\Delta \bar{\varphi}_{\rm 1-loop}}}\!(t, t')\!=\!-\frac{\lambda}{(D\!-\!1)}\frac{\Gamma^2\!\left(D\!-\!\!1\right)}{\Gamma^2\!\left(\frac{D}{2}\right)}
\frac{H^{2D-6}}{2^{2D-1}\pi^{D}}\!
\!\left[\!\frac{2\ln^3(a(t))}{3}\!-\!\ln^2(a(t))\ln(a(t'))\!+\!\frac{\ln^3(a(t'))}{3}\right],
\eneq
which is the same result computed in Ref.\cite{vacuum} for a massless scalar.
\section{Conclusions}
\label{sec:conclusions}

We considered an IR truncated {\it massive} minimally coupled scalar field $\bar\varphi(t, \vec{x})$ with a quartic self-interaction $\lambda\bar\varphi^4(t, \vec{x})/4!$ on a locally de Sitter background of an inflating spacetime, in this paper. We introduced the background spacetime and the Lagrangian of the model in Sec.~\ref{sec:model}. In Sec.~\ref{sec:Quantum}, we obtained the mode expansion of the IR truncated free field $\bar\varphi_0(t, \vec{x})$ and, following Starobinsky's approach, we expressed the interacting field $\bar\varphi(t, \vec{x})$ in terms of the free field $\bar\varphi_0(t, \vec{x})$ up to $\mathcal{O}(\lambda^4)$. Using this expression, we computed the two-point correlation function $\langle\Omega|\bar\varphi(t, \vec{x})\bar\varphi(t'\!, \vec{x}\,')|\Omega\rangle$ of the scalar at tree, one- and two-loop order, in Sec.~\ref{sec:twopointcorrelator}. The tree-order correlator is positive. At a fixed comoving separation, that is at increasing physical distance, it asymptotes to zero for a massive scalar at late times. In the massless limit, however, it asymptotes to a nonzero constant. For a fixed physical distance, it grows logarithmically in the massless limit. The growth is suppressed for massive scalars. As the mass increases, the suppression gets stronger; see Fig.~\ref{fig:tree-corr}.

The one-loop correlator---leading quantum contribution---is negative. At a fixed comoving separation, it asymptotes to zero for a massive scalar but grows like $-\lambda\ln^2(a(t))$ in the massless limit, at late times. For a fixed physical distance, on the other hand, it grows like $-\lambda\ln^2(a(t))\ln(a(t'))$ in the massless limit, at late times. This growth, however, is suppressed for massive scalars. The suppression is very sensitive to an increase in mass; see Fig.~\ref{fig:oneloop-corr}. In fact, for masses larger than $H/2$ it asymptotes to zero.

The two-loop correlator---next to leading quantum contribution---is positive. Adding up tree~(\ref{treecorrseries}), one- and two-loop correlators, (\ref{1loopmassivefinal}) and (\ref{2loopmassivefinal}) yields one- and two-loop corrected two-point correlation function of the IR truncated full field. From its equal spacetime limit~(\ref{eqlspctmexpect}) one obtains the leading terms in the induced self-mass by the curvature of the potential at tree, one- and two-loop order,
\be
&&\hspace{0.5cm} \frac{\lambda}{2}\mathcal{A}\frac{1\!-\!a^{-\delta}(t)}{\delta}\!-\!\frac
{\lambda^2}{4\nu}\frac{{\mathcal{A}}^2}{H^2}\frac{a^{-\delta}(t)}{\delta^2}\!
\left[\frac{a^{\delta}(t)\!-\!a^{-\delta}(t)}{\delta}\!-\!2\ln(a(t))\right]\!\!+\!\frac
{\lambda^3}{(2\nu)^2}\frac{{\mathcal{A}}^3}{H^4}\frac{a^{-\delta}(t)}{\delta^3}\nonumber\\
&&\hspace{-0.7cm}\times\!\Bigg\{\!\frac{4}{3}\!\!
\left[\frac{a^{\delta}(t)\!-\!1}{\delta^2}\right]\!\!-\!\ln^2(a(t))
\!-\!\!\left[\frac{3\!+\!4a^{-\delta}(t)}{\delta}\right]\!\ln(a(t))\!+\!4\!\!
\left[\frac{1\!\!-\!a^{-\delta}(t)}{\delta^2}\right]\!\!\!+\!\frac{5}{6}\!\!
\left[\frac{1\!\!-\!a^{-2\delta}(t)}{\delta^2}\right]\!\!\Bigg\}\!\!+\!\mathcal{O}(\lambda^4)\;.\nonumber
\ee
For the massless scalar, the $\delta\!\!\rightarrow\!0$ limit of this result yields the induced self-mass by the curvature of the potential. It is equal to $\frac{\lambda}{2}$ times the limit given on the right
side of~Eq.~(\ref{limeqst}).

In Sec.~\ref{sec:variance}, we computed the mean squared fluctuations (variance) of field variation. At late times,  the tree-order spatial variance at a fixed comoving separation grows logarithmically in the massless limit, whereas it grows like $\frac{1-a^{-\delta}(t)}{\delta}$ in the massive case. Because $\frac{1-a^{-\delta}(t)}{\delta}\!\leq\!\ln(a(t))$, where the equality holds in the massless limit, the growth is suppressed in the massive case. The larger the mass of the scalar, the stronger the suppression. This positive growth implies that the {\it magnitude} of the spatial variation {\it increases} with time. This fact is the main reason for some cosmologists to doubt about the physical implications of expectation values. They argue that any local observer perceives either an increase or a decrease in the field strength as fluctuations happen. An expectation value is merely a measure of average effect and, therefore, can be misleading. There are, however, counterarguments \cite{W0,Tegmark} claiming that while there is spatial and temporal variation in the actual effects, one can roughly trust expectation values under certain circumstances. We showed that the tree-order variance of field variation decreases if the quantum corrections are taken into account. Hence, the difference between the actual effect perceived by a local observer and the average effect, as the field fluctuates, is not as large as the tree-order variance implies.

\begin{appendix}

\section{Special functions}
\label{App:expintincopletegamma}

In this appendix, we define various special functions we use in the manuscript.

\subsection{Exponential integral function $E_\beta(z)$}

The exponential integral function   \beeq
E_\beta(z)\!\equiv\!\!\int_1^\infty\!\! dt\, t^{-\beta}\, e^{-tz}\; ,\label{expintdefn}
\eneq
where $\beta$ and $z$ are $c$-numbers.
\subsection{Incomplete gamma function $\Gamma(\beta, z)$}
The incomplete gamma function
\beeq
\Gamma(\beta, z)\!\equiv\!\!\int_z^\infty\!\! dt\, t^{\beta-1}\, e^{-t}\; ,\label{incompgammadefn}
\eneq
satisfies the recurrence relation
\beeq
\Gamma(\beta\!+\!1, z)\!=\!\beta\Gamma(\beta, z)\!+\!z^{\beta} e^{-z}\; .\label{incompgammarecurr}
\eneq It is useful to express the incomplete gamma function in terms of the ordinary gamma function plus an alternating power series as
\beeq
\Gamma(\beta, z)\!=\!\Gamma(\beta)\!-\!\!\sum_{n=0}^\infty\frac{(-1)^nz^{\beta+n}}{n!(\beta\!+\!n)}\; .\label{gammaaspowerseries}
\eneq
The right side of Eq.~(\ref{gammaaspowerseries}) is replaced by its limiting value if $\beta$ is a negative integer or zero.

Asymptotic expansion of the $\Gamma(\beta, z)$, as $|z|\!\rightarrow\!\infty$, and for $-\frac{3\pi}{2}\!<\!{\rm arg}(z)\!<\!\frac{3\pi}{2}$, can be represented as
\be
\Gamma(\beta, z)\!\!&\rightarrow&\!\!z^{\beta-1}e^{-z}\!\sum_{n=0}^{\infty}\frac{(-1)^n\Gamma(1\!-\!\beta\!+\!n)}
{z^n\Gamma(1\!-\!\beta)}\;  .\label{asympincomplgamma}
\ee

\subsection{Generalized hypergeometric function ${}_p\mathcal{F}_q(\alpha_1,\ldots,\alpha_p;\beta_1,\ldots,\beta_q;z)$}

Generalized hypergeometric function,
\beeq
{}_p\mathcal{F}_q(\alpha_1,\ldots,\alpha_p;\beta_1,\ldots,\beta_q;z)\!=\!\frac{\Gamma(\beta_1)\cdots\Gamma(\beta_q)}
{\Gamma(\alpha_1)\cdots\Gamma(\alpha_p)}\!
\sum_{n=0}^\infty\frac{\Gamma(\alpha_1\!\!+\!n)\cdots\Gamma(\alpha_p\!\!+\!n)}
{\Gamma(\beta_1\!\!+\!n)\cdots\Gamma(\beta_q\!\!+\!n)}\left(\frac{z^n}{n!}\right)\; ,\label{hypergpq}
\eneq
where $p$ and $q$ are integers, $\alpha_i$, $\beta_j$ and $z$ are $c$-numbers.

\subsection{Cosine integral function $ci(z)$}

The cosine integral function \be
{\rm ci}(z) \!\equiv\! -\!\int_z^{\infty} \!\!dt {\cos(t)
\over t} \!=\! \gamma \!+\! \ln(z) \!+\!\!\int_0^z \!\!dt {\cos(t)
\!-\!1 \over t} \; ,\label{cosintdefn}
\ee where $\gamma \approx .577$ is the Euler's constant. The following identity involving
${\rm ci}(z)$ and $\sin(z)$ is useful\beeq
{\rm ci}(z)\!-\!\frac{\sin(z)}{z}\!=\!\gamma\!-\!1\!+\!\ln(z)\!+\!\!\sum_{n=1}^\infty
                 \frac{(-1)^{n}z^{2n}}{2n (2n\!\!+\!\!1)!}\; .\label{cieksisin}
\eneq The function ${\rm ci}(z)$ has the asymptotic expansion,
\be
&&\hspace{-0.5cm}{\rm ci}(z)\!\longrightarrow\!\frac{\sin(z)}{z}
\!\sum_{n=0}^\infty\frac{(-1)^n\Gamma(2n\!\!+\!\!1)}{z^{2n}}
\!-\!\frac{\cos(z)}{z}\!\sum_{n=0}^\infty\frac{(-1)^n\Gamma(2n\!\!+\!\!2)}{z^{2n+1}}\label{sumcisasym}\\
&&\hspace{0.4cm}\longrightarrow\!\frac{\sin(z)}{z}\!\left(\!1\!-\!\frac{2!}{z^2}
\!+\!\frac{4!}{z^4}\dots\!\right)
\!-\!\frac{\cos(z)}{z}\!\left(\!\frac{1}{z}\!-\!\frac{3!}{z^3}
\!+\!\frac{5!}{z^5}\dots\!\right)\; ,\label{cisasym}
\ee
for large argument, as $|z|\!\rightarrow\!\infty$, and for $-\frac{3\pi}{2}\!<\!{\rm arg}(z)\!<\!\frac{3\pi}{2}$.

\subsection{Digamma function $\psi(z)$}

The digamma function $\psi(z)$ is the logarithmic derivative of the gamma function,
\beeq
\psi(z)\!=\!\frac{\Gamma^\prime(z)}{\Gamma(z)}\; .\label{defndigamma}
\eneq

\section{Sum $\sum_{\vec{n}\neq 0}\!\frac{\Theta(H\!a(t)\!-\! H2\pi n)\,\Theta(H\!a(t')\!-\!H2\pi n)}{n^{2\nu}}e^{i2\pi H\vec{n}\cdot(\vec{x}-\vec{x}\,')}$ in terms of $\Gamma(\beta,z)$} \label{App:sumandtreecorrexpint}

The sum that appears in tree-order correlator~(\ref{sumtheta}),\be
&&\hspace{1.15cm}\sum_{\vec{n}\neq 0}\!\frac{\Theta(H\!a(t)\!-\! H2\pi n)\,\Theta(H\!a(t')\!-\!H2\pi n)}{n^{2\nu}}e^{i2\pi H\vec{n}\cdot(\vec{x}-\vec{x}\,')}\nonumber\\
&&\hspace{-1cm}\simeq\!2^{2\nu-1}\pi^{2\nu-\frac{D}{2}}\!
\frac{\Gamma\!\left(\frac{D}{2}\right)}{\Gamma\!\left(D\!-\!\!1\right)}\Big[\!-\!\!(H\!\Delta x)^2\Big]^{-\frac{\delta}{2}}\!\Bigg\{\!\Gamma\!\left(-\!1\!\!+\!\delta, i \alpha(t')\right)\!-\!\Gamma\!\left(-\!1\!\!+\!\delta, iH\!\Delta x\right)\nonumber\\
&&\hspace{1.30cm}+(-1)^{-\delta}\Big[\Gamma\!\left(-\!1\!\!+\!\delta, -i\alpha(t')\right)\!-\!\Gamma\!\left(-\!1\!\!+\!\delta, -iH\!\Delta x\right)\Big]\!\Bigg\}\; ,\label{sumthetaincompGamma}
\ee
where $\alpha(t')\!\equiv\!a(t')H\!\Delta x$. This form of the sum, when employed in Eq.~(\ref{sumtheta}),
leads to Eq.~(\ref{treecorrgammaf}).  Using Eq.~(\ref{gammaaspowerseries}) in Eq.~(\ref{sumthetaincompGamma}), on the other hand, gives an alternative expression for the sum,\be
&&\hspace{-0.75cm}\sum_{\vec{n}\neq 0}\!\frac{\Theta(H\!a(t)\!-\! H2\pi n)\,\Theta(H\!a(t')\!-\!H2\pi n)}{n^{2\nu}}e^{i2\pi H\vec{n}\cdot(\vec{x}-\vec{x}\,')}\nonumber\\
&&\hspace{-0.75cm}\simeq\!2^{2\nu}\pi^{2\nu-\frac{D}{2}}\!
\frac{\Gamma\!\left(\frac{D}{2}\right)}{\Gamma\!\left(D\!-\!\!1\right)}
\!\sum_{n=0}^\infty\frac{(-1)^n (H\!\Delta x)^{2n}}{(2n\!\!+\!\!1)!}\frac{a^{2n+\delta}(t')\!-\!1}{2n\!\!+\!\delta} \; .\label{sumaspowerseries}
\ee
This form of the sum, when employed in Eq.~(\ref{sumtheta}),
leads to Eq.~(\ref{treecorrseries}).
\section{Massless limit of the tree-order correlator}
\label{App:zeromasscorr}

Massless limit of Eq.~(\ref{treecorrgamma}) yields, \beeq
\lim_{m\rightarrow 0}\langle\Omega|{\bar\varphi}_0(t,\vec{x}) {\bar\varphi}_0(t'\!,\vec{x}\,')|\Omega\rangle\simeq\!
\frac{\Gamma\!\left(D\!-\!1\right)}{\Gamma\!\left(\frac{D}{2}\right)}
\frac{H^{D-2}}{2^{D-1}\pi^{\frac{D}{2}}}\lim_{m\rightarrow 0}\emph{f}_0(t,t'\!,\Delta x)\;,\label{C1mless1}
\eneq
where
\be
\hspace{-0.3cm}\lim_{m\rightarrow 0}\emph{f}_0(t,t'\!,\Delta x)\!=\!\frac{1}{2}\Bigg\{\!\Gamma\!\left(-\!1, i\alpha(t')\right)\!+\!\Gamma\!\left(-\!1, -i\alpha(t')\right)\!-\!\!\Big[\Gamma\!\left(-\!1, iH\!\Delta x\right)\!+\!\Gamma\!\left(-\!1, -iH\!\Delta x\right)\!\Big]\!\Bigg\}\;.\label{treecorrmasslesf}
\ee
Recurrence relation~(\ref{incompgammarecurr}) for the incomplete gamma function implies\beeq
\Gamma(-1, z)\!=\!-\Gamma(0, z)\!+\!z^{-1}e^{-z}\; .
\eneq
Thus, for a purely imaginary variable $z\!\!=\!iy$ with $y\!\in\!\mathbb{R}$, we have
\beeq
\Gamma(-1, iy)\!+\!\Gamma(-1, -iy)\!=\!-\Gamma(0, iy)\!-\!\Gamma(0, -iy)\!-\!2\,\frac{\sin(y)}{y}\; .\label{gammamimus}
\eneq
Using the identity
\beeq
{\rm ci}(z)\!=\!\ln(z)\!-\!\frac{1}{2}\Big[\Gamma(0, iz)\!+\!\Gamma(0, -iz)\!+\!\ln(iz)\!+\!\ln(-iz)\Big]\; ,
\eneq
for a purely real $z\!=\!y$, in Eq.~(\ref{gammamimus}) we find
\be
\hspace{0.3cm}\Gamma(-1, iy)\!+\!\Gamma(-1, -iy)\!=\!2\!\left[{\rm ci}(y)\!-\!\frac{\sin(y)}{y}\!-\!\ln(y)\right]\!\!+\!\ln(iy)\!+\!\ln(-iy)\;.\label{gammaidanalytic}
\ee
Finally, employing Eq.~(\ref{gammaidanalytic}) for $y\!=\!\alpha(t')$ and for $y\!=\!H\!\Delta x$ in Eq.~(\ref{treecorrmasslesf})
gives\be
\lim_{m\rightarrow 0}\emph{f}_0(t,t'\!,\Delta x)
\!=\!{\rm ci}(\alpha(t'))\!-\!\frac{\sin(\alpha(t'))}{\alpha(t')}\!-\!{\rm ci}(H\!\Delta x)
\!+\!\frac{\sin(H\!\Delta x)}{H\!\Delta x}\;.\label{masslessfzero}
\ee
The power series representation of the tree-order correlator~(\ref{treemasslesspower}) can be obtained using
Eq.~(\ref{cieksisin}) in Eq.~(\ref{masslessfzero}). We find\be
\hspace{-0.35cm}\lim_{m\rightarrow 0}\langle\Omega|{\bar\varphi}_0(t,\vec{x}) {\bar\varphi}_0(t'\!,\vec{x}\,')|\Omega\rangle
\!\simeq\!\!
\frac{\Gamma\!\left(D\!-\!\!1\right)}{\Gamma\!\left(\frac{D}{2}\right)}
\frac{H^{D-2}}{2^{D-1}\pi^{\frac{D}{2}}}\!\left[\ln(a(t'))\!+\!\!\!\sum_{n=1}^\infty\frac{\!(-1)^n (H\!\Delta x)^{2n}}{(2n\!\!+\!\!1)!}\frac{a^{2n}(t')\!-\!\!1}{2n}\right]\; .\label{treemasslessseriesrep}
\ee
This massless limit can also be obtained from Eq.~(\ref{treecorrseries}). Singling out the $n\!=\!0$ term in the infinite sum in Eq.~(\ref{treecorrseries})---that is $\frac{a^{\delta}(t')-1}{\delta}$---which causes a ${0}/{0}$ ambiguity in the $\delta\!\!\rightarrow\! 0$ limit, expanding  $a^{\delta}(t')\!=\!e^{\delta\ln(a(t'))}$ around $\delta\!=\!0$ and then taking the $\delta\!\rightarrow \!0$ limit reproduces Eq.~(\ref{treemasslessseriesrep}). Note also that, for comoving separation $\Delta x\!=\!\frac{K}{Ha(t')}$, Eq.~(\ref{treemasslessseriesrep}) agrees with Eq.~(\ref{treemasslesscompare}).

The equal space limit of Eq.~(\ref{treemasslessseriesrep}) [or, the zero mass limit of Eq.~(\ref{massiveequalspaceVEV})] gives
\beeq
\lim_{m\rightarrow 0}\langle\Omega|{\bar\varphi}_0(t,\vec{x}) {\bar\varphi}_0(t'\!,\vec{x})|\Omega\rangle
\!\simeq\!
\frac{\Gamma\!\left(D\!-\!\!1\right)}{\Gamma\!\left(\frac{D}{2}\right)}
\frac{H^{D-2}}{2^{D-1}\pi^{\frac{D}{2}}}\ln(a(t'))\; .\label{masslesseqspace}
\eneq
Finally, let us note that the equal spacetime limit of the tree-order correlator is obtained by taking $t'\!=\!t$ in Eq.~(\ref{masslesseqspace}).

\section{Asymptotic form of the tree-order correlator}
\label{App:asympttreecorr}

Asymptotic form of the time dependent terms inside the curly brackets of tree-order correlator~(\ref{treecorrgamma}) can be obtained using Eq.~(\ref{asympincomplgamma}). We find
\be
&&\hspace{-0.3cm}\left[-\alpha(t)\,\alpha(t')\right]^{-\frac{\delta}{2}}\!\Bigg\{\!\Gamma\!\left(-\!1\!\!+\!\delta, i\alpha(t')\right)\!+\!(-1)^{-\delta}\Gamma\!\left(-\!1\!\!+\!\delta, -i\alpha(t')\right)\!\Bigg\}\nonumber\\
&&\hspace{-2cm}\rightarrow\! -\frac{2}{\Gamma(2\!-\!\delta)}\!\!\left[\frac{a(t')}{a(t)}\right]^\frac{\delta}{2}
\!\!\sum_{n=0}^\infty
\!\frac{(-1)^n\Gamma\!\left(2n\!\!+\!\!2\!-\!\delta\right)}{\alpha^{2n}(t')}\!\!
\left[\frac{\cos(\alpha(t'))}{\alpha^2(t')}\!+\!(2n\!\!+\!\!2\!-\!\delta)\frac{\sin(\alpha(t'))}{\alpha^3(t')}  \right]\; ,\label{asymptdepend}
\ee
which asymptotes to zero for both massive and massless cases. (For $a(t')<a(t)$, the larger the mass, the faster it approaches to zero.) In fact, as $m\rightarrow 0$, Eq.~(\ref{asymptdepend}) limits to
\beeq
2\!\left[{\rm ci}(\alpha(t'))\!-\!\frac{\sin(\alpha(t'))}{\alpha(t')}\right]\!\!\longrightarrow 0\!+\!\mathcal{O}\left(\alpha^{-2}(t')\right)\; .\label{asypmasles}
\eneq
We used Eq.~(\ref{sumcisasym}) in obtaining Eq.~(\ref{asypmasles}). Hence, at late times, tree-order correlator~(\ref{treecorrgamma}),
\be
\hspace{-0.4cm}\langle\Omega|\!{\bar\varphi}_0(t,\vec{x}) {\bar\varphi}_0(t'\!,\vec{x}\,')|\Omega\rangle\!\rightarrow\!
-\frac{\mathcal{A}}{2}\Big[\!\!-\!\alpha(t)\alpha(t')\Big]^{-\frac{\delta}{2}}\!\Bigg\{\!\Gamma\!\left(-\!1\!\!+\!\delta, iH\!\Delta x\!\right)\!+\!(-1)^{-\delta}\Gamma\!\left(-\!1\!\!+\!\delta, \!-iH\!\Delta x\!\right)\!\!\Bigg\} .\label{treecorrgasym}
\ee
Equation~(\ref{treecorrgasym}) asymptotes to zero for the massive case with an increasing rate as the mass increases. In the massless case, however, it asymptotes to
a nonzero constant,
\be
&&\hspace{-1cm}\lim_{m\rightarrow 0}\langle\Omega|{\bar\varphi}_0(t,\vec{x}) {\bar\varphi}_0(t'\!,\vec{x}\,')|\Omega\rangle\longrightarrow\frac{\Gamma\!\left(D\!-\!1\right)}{\Gamma\!\left(\frac{D}{2}\right)}
\frac{H^{D-2}}{2^{D-1}\pi^{\frac{D}{2}}}
\Bigg\{\!\frac{\sin(H\!\Delta x)}{H\!\Delta x}\!-\!{\rm ci}(H\!\Delta x)\!\Bigg\}\; .\label{treemasslessasymp}
\ee
We used Eq.~(\ref{gammaidanalytic}) in obtaining Eq.~(\ref{treemasslessasymp}).

\section{Massless limit of the one-loop correlator} \label{App:oneloopcorrmassless}

Taking the $m\!\rightarrow\!0$ limit of Eq.~(\ref{1loopmassivefinal}) gives the one-loop correlator for the massless scalar,\be
\lim_{m\rightarrow 0}\langle\Omega|
\bar{\varphi}(t,\vec{x})\bar{\varphi}(t'\!,\vec{x}\,')|\Omega\rangle_{\rm 1-loop}\!\simeq\!-\frac
{\lambda}{(D\!-\!\!1)}\frac{H^{2D-6}}{2^{2D-2}\pi^{D}}
\frac{\Gamma^2\!\left(D\!-\!\!1\right)}{\Gamma^2\!\left(\frac{D}{2}\right)} \lim_{m\rightarrow 0}\emph{f}_1(t,t'\!,\Delta x)\;,\label{1loopmassless1}
\ee
where the massless limit of $\emph{f}_1$, defined in Eq.~(\ref{fmassive}), is
\be
&&\hspace{-0.45cm}\lim_{m\rightarrow 0}\emph{f}_1(t,t'\!,\Delta x)\!=\!\frac{1}{4}\Bigg\{\!\!\ln^2(a(t))\ln(a(t'))\!+\!\frac{\ln^3(a(t'))}{3}
\!+\!\frac{1}{2}\!\sum_{n=1}^\infty
\!\frac{(\!-1)^{n}(H\!\Delta x)^{2n}}{n(2n\!\!+\!\!1)!}\nonumber\\
&&\hspace{-1cm}
\times\!\Bigg[a^{2n}(t')\!\Big[\!\ln^2(a(t))\!-\!\ln^2(a(t'))
\!+\!\frac{2\ln(a(t'))}{n}\!-\!\frac{1}{n^2}\Big]\!\!-\!\ln^2(a(t))\!-\!\ln^2(a(t'))\!+\!\frac{1}{n^2}\Bigg]\!\Bigg\}\; .\label{1loopmassless2}
\ee

\section{An analytic form for the one-loop correlator} \label{App:oneloopcorrexpanalyt}

One-loop correlator for the IR truncated massive scalar in our model is given in Eqs.~(\ref{1loopmassivefinal}) and  (\ref{fmassive}) as a power series expansion. An alternative analytic form for one-loop correlator can be obtained by summing up the infinite sum in Eq.~(\ref{fmassive}). The result is
\be
&&\hspace{-0.5cm}\emph{f}_1\!=\!\frac{\left[a(t)\,a(t')\right]^{-\frac{\delta}{2}}}
{\delta^2}\Bigg\{\!\frac{1}{2}\!\left\{a^{\delta}(t')\!\!\left[\ln(a(t))\!-\!\ln(a(t'))
\!+\!\frac{a^{-\delta}(t)}{\delta}\right]
\!\!-\!\frac{1}{\delta}\right\}
{}_1\mathcal{F}_2\Big(\!\frac{\delta}{2};\frac{3}{2},1\!\!+\!\frac{\delta}{2};-\frac{\alpha^2(t')}{4}\!\Big)\nonumber\\
&&\hspace{0.3cm}-\frac{1}{2}\!\left\{\ln(a(t))\!+\!\ln(a(t'))
\!-\!\frac{1\!-\!a^{-\delta}(t)}{\delta}\!-\!\frac{1\!-\!a^{-\delta}(t')}{\delta}\right\}
{}_1\mathcal{F}_2\Big(\!\frac{\delta}{2};\frac{3}{2},1\!\!+\!\frac{\delta}{2};-\frac{(H\!\Delta x)^2}{4}\!\Big)\nonumber\\
&&\hspace{-0.1cm}+\frac{1}{\delta}\!
\left\{a^{\delta}(t')\,{}_2\mathcal{F}_3\Big(\frac{\delta}{2},\frac{\delta}{2}
;\frac{3}{2},\!1\!\!+\!\frac{\delta}{2},\!1\!\!+\!\frac{\delta}{2};\!-\frac{\alpha^2(t')}{4}\!\Big)
\!-\!{}_2\mathcal{F}_3\Big(\frac{\delta}{2},\frac{\delta}{2}
;\frac{3}{2},\!1\!\!+\!\frac{\delta}{2},\!1\!\!+\!\frac{\delta}{2};\!-\frac{(H\!\Delta x)^2}{4}\!\Big)\!\right\}\nonumber\\
&&\hspace{1cm}-\Big\{{\rm ci}(\alpha(t'))\!-\!{\rm ci}(H\!\Delta x)\Big\}\!\!+\!\!\left[1\!\!+\!\frac{1}{\delta}\right]\!\!\!\left[\frac{\sin(\alpha(t'))}{\alpha(t')}
\!-\!\frac{\sin(H\!\Delta x)}{H\!\Delta x}\right]\!\!+\!\frac{1}{3}\frac{(H\!\Delta x)^2}{\delta(2\!+\!\delta)}
\!\nonumber\\
&&\hspace{0.85cm}\times\!\Bigg[a^{2}(t'){}_1\mathcal{F}_2\Big(\!1\!\!+\!\frac{\delta}{2};
\frac{5}{2},2\!+\!\!\frac{\delta}{2};-\frac{\alpha^2(t')}{4}\!\Big)\!-\!{}_1\mathcal{F}_2\Big(\!1\!\!+\!\frac{\delta}{2};\frac{5}{2},
2\!+\!\!\frac{\delta}{2};-\frac{(H\!\Delta x)^2}{4}\!\Big)\!\Bigg]\!\Bigg\}\;.\label{1loopanalyt}
\ee

\section{Computing the two-loop contribution to the two-point correlation function $\langle\Omega|
\bar{\varphi}(t,\vec{x})\bar{\varphi}(t'\!,\vec{x}\,')\!|\Omega\rangle$} \label{App:correlationfunc}

Two-loop contribution to the two-point correlation function of the IR truncated scalar full field consists of three terms each of which is proportional to a VEV; see Eq.~(\ref{correlation}). The VEVs are computed in Sec.~\ref{subsect:2loopcorr}. In this Appendix, we give the details of the computation for each of the VEVs in sections~\ref{App:VEV1}, \ref{App:VEV2} and \ref{App:VEV3}, respectively.

\subsection{Computing the VEV $\langle\Omega|
\bar{\varphi}_0(t,\vec{x})\!\int_0^{t'}\!\!d\tilde{t}\,{\bar{\varphi}}^2_0(\tilde{t},\vec{x}\,')
\!\int_0^{\tilde{t}}\!d\tilde{\tilde{t}}\,a^\frac{\delta}{2}(\tilde{\tilde{t}})  {\bar{\varphi}}^3_0(\tilde{\tilde{t}},\vec{x}\,')
|\Omega\rangle$}
\label{App:VEV1}

In Eq.~(\ref{2loop1stequaltime}) computation of the first VEV in two-loop correlator~(\ref{correlation}) is reduced to the evaluations of three double integrals. The first double integral is
\be
&&\hspace{-1.3cm}\int_0^{t'}\!\!\!d\tilde{t}
                 \Big[a^{2n}(\tilde{t})\!-\!a^{-\delta}(\tilde{t})\Big]
                 \!\!\int_0^{\tilde{t}}\!\!d\tilde{\tilde{t}}
                 \Big[a^{\delta}(\tilde{\tilde{t}})\!+\!a^{-\delta}(\tilde{\tilde{t}})\!-\!2\Big]
                \!\!=\!-H^{-2}\Bigg\{\!\!\left[\frac{a^{2n}(t')}{n}\!+\!2\frac{a^{-\delta}(t')}{\delta}
\!+\!\frac{1}{\delta}\right]\!\ln(a(t'))\nonumber\\
&&\hspace{-1.3cm}+\frac{1}{\delta}\!\!\left[\frac{1\!-\!a^{2n+\delta}(t')}{2n\!\!+\!\delta}\right]
\!\!+\!\frac{1}{n}\!\!\left[\frac{1\!-\!a^{2n}(t')}{2n}\right]
\!\!-\!\frac{1}{\delta}\!\!\left[\frac{1\!-\!a^{2n-\delta}(t')}{2n\!\!-\!\delta}\right]
\!\!-\!\frac{2}{\delta}\!\!\left[\frac{1\!-\!a^{-\delta}(t')}{\delta}\right]
\!\!-\!\frac{1}{\delta}\!\!\left[\frac{1\!-\!a^{-2\delta}(t')}{2\delta}\right]\!\!\Bigg\}\; .\label{2loop1st1stintequaltime}
\ee
The second double integral in Eq.~(\ref{2loop1stequaltime}) is
\be
&&\hspace{-0.65cm}\frac{1}{2}\!\int_0^{t'}\!\!\!d\tilde{t}
                 \Big[1\!-\!a^{-\delta}(\tilde{t})\Big]
                 \!\!\int_0^{\tilde{t}}\!\!d\tilde{\tilde{t}}
                 \Big[a^{2n+\delta}(\tilde{\tilde{t}})\!-\!a^{2n}(\tilde{\tilde{t}})
                 \!+\!a^{-\delta}(\tilde{\tilde{t}})\!-\!1\Big]\!\!=\!-\frac{H^{-2}}{4}\Bigg\{\!\!\ln^2(a(t'))
                 \!-\!2\Big[\frac{1\!\!-\!a^{-\delta}(t')}{\delta}
\!+\!\frac{1}{2n}\nonumber\\
&&\hspace{0.1cm}-\frac{1}{2n\!\!+\!\delta}\Big]\!\ln(a(t'))
\!+\!\frac{2}{2n\!\!+\!\delta}\!\left[\frac{1\!-\!a^{2n+\delta}(t')}{2n\!\!+\!\delta}\right]
\!\!-\!\!\left[\frac{1}{n}\!+\!\frac{2}{2n\!\!+\!\delta}\right]
\!\!\left[\frac{1\!-\!a^{2n}(t')}{2n}\right]
\!\!+\!\frac{1}{n}\!\left[\frac{1\!-\!a^{2n-\delta}(t')}{2n\!\!-\!\delta}\right]\nonumber\\
&&\hspace{3cm}
\!+\!\left[\frac{2}{\delta}
\!+\!\frac{1}{n}\!-\!\frac{2}{2n\!\!+\!\delta}\right]\!\!\left[\frac{1\!-\!a^{-\delta}(t')}{\delta}\right]
\!\!-\!\frac{2}{\delta}\!\left[\frac{1\!-\!a^{-2\delta}(t')}{2\delta}\right]
\!\!\Bigg\}\; .\label{2loop1st2ndintequaltime}
\ee
The third double integral in Eq.~(\ref{2loop1stequaltime}) is
\be
&&\hspace{-0.6cm}\int_0^{t'}\!\!\!d\tilde{t}a^{-\delta}(\tilde{t})
                 \!\!\int_0^{\tilde{t}}\!\!d\tilde{\tilde{t}}
                 \Big[a^{2n}(\tilde{\tilde{t}})\!-\!a^{-\delta}(\tilde{\tilde{t}})\Big]
                 \Big[a^{\delta}(\tilde{\tilde{t}})\!-\!1\Big]^2
                 \!\!\!=\!-H^{-2}\Bigg\{\!\Big[2\frac{a^{-\delta}(t')}{\delta}
\!+\!\frac{1}{\delta}\Big]\!\ln(a(t'))\nonumber\\
&&\hspace{0.3cm}
+\frac{1}{2(n\!\!+\!\delta)}\!\left[\frac{1\!-\!a^{2n+\delta}(t')}{2n\!\!+\!\delta}\right]
\!\!-\!\frac{2}{2n\!\!+\!\delta}\!\left[\frac{1\!-\!a^{2n}(t')}{2n}\right]\!\!+\!
\frac{1}{2n}\!\left[\frac{1\!-\!a^{2n-\delta}(t')}{2n\!\!-\!\delta}\right]\nonumber\\
&&\hspace{0.6cm}\!-\!\left[\frac{2}{\delta}
\!-\!\frac{1}{2n}\!-\!\frac{1}{2(n\!\!+\!\delta)}\!+\!\frac{2}{2n\!+\!\delta}\right]
\!\!\left[\frac{1\!-\!a^{-\delta}(t')}{\delta}\right]
\!\!-\!\frac{1}{\delta}\!\left[\frac{1\!-\!a^{-2\delta}(t')}{2\delta}\right]\!\!\Bigg\}\; .\label{2loop1st3rdintequaltime}
\ee
Employing (\ref{2loop1st1stintequaltime}), (\ref{2loop1st2ndintequaltime}) and (\ref{2loop1st3rdintequaltime}) in Eq.~(\ref{2loop1stequaltime}) yields the first VEV in two-loop correlator~(\ref{correlation}) given in Eq.~(\ref{2loop1stequaltimeresult}). Notice that the infinite sum in Eq.~(\ref{2loop1stequaltimeresult}) runs from $n\!=\!0$ and includes terms proportional to $n^{-1}$ and $n^{-2}$. The $n\!=\!0$ term in the summand of infinite sum,\beeq
\frac{3}{2\delta}\ln^2(a(t'))\!+\!\frac{6}{\delta^2}\!\left[a^{-\delta}(t')\!+\!\frac{3}{4}\right]\!\ln(a(t'))
\!-\!\frac{2}{\delta^3}\!\left[a^{\delta}(t')\!-\!3a^{-\delta}(t')
\!-\!\frac{5}{8}a^{-2\delta}(t')\!+\!\frac{21}{8}\right]\;,\label{n0term1stvev}
\eneq
however, is finite.

\subsection{Computing the VEV $\langle\Omega|\!
\int_0^{t}\!dt''{\bar{\varphi}}^2_0(t''\!,\vec{x})\!
\int_0^{t''}\!\!dt'''a^\frac{\delta}{2}(t''') {\bar{\varphi}}^3_0(t'''\!,\vec{x})
\bar{\varphi}_0(t'\!,\vec{x}\,')
|\Omega\rangle$}
\label{App:VEV2}

In Eq.~(\ref{twoloopsecondexpandequaltime}) computation of this VEV is reduced to the evaluations of three double integrals. Let us evaluate the first one. Because $t'\!\leq\!t$, we can break up the double integral into two [to be able to use Eqs.~(\ref{treecorrgamma}), (\ref{Acons}) and (\ref{treecorrseries})]: \be\int_0^{t}\!\!dt''\!\!\int_0^{t''}\!\!\!dt'''\!=\!\int_0^{t'}\!\!\!dt''\!\!\int_0^{t''}\!\!\!dt'''
\!+\!\!\int_{t'}^{t}\!\!dt''\!\!\int_0^{t''}\!\!\!dt'''\; .\label{divideint}\ee
In the first double integral on the right we have $0\!\leq\!t''\!\leq\!t'$, whereas $t'\!\leq\!t''\!\leq\!t$ in the second. Thus, using Eqs.~(\ref{treecorrgamma}), (\ref{Acons}), (\ref{treecorrseries}) and (\ref{divideint}) we can express the first double integral in Eq.~(\ref{twoloopsecondexpandequaltime})
as\be
&&\hspace{-0.3cm}\mathcal{A}a^{-\frac{\delta}{2}}(t')\!\sum_{n=0}^\infty
\!\frac{(-1)^{n}(H\!\Delta x)^{2n}}{(2n\!\!+\!\!1)!(2n\!\!+\!\delta)}
\Bigg\{\!\!\int_0^{t'}\!\!\!dt'' a^{-\delta}(t'')\!\Big[a^{2n+\delta}(t'')\!-\!\!1\Big]\!\int_0^{t''}\!\!\!\!dt'''\!
\Big[a^\delta(t''')\!-\!\!1\Big]\Big[1\!-\!a^{-\delta}(t''')\Big]\nonumber\\
&&\hspace{1.7cm}+\Big[a^{2n+\delta}(t')\!-\!\!1\Big]\!\int_{t'}^t\!\!dt''\!a^{-\delta}(t'')
\!\int_0^{t''}\!\!\!\!dt'''\Big[a^\delta(t''')\!-\!\!1\Big]\Big[1\!-\!a^{-\delta}(t''')\Big]\!\Bigg\}\;.
\ee
Evaluating the integrals we find,
\be
&&\hspace{-0.3cm}=\!-\frac{\mathcal{A}}{H^2}a^{-\frac{\delta}{2}}(t')\!\sum_{n=0}^\infty
\!\frac{(-1)^{n}(H\!\Delta x)^{2n}}{(2n\!\!+\!\!1)!(2n\!\!+\!\delta)}
\Bigg\{\!\Big[1\!\!+\!2a^{-\delta}(t)\Big]\!\!\left[\frac{1\!-\!a^{2n+\delta}(t')}{\delta}\right]\!\ln(a(t))
\!+\!\!\left[\frac{a^{\delta}(t')}{\delta}\!+\!\frac{1}{n}\!+\!\frac{2}{\delta}\right]\nonumber\\
&&\hspace{0cm}\times a^{2n}(t')\ln(a(t'))\!+\!\!\left[\frac{1}{2n\!\!+\!\delta}\!+\!\frac{2a^{-\delta}(t)}{\delta}
\!+\!\frac{a^{-2\delta}(t)}{2\delta}\right]\!\!\left[\frac{1\!-\!a^{2n+\delta}(t')}{\delta}\right]
\!\!+\!\!\left[\frac{\delta}{2n^2}\!-\!\frac{2}{\delta}
\right]\!\!\left[\frac{1\!-\!a^{2n}(t')}{\delta}\right]\nonumber\\
&&\hspace{4.6cm}\!-\!\!\left[\frac{1}{2n\!-\!\delta}\!+\!\frac{1}{2\delta}
\right]\!\!\left[\frac{1\!-\!a^{2n-\delta}(t')}{\delta}\right]\!\!\Bigg\}\label{firstint}\; .
\ee

Let us now evaluate the second double integral in Eq.~(\ref{twoloopsecondexpandequaltime}). Breaking up the integrals~as
\be\int_0^{t}\!dt''\!\!\int_0^{t''}\!\!\!dt'''\!=\!\int_0^{t'}\!\!dt''\!\!\int_0^{t''}\!\!\!dt'''
\!+\!\!\int_{t'}^{t}\!dt''\!\!\int_0^{t'}\!\!\!dt'''
\!+\!\!\int_{t'}^{t}\!dt''\!\!\int_{t'}^{t''}\!\!\!dt'''\; ,\label{divisonfinal}\ee
and using Eqs.~~(\ref{treecorrgamma}), (\ref{Acons}) and (\ref{treecorrseries}) the second double integral in Eq.~(\ref{twoloopsecondexpandequaltime}) can be written~as
\be
&&\hspace{-0.5cm}\frac{\mathcal{A}}{2}\,a^{-\frac{\delta}{2}}(t')\!\sum_{n=0}^\infty
\!\frac{(-1)^{n}(H\!\Delta x)^{2n}}{(2n\!\!+\!\!1)!(2n\!\!+\!\delta)}
\Bigg\{\!\!\int_0^{t'}\!\!\!dt''\!\left[1\!-\!a^{-\delta}(t'')\right]
\!\!\int_0^{t''}\!\!\!\!dt'''\Big[a^{2n+\delta}(t''')\!-\!a^{2n}(t''')\!+\!a^{-\delta}(t''')\!-\!1\Big]\nonumber\\
&&\hspace{2cm}+\!\!\int_{t'}^t\!\!dt''\!\left[1\!-\!a^{-\delta}(t'')\right]
\!\!\int_0^{t'}\!\!\!dt'''\Big[a^{2n+\delta}(t''')\!-\!a^{2n}(t''')\!+\!a^{-\delta}(t''')\!-\!1\Big]\nonumber\\
&&\hspace{2.3cm}+\Big[a^{2n+\delta}(t')\!-\!1\Big]\!\!\int_{t'}^t\!\!dt''\Big[1\!-\!a^{-\delta}(t'')\Big]
\!\!\int_{t'}^{t''}\!\!\!dt'''\Big[1\!-\!a^{-\delta}(t''')\Big]\!\Bigg\}\; .\label{2loop2ndvev2ndint}
\ee
Evaluating the integrals in Eq.~(\ref{2loop2ndvev2ndint}) yields
\be
&&\hspace{-0.65cm}-\frac{\mathcal{A}}{2H^2}\,a^{-\frac{\delta}{2}}(t')\!\sum_{n=0}^\infty
\!\frac{(-1)^{n}(H\!\Delta x)^{2n}}{(2n\!\!+\!\!1)!(2n\!\!+\!\delta)}
\Bigg\{\!\!\!\left[\frac{1\!-\!a^{2n+\delta}(t')}{2}\right]\!\ln^2(a(t))
\!+\!a^{2n+\delta}(t')\ln(a(t))\!\ln(a(t'))\nonumber\\
&&\hspace{-0.2cm}-\frac{a^{2n+\delta}(t')}{2}\ln^2(a(t'))
\!+\!\!\left[\frac{1\!-\!a^{2n+\delta}(t')}{\delta}\!\left[a^{-\delta}(t)
\!+\!\frac{\delta}{2n\!\!+\!\delta}\right]\!-\!\frac{1\!-\!a^{2n}(t')}{\delta}\!
\left[1
\!+\!\frac{\delta}{2n}\right]\right]\!\ln(a(t))\nonumber\\
&&\hspace{-0.6cm}+\!\left[\frac{a^{2n+\delta}(t')}{\delta}\!\left[a^{-\delta}(t)
\!+\!\frac{\delta}{2n\!\!+\!\delta}\right]\!\!-\!\frac{a^{2n}(t')}{\delta}\!\!
\left[1
\!+\!\frac{\delta}{2n}\right]\right]\!\ln(a(t'))\!+\!\frac{1\!-\!a^{2n+\delta}(t')}{\delta}
\Bigg[\frac{a^{-\delta}(t)}{2n\!\!+\!\delta}\!+\!\frac{a^{-2\delta}(t)}{2\delta}\nonumber\\
&&\hspace{-0.7cm}-\frac{a^{-\delta}(t)\,a^{-\delta}(t')}{\delta}
\!-\!\frac{a^{-\delta}(t')}{2n\!\!+\!\delta}\!+\!\frac{a^{-2\delta}(t')}{2\delta}
\!+\!\frac{\delta}{(2n\!\!+\!\delta)^2}\Bigg]
\!\!-\!\frac{1\!-\!a^{2n}(t')}{2n}
\Bigg[\frac{a^{-\delta}(t)}{\delta}\!-\!\frac{a^{-\delta}(t')}{\delta}
\!+\!\frac{1}{2n}\!+\!\frac{1}{2n\!\!+\!\delta}\Bigg]\nonumber\\
&&\hspace{1.1cm}+\frac{1\!-\!a^{2n-\delta}(t')}{2n\!-\!\delta}
\frac{1}{2n}\!-\!\frac{1\!-\!a^{-\delta}(t')}{\delta}\!\left[\frac{a^{-\delta}(t)}{\delta}
\!-\!\frac{1}{2n}\!+\!\frac{1}{2n\!\!+\!\delta}\right]
\!\!+\!\frac{1\!-\!a^{-2\delta}(t')}{2\delta^2}\Bigg\}\; .\label{secondint}
\ee

The third double integral in Eq.~(\ref{twoloopsecondexpandequaltime}) is obtained similarly. Employing Eqs.~(\ref{treecorrgamma}), (\ref{Acons}), (\ref{treecorrseries}) and (\ref{divisonfinal}) we express it as
\be
&&\hspace{-0.8cm}\mathcal{A}\,a^{-\frac{\delta}{2}}(t')\!\sum_{n=0}^\infty
\!\frac{(-1)^{n}(H\!\Delta x)^{2n}}{(2n\!\!+\!\!1)!(2n\!\!+\!\delta)}
\Bigg\{\!\!\int_0^{t'}\!\!\!dt''a^{-\delta}(t'')
\!\!\int_0^{t''}\!\!\!\!dt'''a^{-\delta}(t''')\Big[a^{2n+\delta}(t''')\!-\!1\Big]
\Big[a^{\delta}(t''')\!-\!1\Big]^2\nonumber\\
&&\hspace{1.7cm}+\!\!\int_{t'}^t\!\!dt''\!a^{-\delta}(t'')
\!\!\int_0^{t'}\!\!\!dt'''a^{-\delta}(t''')\Big[a^{2n+\delta}(t''')\!-\!1\Big]
\Big[a^{\delta}(t''')\!-\!1\Big]^2\nonumber\\
&&\hspace{1.6cm}+\Big[a^{2n+\delta}(t')\!-\!1\Big]\!\!\int_{t'}^t\!\!dt''a^{-\delta}(t'')
\!\!\int_{t'}^{t''}\!\!\!\!dt'''a^{-\delta}(t''')\Big[a^{\delta}(t''')\!-\!1\Big]^2\Bigg\}\; .\label{thrddbint2loop}
\ee
Evaluating the integrals in Eq.~(\ref{thrddbint2loop}) yields\be
&&\hspace{-0.4cm}-\frac{\mathcal{A}}{H^2}a^{-\frac{\delta}{2}}(t')\!\sum_{n=0}^\infty
\!\frac{(-1)^{n}(H\!\Delta x)^{2n}}{(2n\!\!+\!\!1)!(2n\!\!+\!\delta)}
\Bigg\{\!\!\left[1\!-\!a^{2n+\delta}(t')\right]\!\!\left[\frac{1\!\!+\!\!2a^{-\delta}(t)}{\delta}\right]\!\ln(a(t))
\!+\!a^{2n+\delta}(t')\!\!\left[\frac{1\!\!+\!\!2a^{-\delta}(t)}{\delta}\right]\nonumber\\
&&\hspace{-0.5cm}\times\!\ln(a(t'))\!-\!\frac{1\!-\!a^{2n+2\delta}(t')}{\delta}\!\left[a^{-\delta}(t)\!\left[
\frac{1}{2(n\!+\!\delta)}\!-\!\frac{1}{\delta}\right]\!\!-\!\frac{a^{-\delta}(t')}{2(n\!+\!\delta)}\right]
\!\!+\!\frac{1\!-\!a^{2n+\delta}(t')}{\delta}\!\Bigg[a^{-\delta}(t)\!\left[
\frac{2}{2n\!\!+\!\delta}\!+\!\frac{2}{\delta}\right]\nonumber\\
&&\hspace{-0.5cm}-\frac{1\!\!-\!a^{-2\delta}(t)}{2\delta}
\!-\!\frac{2a^{-\delta}(t')}{2n\!\!+\!\delta}\!-\!\frac{1\!\!-\!a^{-2\delta}(t')}{2\delta}
\!+\!\frac{\delta}{2(n\!\!+\!\delta)(2n\!\!+\!\delta)}\Bigg]
\!\!-\!\frac{1\!\!-\!a^{2n}(t')}{\delta}\Bigg[a^{-\delta}(t)\!\!\left[\frac{1}{2n}\!+\!
\frac{1}{\delta}\right]\!\!-\!\frac{a^{-\delta}(t')}{2n}\nonumber\\
&&\hspace{-0.1cm}+\frac{\delta}{n(2n\!\!+\!\delta)}
\!+\!\frac{2}{\delta}\Bigg]\!\!+\!\frac{1\!\!-\!a^{2n-\delta}(t')}{2n(2n\!\!-\!\delta)}
\!+\!\frac{1\!\!-\!a^{-\delta}(t')}{\delta}
\Bigg[\frac{1}{2n}\!+\!\frac{1}{2(n\!\!+\!\delta)}\!-\!\frac{2}{2n\!\!+\!\delta}\Bigg]
\!\!+\!\frac{1\!\!-\!a^{-2\delta}(t')}{2\delta^2}\Bigg\}\; .\label{thirdint}
\ee

Combining (\ref{firstint}), (\ref{secondint}) and (\ref{thirdint}) in Eq.~(\ref{twoloopsecondexpandequaltime}) yields the second VEV in two-loop correlator~(\ref{correlation}) as a power series expansion. The outcome is given in Eq.~(\ref{2loop2ndresult}). Notice that the infinite sum in Eq.~(\ref{2loop2ndresult}) runs from $n\!=\!0$ and includes terms proportional to $n^{-1}$ and $n^{-2}$. However, the $n\!=\!0$ term in the summand in Eq.~(\ref{2loop2ndresult}) is finite. In fact, it is
\be
&&\hspace{1.1cm}\left[\frac{1\!\!-\!a^{\delta}(t')}{\delta}\right]
\!\!\frac{\ln^2(a(t))}{2}
\!+\!\!\left[\frac{1\!\!+\!a^{\delta}(t')}{\delta}\right]\!\ln(a(t))\ln(a(t'))\!+\!
\!\left[\frac{1\!\!-\!a^{\delta}(t')}{\delta}\!+\!\frac{2}{\delta}\right]\!\!\frac{\ln^2(a(t'))}{2}\nonumber\\
&&\hspace{-0.4cm}
\!+\!\left\{\!\frac{1\!\!-\!a^{\delta}(t')}{\delta}\!\!\left[5\!\!\left[\frac{1\!\!+\!a^{-\delta}(t)}{\delta}\right]
\!\!\!+\!4\frac{a^{-\delta}(t)}{\delta}\right]\!\right\}\!\ln(a(t))
\!\!+\!\!\left\{\!\frac{1\!\!+\!a^{-\delta}(t)}{\delta}\!\!\left[3\!\!\left[\frac{1\!\!+\!a^{\delta}(t')}{\delta}\right]
\!\!\!+\!2\frac{a^{\delta}(t')}{\delta}\right]\!\!\!+\!\frac{5}{\delta^2}\!\right\}\!\ln(a(t'))\nonumber\\
&&\hspace{1.65cm}+\frac{5}{\delta^2}\!\left[\frac{1\!\!+\!a^{-2\delta}(t)}{2\delta}\right]
\!\!+\!\frac{a^{-\delta}(t)}{\delta^2}\!\left[13\!\left[\frac{1\!\!-\!a^{\delta}(t')}{\delta}\right]
\!\!+\!\frac{1\!\!-\!a^{2\delta}(t')}{\delta}\!-\!\frac{5}{2\delta}a^{-\delta}(t)a^{\delta}(t')\right]\nonumber\\
&&\hspace{5.5cm}-\frac{a^{\delta}(t')}{\delta^2}\!\left[\frac{1\!\!-\!a^{-2\delta}(t')}{2\delta}
\!+\!\frac{5}{2\delta}\right]\;.\label{n0term2ndvev}
\ee
Thus, the second VEV in the two-loop correlator is finite.

Computation of the remaining VEV in two-loop correlator~(\ref{correlation}) is similar to the first two ones given in Secs.~(\ref{App:VEV1})-(\ref{App:VEV2}). Some details are outlined in the next section.

\subsection{Computing the VEV $\langle\Omega|\!
\int_0^{t}\!dt'' a^{\frac{\delta}{2}}(t''){\bar{\varphi}}^3_0(t''\!,\vec{x})\!
\int_0^{t'}\!\!d\tilde{t} \, a^{\frac{\delta}{2}}(\tilde{t}){\bar{\varphi}}^3_0(\tilde{t},\vec{x}\,')
|\Omega\rangle$}
\label{App:VEV3}

In Eq.~(\ref{2loop3rdaftmidequaltime}) computation of this VEV is reduced to the evaluations of four double integrals. The first double integral in Eq.~(\ref{2loop3rdaftmidequaltime}) yields,
\be
&&\hspace{-0.5cm}\int_0^{t'}\!\!\!d\tilde{t}
\!\left[1\!\!-\!a^{-\delta}(\tilde{t})\right]\!\!\!\int_0^{\tilde{t}}\!\!dt''\!\left[1\!\!-\!a^{-\delta}(t'')\right]\!
\!\left[a^{2n+\delta}(t'')\!-\!1\right]\!\!=\!-H^{-2}\Bigg\{\!\frac{\ln^2(a(t'))}{2}
\!-\!\!\Bigg[\!\frac{1\!\!-\!a^{-\delta}(t')}{\delta}\!+\!\frac{1}{2n}\!-\!\frac{1}{2n\!\!+\!\delta}\Bigg]\nonumber\\
&&\hspace{-0.4cm}\times\!\ln(a(t'))
\!+\!\frac{1\!\!-\!a^{2n+\delta}(t')}{(2n\!\!+\!\delta)^2}\!-\!\frac{1\!\!-\!a^{2n}(t')}{2n}\!\left[
\frac{1}{2n}\!+\!\frac{1}{2n\!\!+\!\delta}\right]\!\!+\!\frac{1\!\!-\!a^{2n-\delta}(t')}
{(2n\!\!-\!\delta)2n}\!+\!\frac{1\!\!-\!a^{-\delta}(t')}
{\delta}\!\left[\frac{1}
{2n}\!-\!\frac{1}{2n\!\!+\!\delta}\!+\!\frac{1}{\delta}\right]\nonumber\\
&&\hspace{12cm}-\frac{1\!\!-\!a^{-2\delta}(t')}
{2\delta^2}\!\Bigg\}\;.\label{firstdbint}
\ee
The second double integral in Eq.~(\ref{2loop3rdaftmidequaltime}) yields,
\be
&&\hspace{-0.3cm}\int_0^{t'}\!\!\!d\tilde{t}
\!\left[1\!-\!a^{-\delta}(\tilde{t})\right]\!\!
\left[a^{2n+\delta}(\tilde{t})\!-\!1\right]\!\!
\int_{\tilde{t}}^t\!\!dt''\!\left[1\!-\!a^{-\delta}(t'')\right]\!\!=\!-H^{-2}\Bigg\{\!\!\ln(a(t))\ln(a(t'))
\!-\!\frac{\ln^2(a(t'))}{2}\nonumber\\
&&\hspace{-0.7cm}+\Bigg[\!\frac{1\!\!-\!a^{2n+\delta}(t')}{2n\!\!+\!\delta}\!-\!\frac{1\!\!-\!a^{2n}(t')}{2n}
\!-\!\frac{1\!\!-\!a^{-\delta}(t')}{\delta}
\Bigg]\!\ln(a(t))
\!+\!\Bigg[\!\frac{a^{2n+\delta}(t')}{2n\!\!+\!\delta}\!-\!\frac{a^{2n}(t')}{2n}
\!+\!\frac{a^{-\delta}(t)}{\delta}\!-\!\frac{a^{-\delta}(t')}{\delta}
\Bigg]\nonumber\\
&&\hspace{-0.2cm}\times\!\ln(a(t'))
\!+\!\frac{1\!\!-\!a^{2n+\delta}(t')}{2n\!\!+\!\delta}\!
\left[\frac{a^{-\delta}(t)}{\delta}\!+\!\frac{1}{2n\!\!+\!\delta}\right]\!\!-\!\frac{1\!\!-\!a^{2n}(t')}{2n}\!\left[
\frac{1\!\!+\!a^{-\delta}(t)}{\delta}\!+\!\frac{1}{2n}\right]\!\!+\!\frac{1\!\!-\!a^{2n-\delta}(t')}
{(2n\!\!-\!\delta)\delta}\nonumber\\
&&\hspace{4.5cm}-\frac{1\!\!-\!a^{-\delta}(t')}
{\delta}\,\frac{a^{-\delta}(t)}{\delta}\!+\!\frac{1\!\!-\!a^{-2\delta}(t')}
{2\delta^2}\!\Bigg\}\;.\label{seconddbint}
\ee The third double integral in Eq.~(\ref{2loop3rdaftmidequaltime}) yields,
\be
&&\hspace{1.5cm}\int_0^{t'}\!\!\!d\tilde{t}\,a^{-\delta}(\tilde{t})
\!\!\int_0^{\tilde{t}}\!\!\!dt''a^{-\delta}(t'')\!\left[\sum_{n=0}^\infty
\!\frac{(-1)^{n}(H\!\Delta x)^{2n}}{(2n\!\!+\!\!1)!(2n\!\!+\!\delta)}\!\left[a^{2n+\delta}(t'')\!-\!1\right]\!\right]^3\nonumber\\
&&\hspace{-1cm}\!=\!-
H^{-2}\!\sum_{q=0}^\infty\!\sum_{p=0}^q\!\sum_{n=0}^p
\Gamma_{qp\,n}(H\!\Delta x)^{2q}\Bigg\{\!\frac{1}{2(q\!+\!\delta)}\!\left[\frac{1\!\!-\!a^{2q+\delta}(t')}
{2q\!+\!\delta}\!+\!\frac{1\!\!-\!a^{-\delta}(t')}
{\delta}\right]\!\!-\!\frac{3}{2p\!+\!\delta}\!\Bigg[\frac{1\!\!-\!a^{2p}(t')}
{2p}\nonumber\\
&&\hspace{-1cm}+\frac{1\!\!-\!a^{-\delta}(t')}
{\delta}\Bigg]\!\!+\!\frac{3}{2(q\!-\!p)}\!\left[\frac{1\!\!-\!a^{2(q-p)-\delta}(t')}
{2(q\!-\!p)\!-\!\delta}\!+\!\frac{1\!\!-\!a^{-\delta}(t')}
{\delta}\right]\!\!+\!\frac{1}
{\delta}\!\left[\frac{1\!\!-\!a^{-\delta}(t')}{\delta}
\!-\!\frac{1\!\!-\!a^{-2\delta}(t')}{2\delta}\right]\!\!\Bigg\}
\; ,\label{thirddbint}
\ee
where we define the coefficients\be\Gamma_{qp\,n}\!\equiv\!
\frac{(-1)^q}{[2(q\!-\!p)\!+\!1]![2(q\!-\!p)\!+\!\delta][2(p\!-\!n)\!+\!1]!
[2(p\!-\!n)\!+\!\delta][2n\!+\!1]![2n\!+\!\delta]}\; .\label{coeffgamma}\ee The fourth double integral in Eq.~(\ref{2loop3rdaftmidequaltime}) yields
\be
&&\hspace{2.7cm}\int_0^{t'}\!\!\!\!d\tilde{t}\,a^{-\delta}(\tilde{t})
\!\!\!\int_{\tilde{t}}^t\!\!\!dt''a^{-\delta}(t'')\!\!\left[\sum_{n=0}^\infty
\!\frac{(-1)^{n}(H\!\Delta x)^{2n}}{(2n\!\!+\!\!1)!(2n\!\!+\!\delta)}\!\left[a^{2n+\delta}(\tilde{t})\!-\!1\right]\!\right]^3\nonumber\\
&&\hspace{-0.05cm}\!\!\!\!\!=\!-\frac{H^{-2}}
{\delta}\!\!\sum_{q=0}^\infty\!\sum_{p=0}^q\!\sum_{n=0}^p
\Gamma_{qp\,n}(H\!\Delta x)^{2q}\Bigg\{\!\!\frac{1\!\!-\!a^{2q+\delta}(t')}
{2q\!+\!\delta}\!-\!3\frac{1\!\!-\!a^{2p}(t')}
{2p}\!+\!3\frac{1\!\!-\!a^{2(q-p)-\delta}(t')}
{2(q\!-\!p)\!-\!\delta}\!+\!\frac{1\!\!-\!a^{-2\delta}(t')}
{2\delta}\nonumber\\
&&\hspace{1.5cm}-a^{-\delta}(t)\!\Bigg[\!\frac{1\!\!-\!a^{2(q\!+\!\delta)}(t')}{2(q\!+\!\delta)}\!-\!3\frac{1\!\!-\!a^{2p+\delta}(t')}
{2p\!+\!\delta}\!+\!3\frac{1\!\!-\!a^{2(q-p)}(t')}
{2(q\!-\!p)}\!+\!\frac{1\!\!-\!a^{-\delta}(t')}{\delta}\Bigg]\!\Bigg\}
\label{fourthdbint}\; .
\ee
Employing Eqs.~(\ref{firstdbint})-(\ref{thirddbint}) and (\ref{fourthdbint}) in Eq.~(\ref{2loop3rdaftmidequaltime}), yields the third VEV in two-loop correlator~(\ref{correlation}) given in Eq.~(\ref{2loop3rdequaltimeresult}). Notice that the $n\!=\!0$ term in the summand of the first infinite sum in Eq.~(\ref{2loop3rdequaltimeresult}) is finite. It is
\be
&&\hspace{-1cm}-\frac{1}{\delta^2}\Big[a^{\delta}(t')\!-\!a^{-\delta}(t')\Big]\!\!\left[\ln(a(t))
\!+\!\frac{2\!\!+\!a^{-\delta}(t)}{\delta}\right]\!\!+\!
\frac{2}{\delta}\ln(a(t))\ln(a(t'))\nonumber\\
&&\hspace{1cm}+\frac{1}{\delta^2}
\Big[2a^{-\delta}(t)\!+\!a^{\delta}(t')\!+\!a^{-\delta}(t')\!+\!\!2\Big]\!\ln(a(t'))\;.\label{n0term3rdvev}
\ee
Notice also that, for $p\!=\!0$ (which implies $n\!=\!0$) the terms proportional to $p^{-1}$ in the summand of triple infinite sum in Eq.~(\ref{2loop3rdequaltimeresult}),
\beeq
\Gamma_{qp\,n}(H\!\Delta x)^{2q}\,3\frac{1\!\!-\!a^{2p}(t')}
{2p}\!\left[\frac{1}{2p\!\!+\!\delta}\!+\!\frac{1}{\delta}\right]
\!\longrightarrow\! -6\frac{(-1)^q(H\!\Delta x)^{2q}\ln(a(t'))}{\delta^3[2q\!\!+\!\!1]![2q\!\!+\!\delta]}\; ,
\label{p0term3rdvev}
\eneq
is finite as well. Let us note finally that for $p\!=\!q$, the terms proportional to $(q\!-\!p)^{-1}$ in the summand of triple infinite sum,
\be
&&\hspace{-0.5cm}\Gamma_{qp\,n}(H\!\Delta x)^{2q}\Bigg[\frac{a^{-\delta}(t)}{\delta}\,3\frac{1\!\!-\!a^{2(q-p)}(t')}
{2(q\!-\!p)}\!-\!3\frac{1\!\!-\!a^{2(q-p)-\delta}(t')}
{2(q\!-\!p)\!-\!\delta}\frac{1}{2(q\!-\!p)}\!-\!\frac{1\!\!-\!a^{-\delta}(t')}
{\delta}\frac{3}{2(q\!-\!p)}\!\Bigg]\nonumber\\
&&\hspace{0.5cm}\longrightarrow-3\frac{(-1)^q(H\!\Delta x)^{2q}\!\left\{a^{-\delta}(t')\!\left[\frac{1-a^{\delta}(t')}{\delta}\right]
\!\!+\!\!\left[a^{-\delta}(t)\!+\!a^{-\delta}(t')\right]\!\ln(a(t'))\right\}}
{\delta^2[2(q\!-\!n)\!+\!1]![2(q\!-\!n)\!+\!\delta][2n\!\!+\!\!1]![2n\!\!+\!\delta]}\; .\label{pqterm3rdvev}
\ee
are also finite. Combining all three VEVs computed in Secs.~(\ref{App:VEV1})-(\ref{App:VEV3}) in Eq.~(\ref{correlation}) gives the two-loop correlator.
\end{appendix}


\begin{thebibliography}{99}

\bibitem{vacuum}  V. K. Onemli, Phys. Rev. D {\bf 91}, 103537 (2015).

\bibitem{WR1} R. P. Woodard, in {\it Quantum Field Theory under the Influence of External
Conditions}, edited by K. A. Milton (Rinton Press, Princeton, 2004) p.
325, astro-ph/0310757.

\bibitem{WR2} R. P. Woodard, Rep. Prog. Phys. {\bf 72}, 126002 (2009).

\bibitem{WR3} R. P. Woodard, Int. J. Mod. Phys. D {\bf 23}, 1430020 (2014).

\bibitem{Star} A.~A.~Starobinsky, in {\it Field Theory, Quantum
Gravity and Strings}, edited by H.~J.~de~Vega and N.~Sanchez
(Springer-Verlag, Berlin, 1986), p. 107.

\bibitem{StarYok} A.~A.~Starobinsky and J.~Yokoyama, Phys. Rev.~D~{\bf 50},
6357 (1994).

\bibitem{W1}
R.~P.~Woodard, Nucl. Phys.~B, Proc. Suppl. {\bf 148}, 108 (2005).

\bibitem{W3}
S.~P.~Miao and R.~P.~Woodard, Phys. Rev.~D~{\bf 74},
044019 (2006).

\bibitem{W4}
R.~P.~Woodard, J. Phys. Conf. Ser. {\bf 68}, 012032 (2007).

\bibitem{Wstocsqed}
T.~Prokopec, N.~C.~Tsamis and R.~P.~Woodard, Ann. Phys. (Amsterdam)~{\bf 323}, 1324 (2008).


\bibitem{Wstocqgrav}
N.~C.~Tsamis and R.~P.~Woodard, Nucl. Phys.~{\bf B724}, 295 (2005).

\bibitem{MiaoWood} S.~P.~Miao and R.~P.~Woodard, Classical Quantum Gravity {\bf 25}, 145009 (2008).

\bibitem{Kit1}
H.~Kitamoto and Y.~Kitazawa, Phys. Rev.~D~{\bf 83}, 104043 (2011).

\bibitem{Kit2}
H.~Kitamoto and Y.~Kitazawa, Phys. Rev.~D~{\bf 85}, 044062 (2012).

\bibitem{MC} V. F. Mukhanov and G. V. Chibisov, JETP Lett. {\bf 33}, 532 (1981).

\bibitem{AS1} A. A. Starobinsky, JETP Lett. {\bf 30}, 682 (1979).

\bibitem{sfl} References on quantum theory of scalar field fluctuations include:
S. Weinberg, Phys. Rev. D {\bf 72}, 043514 (2005); Phys. Rev. D {\bf74}, 023508 (2006);
K. Chaicherdsakul, Phys. Rev. D {\bf75}, 063522 (2007); P. Adshead, R. Easther and E. A. Lim, Phys. Rev. D {\bf79},
063504 (2009); D. Boyanovsky, H. J. de Vega and N. G. Sanchez, Nucl.
Phys. {\bf B747}, 25 (2006); Phys. Rev. D {\bf72}, 103006 (2005); M. Sloth, Nucl. Phys. {\bf B748}, 149 (2006); Nucl. Phys. {\bf B775}, 78 (2007); D. Seery, J. E. Lidsey and M. S. Sloth, J.~Cosmol. Astropart. Phys.~01 (2007) 027; M. van der Meulen and J. Smit, J.~Cosmol. Astropart. Phys.~11 (2007) 023; D. H. Lyth, J.~Cosmol. Astropart. Phys.~12 (2007) 016; D. Seery, J.~Cosmol. Astropart. Phys.~11 (2007) 025; J.~Cosmol. Astropart. Phys.~02 (2008) 006; J.~Cosmol. Astropart. Phys.~05 (2009) 021; Classical Quantum Gravity~{\bf 27},
124005 (2010); Y. Urakawa and K. I. Maeda, Phys. Rev. D {\bf78}, 064004 (2008); A. Riotto and M. Sloth,
J.~Cosmol. Astropart. Phys.~04 (2008) 030; J.~Cosmol. Astropart. Phys.~10 (2011) 003; P. Adshead, R. Easther and E. A. Lim, Phys. Rev.
D {\bf 79}, 063504 (2009); Y. Urakawa and T. Tanaka, Prog. Theor. Phys. {\bf 122}, 779 (2009); Prog. Theor.
Phys. {\bf 122}, 1207 (2009); Phys. Rev. D {\bf82}, 121301 (2010);
Prog. Theor. Phys. {\bf 125}, 1067 (2011); J.~Cosmol. Astropart. Phys.~05 (2011)
014; Y. Urakawa, Prog. Theor. Phys. {\bf 126}, 961 (2011); S. B. Giddings and M. S. Sloth,
J.~Cosmol. Astropart. Phys.~07 (2010) 015; J.~Cosmol. Astropart. Phys.~01 (2011) 023; Phys. Rev.
D {\bf84}, 063528 (2011); Phys. Rev. D {\bf 86}, 083538 (2012); C.
P. Burgess, R. Holman, L. Leblond and S. Shandera, J.~Cosmol. Astropart. Phys.~03 (2010) 033;
D. Boyanovsky, Phys. Rev. D {\bf 85}, 123525 (2012); Phys. Rev. D {\bf 86}, 023509 (2012); K. Feng, Y.-F.
Cai and Y.-S. Piao, Phys. Rev. D {\bf 86}, 103515 (2012);
E. T. Akhmedov, Int. J. Mod. Phys. D {\bf 23}, 1430001 (2014); K. Larjo and D. Lowe, Phys. Rev. D {\bf 87},
083506 (2013); J. Serreau and R. Parentani, Phys. Rev. D {\bf 87},
085012 (2013); J. Serreau, Phys. Lett. B {\bf 728}, 380 (2014); L. Lello, D. Boyanovsky and
R. Holman, Phys. Rev. D {\bf 89}, 063533 (2014); M. Herranen, T. Markkanen and A.
Tranberg, J. High Energy Phys. 05 (2014) 026; C. Armendariz-Picon and G. \c{S}eng\"{o}r, J.~Cosmol. Astropart. Phys.~11 (2016) 016; P. Adshead, C. P. Burgess, R. Holman and S. Shandera, arXiv:1708.07443 [hep-th]; E. T. Akhmedov and F. Bascone, arXiv:1710.06118 [hep-th].

\bibitem{TT} J. Tokuda and T. Tokuda, arXiv:1708.01734 [gr-qc].

\bibitem{AMPP} E. T. Akhmedov, U. Moschella, K. E. Pavlenko and F. K. Popov, Phys. Rev. D {\bf96}, 025002 (2017).

\bibitem{MR} I. Moss and G. Rigopoulos, J.~Cosmol. Astropart. Phys.~05 (2017) 009.

\bibitem{R1} G. Rigopoulos, J.~Cosmol. Astropart. Phys.~07 (2016) 035.

\bibitem{DB1} D. Boyanovsky, Phys. Rev. D {\bf93}, 043501 (2016); New J. Phys. {\bf17} 063017 (2015).

\bibitem{CW1} C. Wetterich, J.~Cosmol. Astropart. Phys.~05 (2016) 041; Phys. Rev. D {\bf92}, 083507 (2015).

\bibitem{CWX} X. Chen, Y. Wang, Z.-Z. Xianyu, J. High Energy Phys. 08 (2016) 051.

\bibitem{AFNVW} H. Assadullahi, H. Firouzjahi, M. Noorbala, V. Vennin and D. Wands, J.~Cosmol. Astropart. Phys.~06 (2016) 043.

\bibitem{GS1} F. Gautier and J. Serreau, Phys. Rev. D {\bf92}, 105035 (2015).

\bibitem{DB2} D. Boyanovsky, Phys. Rev. D {\bf92}, 023527 (2015).

\bibitem{GS2} M. Guilleux and J. Serreau, Phys. Rev. D {\bf92}, 084010 (2015).

\bibitem{JSS} R. K. Jain, M. Sandora and M. S. Sloth, J.~Cosmol. Astropart. Phys.~06 (2015) 016.

\bibitem{O2}  V. K. Onemli, arXiv:1510.02272 [gr-qc].

\bibitem{DB3} D. Boyanovsky, Phys. Rev. D {\bf93}, 083507 (2016).

\bibitem{OW1} V. K. Onemli and R. P. Woodard, Classical Quantum Gravity {\bf 19}, 4607 (2002).

\bibitem{OW2} V. K. Onemli and R. P. Woodard, Phys. Rev. D {\bf70}, 107301 (2004).

\bibitem{BOW} T. Brunier, V. K. Onemli and R. P. Woodard, Classical Quantum Gravity {\bf 22}, 59 (2005).

\bibitem{KO} E. O. Kahya and V. K. Onemli, Phys. Rev. D {\bf 76}, 043512 (2007).

\bibitem{KOW1} E. O. Kahya, V. K. Onemli and R. P. Woodard, Phys. Rev. D {\bf 81}, 023508 (2010).

\bibitem{KOW2} E. O. Kahya, V. K. Onemli and R. P. Woodard, Phys. Lett. B {\bf 694}, 101 (2010).

\bibitem{O1}  V. K. Onemli, Phys. Rev. D {\bf 89}, 083537 (2014).

\bibitem{W0} N.~C.~Tsamis, A.~Tzetzias and R.~P.~Woodard, J.~Cosmol. Astropart. Phys.~09 (2010) 016.

\bibitem{Tegmark} M.~Tegmark, Phys. Rev.~D~{\bf 85}, 123517 (2012).




\end{thebibliography}
\end{document}